\documentclass[10pt,twocolumn,letterpaper]{article}

\usepackage{cvpr}              %

\usepackage[dvipsnames]{xcolor}

\usepackage{amsmath,amssymb,amsfonts}
\usepackage{algorithmic}
\usepackage{graphicx}
\usepackage{textcomp}
\usepackage{xcolor}

\def\BibTeX{{\rm B\kern-.05em{\sc i\kern-.025em b}\kern-.08em
    T\kern-.1667em\lower.7ex\hbox{E}\kern-.125emX}}
    
\usepackage[font=small]{caption}
\usepackage{xspace}
\usepackage{microtype}
\usepackage{bm}

\usepackage{multicol}
\usepackage{tabularx}
\usepackage{tabularray}
\usepackage{booktabs}

\usepackage{pifont} %

\newcommand{\encoder}{\ensuremath{\mathcal{E}}\xspace}  %
\newcommand{\decoder}{\ensuremath{D}\xspace}  %
\newcommand{\denoising}{\ensuremath{u}\xspace} %

\newcommand{\model}{\ensuremath{\Theta}\xspace}

\newcommand{\x}{\ensuremath{x}\xspace}  %

\newcommand{\z}{\ensuremath{z}\xspace}  %

\newcommand{\zprop}{\hat{\z}_0^t}

\newcommand{\zzero}{\ensuremath{\z_0}\xspace}  %
\newcommand{\zT}{\ensuremath{\z_T}\xspace}  %

\newcommand{\condemb}{\ensuremath{C}\xspace}  %

\newcommand{\gen}[1]{\ensuremath{\mathcal{G}_{T \rightarrow 0}(#1; \denoising)}\xspace}
\newcommand{\inv}[1]{\ensuremath{\mathcal{I}_{0 \rightarrow T}(#1; \denoising)}\xspace}

\newcommand{\zwat}[1]{\ensuremath{\z_{#1}^{(w)}}\xspace}  %
\newcommand{\xwat}{\ensuremath{\x^{(w)}\xspace}} %

\newcommand{\zwathat}[1]{\ensuremath{\hat{\z}_{#1}^{(w)}}\xspace}  %

\newcommand{\attackermodel}{\ensuremath{\model_A}\xspace}
\newcommand{\attackerencoder}{\ensuremath{\mathcal{E}_A}\xspace}  %
\newcommand{\attackerdecoder}{\ensuremath{D_A}\xspace}  %
\newcommand{\attackerdenoising}{\ensuremath{u_A}\xspace} %

\newcommand{\zhatatt}[1]{\ensuremath{\tilde{z}_{#1}^{(w)}}\xspace}

\newcommand{\invproxyattacker}[1]{\ensuremath{\mathcal{I}_{0 \rightarrow T}(#1; \attackerdenoising)}\xspace}

\newcommand{\xcover}{\ensuremath{x^{(c)}}\xspace} 
\newcommand{\zcover}[1]{\ensuremath{\tilde{\z}^{(c)}_{#1}}\xspace} 

\newcommand{\xout}{\ensuremath{\check{x}^{(c)}}\xspace}  
 
\newcommand{\xoutreprompt}{\ensuremath{\check{x}}\xspace}

\newcommand{\mess}{\ensuremath{s}\xspace} %
\newcommand{\messrecov}{\ensuremath{s'}\xspace} %
\newcommand{\rep}{\ensuremath{\rho}\xspace} %

\newcommand{\imprint}{imprint\xspace}
\newcommand{\imprints}{imprints\xspace}
\newcommand{\imprinting}{imprinting\xspace}
\newcommand{\Imprint}{Imprint\xspace}
\newcommand{\Imprinting}{Imprinting\xspace}

\newcommand{\ImprintForgeLong}{Imprint-Forgery\xspace}
\newcommand{\ImprintForgeShort}{Imprint-F\xspace}

\newcommand{\ImprintRemovalLong}{Imprint-Removal\xspace}
\newcommand{\ImprintRemovalShort}{Imprint-R\xspace}

\newcommand{\reprompt}{reprompt\xspace}
\newcommand{\reprompting}{reprompting\xspace}

\newcommand{\Reprompt}{Reprompt\xspace}
\newcommand{\RepromptShort}{Reprompt\xspace}
\newcommand{\Reprompting}{Reprompting\xspace}

\newcommand{\RepromptPlus}{Reprompt$+$\xspace}
\newcommand{\RepromptPlusShort}{Reprompt$+$\xspace}
\newcommand{\RepromptingPlus}{Reprompt$+$\xspace}

\usepackage{lipsum} %
\usepackage{comment}
\usepackage{placeins}
\usepackage{graphicx}
\usepackage{adjustbox}
\usepackage{subcaption}
\usepackage{booktabs}
\usepackage{array}
\usepackage{multirow}
\usepackage{colortbl}
\usepackage{multicol}

\usepackage[accsupp]{axessibility}  %

\usepackage{xcolor}  %
\definecolor{lightgrey}{rgb}{0.8, 0.8, 0.8} %

\usepackage{siunitx}
\usepackage{titletoc}
\usepackage{bm}

\newlist{paranumeration}{enumerate}{3}
\setlist[paranumeration]{label=\bfseries\Alph*.,wide=0em,itemsep=0em,topsep=0pt}

\renewcommand{\paragraph}[1]{\vspace{0.5em}\noindent\textbf{#1}}

\definecolor{cvprblue}{rgb}{0.21,0.49,0.74}
\usepackage[pagebackref,breaklinks,colorlinks,citecolor=cvprblue]{hyperref}

\title{Black-Box Forgery Attacks on\\Semantic Watermarks for Diffusion Models}

\author{Andreas Müller,
	Denis Lukovnikov,
	Jonas Thietke,
	Asja Fischer\thanks{Equal supervision}~,~%
	Erwin Quiring\footnotemark[1]\vspace{0.3em}\\
	Ruhr University Bochum\\
    {\tt\small \{andreas.mueller-t1x, denis.lukovnikov, jonas.thietke, asja.fischer, erwin.quiring\}@rub.de}
}

\begin{document}
\maketitle

\begin{abstract}
    Integrating watermarking into the generation process of latent diffusion models (LDMs) simplifies detection and attribution of generated content. Semantic watermarks, such as Tree-Rings and Gaussian Shading, represent a novel class of watermarking techniques that are easy to implement and highly robust against various perturbations. However, our work demonstrates a fundamental security vulnerability of semantic watermarks. We show that attackers can leverage unrelated models, even with different latent spaces and architectures (UNet vs DiT), to perform powerful and realistic forgery attacks. Specifically, we design two watermark forgery attacks. The first \imprints a targeted watermark into real images by manipulating the latent representation of an arbitrary image in an unrelated LDM to get closer to the latent representation of a watermarked image. We also show that this technique can be used for watermark removal. The second attack generates new images with the target watermark by inverting a watermarked image and re-generating it with an arbitrary prompt. Both attacks just need a single reference image with the target watermark. Overall, our findings question the applicability of semantic watermarks by revealing that attackers can easily forge or remove these watermarks under realistic conditions.
    
    Github: \href{https://github.com/and-mill/semantic-forgery}{https://github.com/and-mill/semantic-forgery}
\end{abstract}

\section{Introduction}

\begin{figure}
    \centering
    \includegraphics[width=0.92\linewidth]{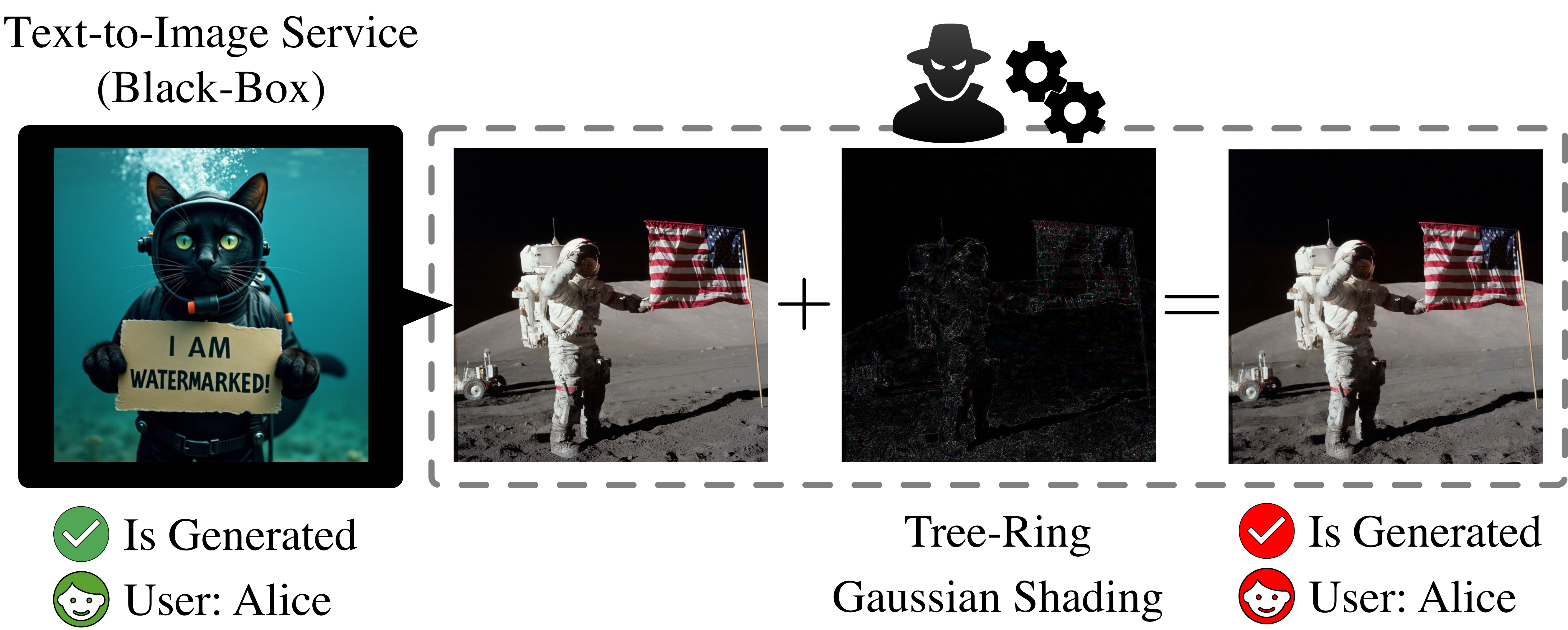}
    \vspace{-0.5em}
    \caption{Semantic watermark forgery. The attacker can transfer the watermark from a watermarked reference image requested by Alice (here: the diving cat) %
    into any cover image (here: the moon landing). 
    The obtained image will be detected as watermarked and attributed to Alice by the service provider, eroding the trust in watermark-based detection and attribution of AI-generated content.
    }
    \label{fig:diagram:frontpage}
\end{figure}

Recent advancements in generative AI, 
especially latent diffusion models (LDMs), 
have made it extremely challenging to distinguish between authentic and AI-generated images. 
Despite a multitude of beneficial applications, this also introduces a considerable range of threats.
For example, so-called deepfakes---compellingly realistic but AI-generated media---are now routinely used to defraud people, distribute misinformation, and to manipulate public opinion~\cite{goldsteinHowDisinformationEvolved2021,Europol2024}. 

Thus, there is an urgent \emph{need to detect and to attribute AI-generated images}, that is, to identify that an image is AI-generated and to determine who used the AI system to generate it.
Watermarking is currently one of the main approaches for achieving these goals, and leading AI companies such as Google and Meta, as well as AI platforms such as Hugging Face and Google Gemini, are either already using it or have made commitments to do so~\cite{techcompaniespledgewatermark2023, cleggLabelingAIgeneratedImages2024, HuggingFaceWatermarking, GoogleSynthID}.

Recently, a novel family of watermarking methods for LDMs has been proposed that relies on the inversion of the denoising process in the diffusion model~\cite{Yang2024GaussianShading, CiYanSon24RingID, Wen2023TreeRing, Gunn2024Undetectable}.  
These \emph{semantic watermarks} work by modifying the initial latent noise to contain a specific watermark pattern which can be recovered through inversion of the denoising process.
Semantic watermarks offer several advantages. First, they assert greater robustness against various image transformations and attacks compared to previous watermarking methods. Moreover, the DM remains unchanged, which makes the watermarking approach easy to deploy. 
However, it is critical that these watermarks are also secure against forgery attacks.
\cref{fig:diagram:frontpage} illustrates this threat where an attacker transfers the watermark of a service provider to any image.
This can lead to an erosion of trust in the watermarking system, as real images are flagged as watermarked and thus AI-generated or regular users are blamed for distributing harmful content.

In this work, we uncover a fundamental security weakness of semantic watermarks.
We demonstrate that highly effective attacks only require an unrelated model, even with a different latent space and architecture, and a single reference image with the target watermark.
Specifically, we propose two forgery strategies:
(1) Our \textit{\Imprinting attack} operates in the latent space of the attacker model where it reduces the distance between the latent representation of a watermarked and a clean cover image. The reduced distance also transfers to the latent space of the target model, effectively carrying over the watermark. 
(2) Our \textit{\Reprompting attack} creates arbitrary images with the target watermark by inverting a watermarked image and re-generating it with another prompt.  
(3) In addition, we introduce a \textit{removal strategy} where we adjust the imprinting method for watermark removal. 

We perform an extensive empirical analysis on different diffusion models to demonstrate effectiveness of our proposed attacks on the two primary semantic watermarking approaches, Tree-Ring~\cite{Wen2023TreeRing} and Gaussian Shading~\cite{Yang2024GaussianShading}. 
A robustness and transferability analysis further shows that adjusting the detection threshold is ineffective as defense.

\vspace{0.4em}\noindent In summary, we make the following contributions:
\begin{itemize}
    \vspace{0.2em}
    \setlength{\itemsep}{0.4em} %
    \item \textbf{Black-Box Attacks.} We propose two novel watermark forgery attacks that can imitate a watermark from a single watermarked reference image by only using a model that is unrelated to the watermarked model. We further introduce  an adaption of the approach for watermark removal.
    \item \textbf{Extensive Evaluation.} We empirically show that our attacks work against two primary semantic watermarks on a range of target models, including SDXL and FLUX.1. We compare our attack with previously proposed attacks. 
    \item \textbf{Analysis.} A thorough robustness and transferability analysis shows that simple defenses are ineffective.
\end{itemize}

\section{Background}
\label{sec:background}
We start by introducing our notation and revisiting diffusion models and watermarking techniques.

\subsection{Diffusion Models and Inverse DDIM}
Diffusion models (DM)~\cite{HoJaiAbb20,SohWeiMah15,SonMenErm21,SonErm19,SonSohKin21} are a class of probabilistic generative models that learn a transport map between some known distribution (e.g.~a standard Gaussian) and the target data distribution as the reversal of a diffusion process.
During training, DMs learn a denoising function $\denoising(\cdot)$ that takes a noisy image $\z_t$ from an arbitrary diffusion step $t \in \{0,\dots,T\}$  and tries to predict the noise-free version $\zprop$, which is then passed to the samplers of the diffusion process to compute a slightly less noisy version of the image, $\z_{t-1}$.
Various samplers have been proposed with the goal to increase sampling speed~\cite{HoJaiAbb20,LuZhoBao22,SalHo22,LiuRenLin22,XiaKreVah22,ZhaChe23,ZhaBaiRao23}, and DDIM~\cite{HoJaiAbb20} is one of the most common and straightforward sampling methods in use.

Currently, most DMs are learned in a lower-resolution latent space, which enables higher computational efficiency and leads to the name LDM.  Formally, given an image $\x$, an LDM uses an encoder \encoder  to project $\x$ to the latent space, i.e., $\zzero = \encoder (\x)$, and a decoder \decoder to project the latent presentation back to pixel space, i.e., $\x' = \decoder (\zzero)$.
The DDIM sampler samples along a deterministic trajectory defined by the following denoising steps:
\begin{align}
\zprop &= \dfrac{\z_t - \sqrt{1 - \alpha_t} \denoising (\z_t, t, \condemb)}{\alpha_t} \enspace , \\
\z_{t-1} &= \sqrt{\alpha_{t-1}} \zprop + \sqrt{1 - \alpha_{t-1}} \, \denoising (\z_t, t, \condemb) \enspace ,
\end{align}
where $\alpha_t = \prod^t_{i=0} ( 1 - \beta_t)$ and $\beta_t$ define the noise schedule. \condemb is the conditioning information (e.g. textual embeddings).
During generation, $\zT$ is typically sampled from $\mathcal{N}(\textbf{O}, \textbf{I})$.
We denote the full denoising process of the trained model mapping $\zT$ back into the latent space of the auto-encoder as 
$\gen{\zT}$.
The full generative model is denoted as \model~=~(\encoder, \denoising, \decoder). 

Inverse DDIM sampling~\cite{MokHerAbe23}  tries to follow the trajectory that (would have) generated a given image in reverse.
It starts from a  latent image $\zzero$, and  adds noise in a step-wise fashion, where the $t$-th step is described by %
\begin{align}
\z_{t+1} &= \sqrt{\alpha_{t+1}} \zprop + \sqrt{1 - \alpha_{t+1}} \, \denoising (\z_t, t, \condemb) \enspace ,
\end{align}
simply inverting the denoising steps.
Note that this is different from the forward diffusion process, where random noise is added to the image.
Instead, this sampler tries to follow the trajectory that generated the given image in reverse, which can be done even without knowing %
$\condemb$ \cite{MokHerAbe23}.
We denote the inverse sampling process as $\inv{\zzero}$.

\subsection{Watermarking}
The powerful generation capabilities of diffusion models have also created an urgent need for techniques to identify and attribute images produced by these models~\cite{techcompaniespledgewatermark2023,Biden2023ExecutiveOrderAI,EU2024AIAct}.
In the following, we first study the use case of watermarking for generative AI and then examine watermarking approaches.

\paragraph{Application.}
Watermarking supports a safe and trustworthy usage of AI-generated content by enabling (1) detection and (2) attribution of generated content. 
\textbf{Detection} means determining whether a certain image was generated by the service provider (SP). The watermark allows the SP to claim copyright or to mark an image as AI-generated, which can be leveraged for deepfake detection.
In \textbf{attribution}, the goal is to identify the user who generated a certain image using the SP.
Watermarking enables the SP to identify who is violating the copyright or is distributing NSFW content.

\paragraph{Watermarking methods.}
Various post-processing watermarking techniques for images have been proposed~\cite{CoxMilBlo07,al2007combined,Zhu2018hidden,tancik2020stegastamp,rivagan,lu2024}, that are applied in a post-hoc manner on existing images.
Recently, watermarking methods have been developed specifically for use in large-scale text-to-image diffusion models~\cite{Fernandez2023Stable, CiSonYan2024wmadapter, Yang2024GaussianShading, Wen2023TreeRing, CiYanSon24RingID}.
For example, it is possible to fine-tune the decoder of an LDM to always produce watermarked images from the latents~\cite{Fernandez2023Stable, XioQinFen23Flexible, CiSonYan2024wmadapter}.

\begin{figure}
    \centering
    \includegraphics[width=0.85\linewidth]{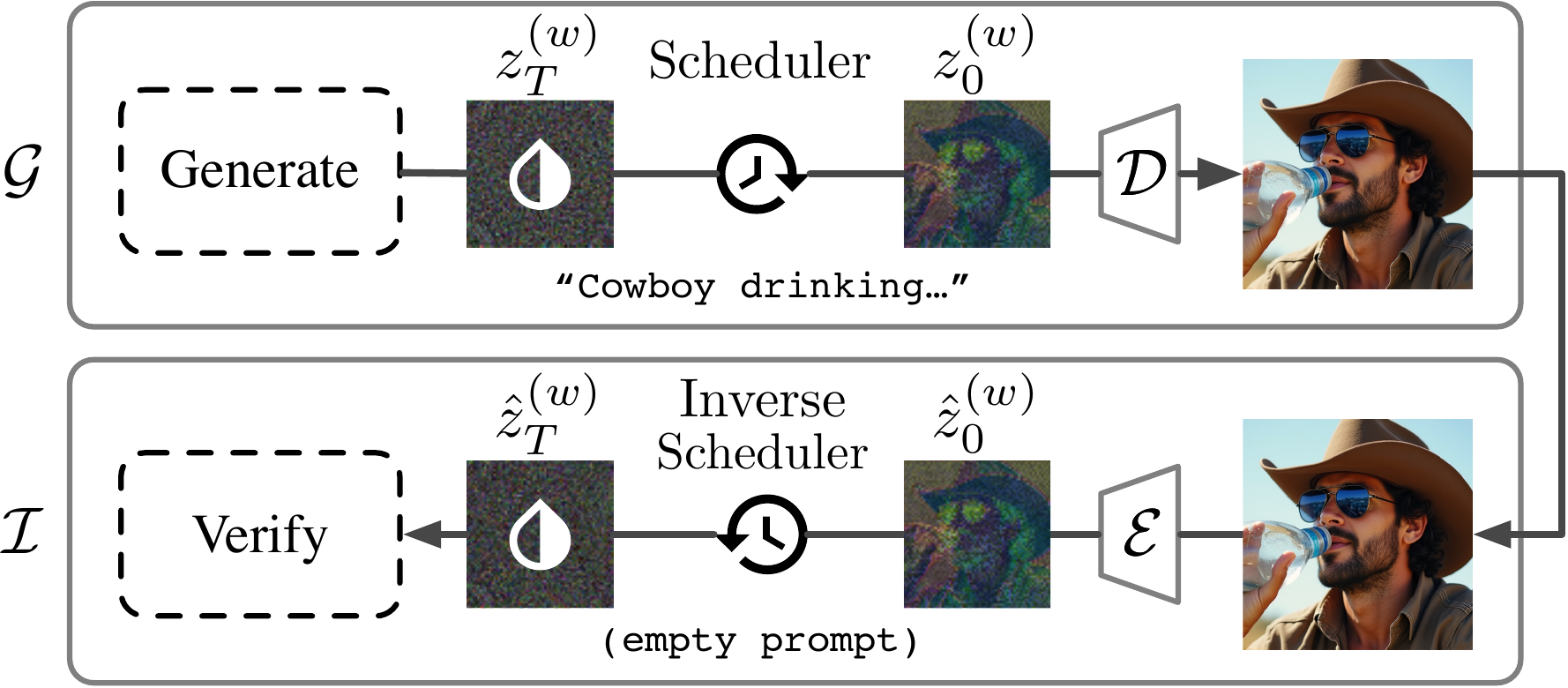}
    \vspace{-0.5em}
    \caption{Concept of Semantic Watermarking. The initial latent contains a decodable watermark pattern, which can be reconstructed after inversion.}
    \label{fig:diagram:basic}
\end{figure}

\paragraph{Semantic Watermarking.}
\label{sec:watermarking-schemes}
In this work, we focus on a specific family of watermarking methods for diffusion models that rely on the inversion of the denoising process in the diffusion model~\cite{Wen2023TreeRing, Yang2024GaussianShading, CiYanSon24RingID, Gunn2024Undetectable}.
They are easy to integrate, do not require additional training, and claim to offer significantly higher robustness to various image perturbations and targeted attacks, that earlier post-hoc-style methods (like Stable Signature~\cite{Fernandez2023Stable}) are vulnerable to.

These inversion-based watermarking methods operate by modifying the initial latent $\zT$ to have a certain structure that is recoverable using inversion and subsequently verifiable.
Because only the initial latent $\zT$ is modified, the effect of the watermark is realized as higher-level variations in the final image (as opposed to imperceptible noise patterns from post-hoc-style methods).
These techniques thus drop the imperceptibility constraint from post-hoc-style methods and exploit the fact that there is a myriad of different ways to generate an image that still conforms to user specifications (e.g.~the prompt).
Because the watermarks are realized through higher-level image variations (such as specific details of objects and their exact arrangement), we refer to them as \textit{semantic watermarks}~\cite{CiYanSon24RingID}.

\cref{fig:diagram:basic} illustrates their general technical realization.
During image generation, first a watermarked latent~$\zwat{T}$ is prepared and the image is generated as usual: $\xwat = \decoder (\mathcal{G}_{T \rightarrow 0}(\zwat{T};\denoising) )$.
In order to verify a watermark, the image is inverted ($\zwathat{T} = \mathcal{I}_{0 \rightarrow T}(\encoder (\xwat); \denoising)$) and the watermark is extracted from $\zwathat{T}$.
The different watermarking methods differ in how $\zwat{T}$ is generated and verified.
\textbf{Tree-Ring}~\cite{Wen2023TreeRing} embeds circular patterns into the frequency representation of $\zwat{T}$ and verifies the pattern by checking if the frequency representation of $\zwathat{T}$ is close enough to the pattern. 
This method works in the previously described detection scenario of watermarking.
\textbf{Gaussian Shading}~\cite{Yang2024GaussianShading} takes a message bit string~$\mess$, encrypts it, and uses it to select which bins of $\mathcal{N}(0, I)$ to sample the entries of $\zwat{T}$ from. The process is inverted during verification and the recovered bit string is compared to registered bit strings.
Gaussian Shading allows for both detection and attribution.
We describe both methods in more detail in Sec.~\ref{sec:appendix:semantic-watermarks} in the Supplementary Material.
Both methods have been further improved by follow-up works~\cite{CiYanSon24RingID, Gunn2024Undetectable}.  
As the main principle relevant to our attack remains the same, we focus on the two primary approaches Tree-Ring and Gaussian Shading.

\section{Black-Box Attacks on Semantic Watermarks}
\label{sec:approach}
\begin{figure}[t!]
    \centering
    \includegraphics[width=\linewidth]{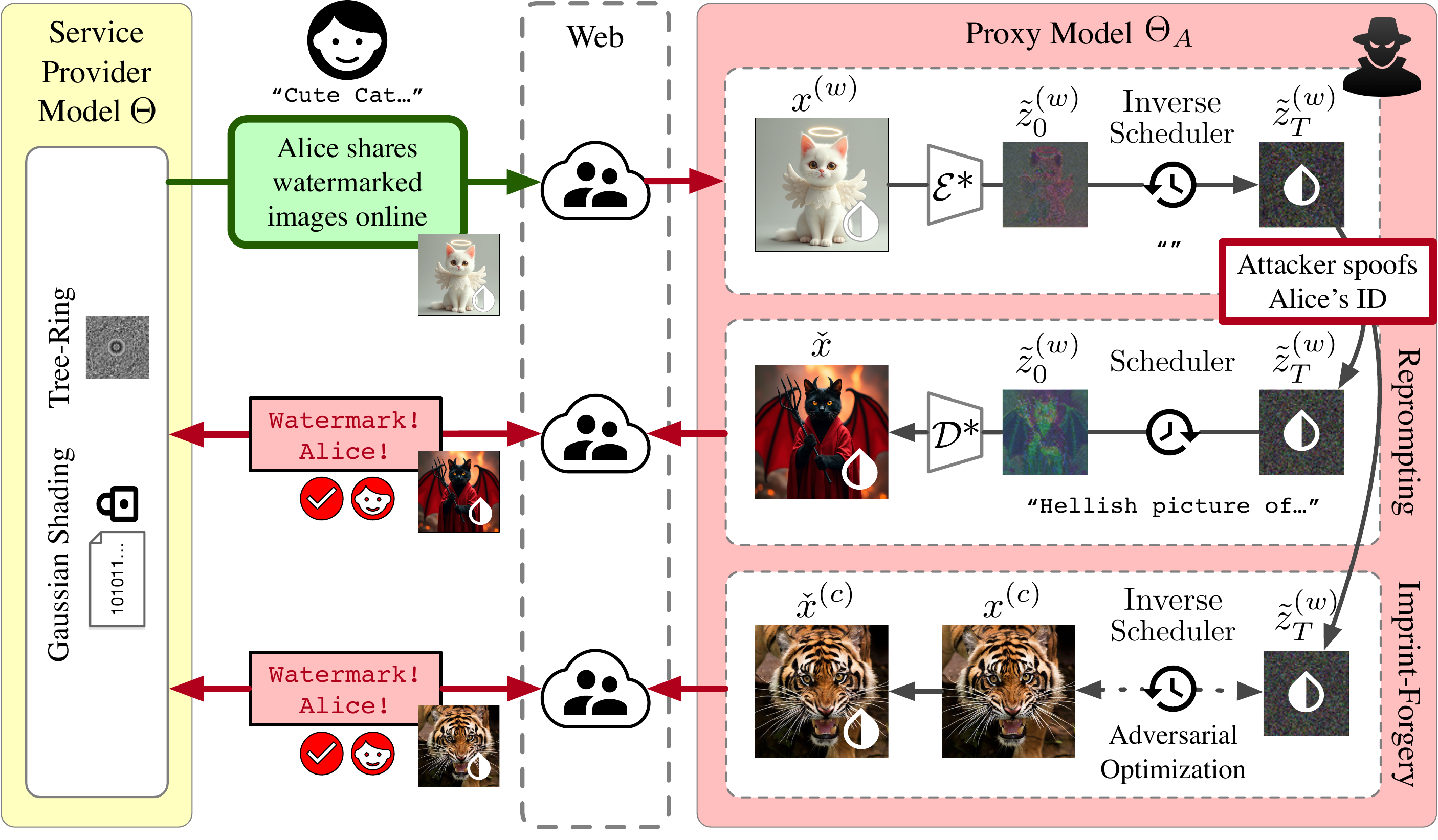}
    \caption{Illustration of our watermark forgery attacks. Using a proxy model~\attackermodel, the attacker first computes $\zhatatt{0}$ and $\zhatatt{T}$ (1st row). 
    The \Reprompting attack (2nd row) takes $\zhatatt{T}$ and generates a new image from another prompt.
    Alternatively, the \ImprintForgeLong attack (3rd row) takes an unwatermarked cover image $\xcover$ and finds a modification so that its inverted initial latent noise is similar to $\zhatatt{T}$.  
    The watermark verification on the target model~\model detects the original watermark in the attack images, even though these images were computed with the proxy model~\attackermodel.
    }
    \label{fig:forgery-attacks-principle}
\end{figure}

Semantic watermarks require keeping the original generative model~\model secret---with the assumption that this prevents an attacker from forging (and removing) the watermark.
In this section, however, we propose attacks that utilize a proxy model~$\attackermodel = (\attackerencoder, \attackerdenoising, \attackerdecoder)$ for watermark forgery (and removal). 
In the following, we describe two forgery strategies and also discuss a modification for watermark removal. 
\cref{fig:forgery-attacks-principle} illustrates both forgery attacks. %

\subsection{\ImprintForgeLong Attack}
\label{sec:attack2}
Our first attack takes a clean cover image (real or generated) and slightly modifies it such that it is verified as watermarked by the SP.
The procedure is as follows:

\begin{paranumeration}[label=\bfseries(\Roman*)]
\item \label{imprint:one} The attacker takes a watermarked target image $\xwat$ that another user has generated using the SP's model~\model. 

\item \label{imprint:two} The attacker uses the encoder~\attackerencoder of the proxy model to map the watermarked target image to the latent space of the attacker's auto-encoder: $\zhatatt{0} = \attackerencoder(\xwat)$.
Next, the attacker estimates the latent noise  $\zhatatt{T}$ by running the inverse DDIM sampler $\invproxyattacker{\zhatatt{0}}$ of the proxy model. 

\item \label{imprint:three} 
For a given clean cover image $\xcover$, the attacker also computes its latent representation $\zcover{0} = \attackerencoder(\xcover)$. 
Next, the attacker finds a difference vector $\delta$ %
that decreases the Euclidean distance between $\zhatatt{T}$ and the inversion of $\zcover{0} + \delta$ by minimizing the following loss function: %
\begin{align}
    \mathcal{L}_{\mathrm{forgery}}(\delta) &= | \mathcal{I}_{0 \rightarrow T} (\zcover{0} + \delta; \attackerdenoising) - \zhatatt{T} |_2  \enspace. 
      \label{eq:attack2:loss}
\end{align}
This loss is minimized by applying up to $N$ gradient descent updates. Due to limited GPU memory, we apply gradient checkpointing to backpropagate through the entire inverse DDIM sampler. We stop when the desired watermark detection accuracy is reached.
The final image $\xout$ containing the stolen watermark is obtained by decoding the obtained $(\zcover{0} + \delta)$ back to the pixel space %
using \attackerdecoder.

\item \label{imprint:four} When $\xout$ is sent to the SP's watermark decoding API, 
it gets verified as generated by the service or attributed to the user from whom $\xwat$ was collected.
\end{paranumeration}

\vspace{0.5em}
\noindent Note that the inversion (Step \ref{imprint:two}) and the \imprinting (Step~\ref{imprint:three}) take place in the latent space of the proxy model~\attackermodel, and do not require the target model.
Still, this allows us to forge the watermark from the target model, as demonstrated by our experiments in Sec.~\ref{sec:evaluation}.

\begin{figure}
    \centering
    \includegraphics[width=1.0\linewidth]{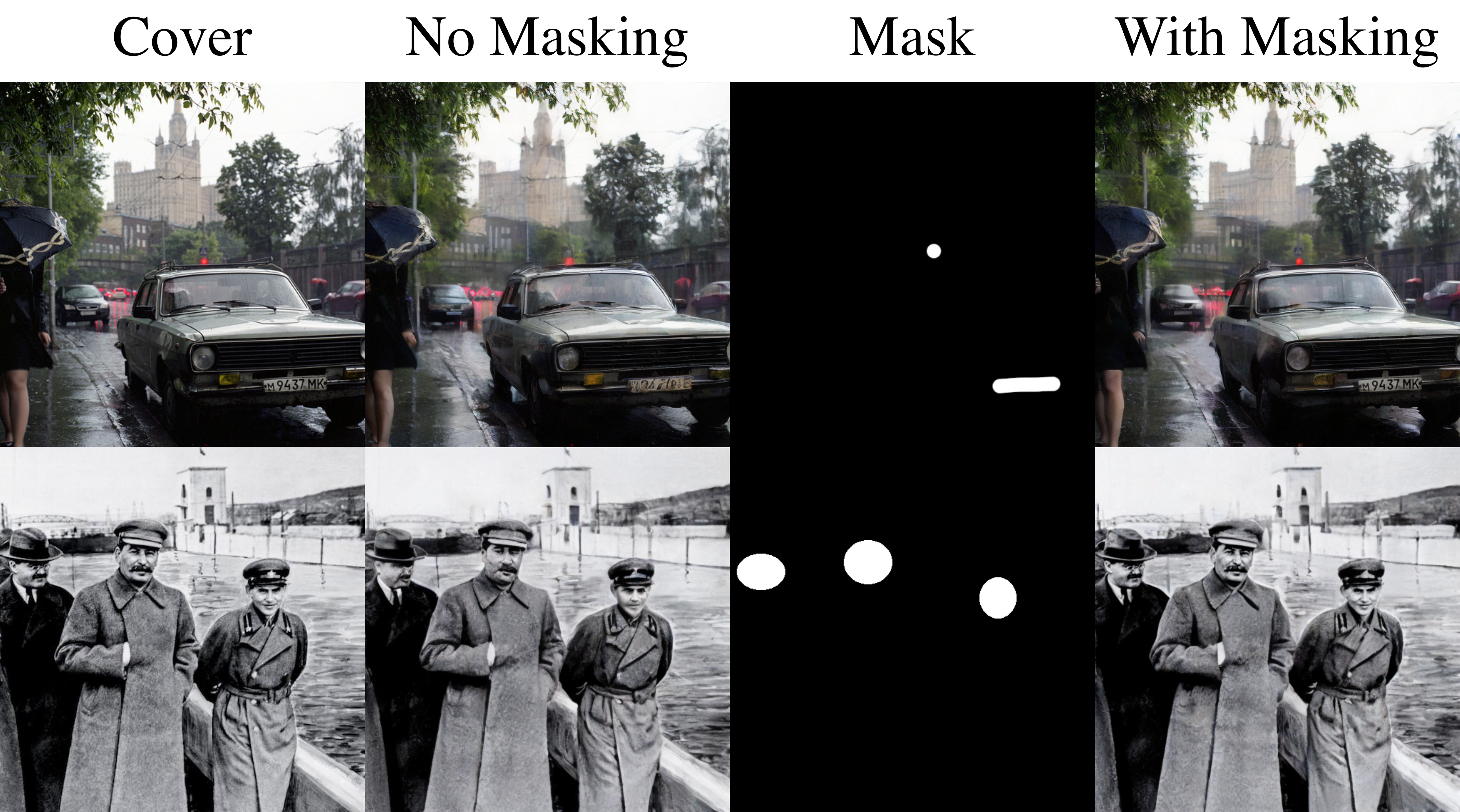}
    \caption{\ImprintForgeLong attack with and without masking. %
    Masking is a way to preserve crucial details that are distorted by the auto-encoder, such as text and faces.
    The attacks are performed using SD2.1 as proxy model, against an image of size $512\times512$ generated by FLUX.1 with Gaussian Shading watermarking.}
    \label{fig:diagram:masking}
\end{figure}

\paragraph{Extension: Masking to preserve important details.} 
For small image sizes, certain objects such as text and human faces are susceptible to small changes in the latent variables. Simply encoding and decoding a real image with common auto-encoders already leads to noticeable distortions.
In order to preserve crucial details like faces and text, we introduce a simple, yet effective extension of the procedure described above. This extension is illustrated in~\cref{fig:diagram:masking}.
Each gradient update on $\delta$ uses a downsampled mask $m_z$ to keep certain parts of the latent the same, i.e., the $\delta$  resulting from the gradient update is replaced by $\delta * (1 - m_z)$. 
Additionally, after optimization and decoding, we paste the original pixels from $\xcover$ corresponding to the masked region to the output of the \imprinting procedure $\xout$ to prevent distortion introduced by the decoder \attackerdecoder.

\subsection{\ImprintRemovalLong Attack}
\label{sec:removal}
It is also possible to remove watermarks using the same methodology.
In the removal set-up, instead of a cover image, we have a watermarked image $\xwat$ that we aim to modify so that it no longer contains its original watermark.
To accomplish this, we simply change the target $\zhatatt{T}$ in \cref{eq:attack2:loss} to its negation $-\zhatatt{T}$.
This leads to the following loss, which encourages erasing the watermark pattern encoded in the initial latent noise of the watermarked image: 
\begin{align}
    \mathcal{L}_{\mathrm{removal}}(\delta) &= | \mathcal{I}_{0 \rightarrow T} (\zhatatt{0} + \delta; \attackerdenoising) + \zhatatt{T} |_2  \enspace  .
      \label{eq:removal:loss}
\end{align}

\noindent Images obtained through this procedure should no longer be verified as watermarked.

\subsection{\Reprompting Attack}
\label{sec:attack}
Our second attack strategy aims at generating new (potentially harmful) images that are verified to have the target watermark. 
To this end, we replace Step~\ref{imprint:three} of the previously described \ImprintForgeLong attack: Starting from the extracted $\zhatatt{T}$, the attacker just generates a different image $\xoutreprompt$ with the desired target prompt using the proxy model $\attackermodel$. 

\paragraph{Multiple Prompts and Bin Resampling.}
To improve the attack effectiveness, we propose two simple, yet effective augmentations. First, the attacker can try multiple prompts (all of which may prompt for harmful content). Second, for Gaussian Shading, the attacker can also resample $\zhatatt{T}$ such that the values in $\zhatatt{T}$ remain in the same bin. In the default setup of Gaussian Shading, it means each value must retain its sign. This procedure preserves the encoded watermark in $\zhatatt{T}$, but provides a novel seed for the diffusion process.

\section{Evaluation}
\label{sec:evaluation}
We proceed with an empirical evaluation to demonstrate the effectiveness of the proposed attacks.

\begin{figure*}[t]
\centering
\begin{minipage}[t]{0.550\textwidth}
   
        \centering
        \resizebox{1.0\linewidth}{!}{%
        \begin{tabular}{@{}lllrrrrrrr@{}}
\toprule
& & & \multicolumn{4}{c}{Gaussian Shading} & \multicolumn{3}{c}{Tree-Ring} \\
\cmidrule(lr){4-7} \cmidrule(lr){8-10}
& & & \multicolumn{4}{c}{\small (FPR=$10^{-6}$)} & \multicolumn{3}{c}{\small (FPR=$1\%$)} \\
Model & Attack & Step & Det. & Attr. & PSNR & LPIPS & Det. & PSNR & LPIPS \\ 
\midrule
\multirow{5}{*}{\parbox{1px}{SD2.1-Anime}} 
& \multirow{4}{*}{\ImprintForgeShort} 
  & 20 & 1.00 & 1.00 & 23.58 & 0.116 & 0.94 & 23.59 & 0.116 \\
& & 50 & 1.00 & 1.00 & 22.13 & 0.160 & 1.00 & 22.16 & 0.160 \\
& & 100 & 1.00 & 1.00 & 20.90 & 0.208 & 1.00 & 20.95 & 0.209 \\
& & 150 & 1.00 & 1.00 & 20.11 & 0.246 & 1.00 & 20.16 & 0.248 \\
\cline{2-10} \noalign{\vskip 0.2em}
& Avg-5000 & - & 0.00 & 0.00 & 32.78 & 0.010 & 1.00 & 29.27 & 0.021 \\
\midrule
\multirow{5}{*}{SDXL}                                          
& \multirow{4}{*}{\ImprintForgeShort} 
  & 20 & 0.99 & 0.99 & 23.27 & 0.125 & 0.43 & 23.55 & 0.117 \\
& & 50 & 1.00 & 1.00 & 21.86 & 0.168 & 0.72 & 22.09 & 0.161 \\
& & 100 & 1.00 & 1.00 & 20.51 & 0.219 & 0.95 & 20.86 & 0.207 \\
& & 150 & 1.00 & 1.00 & 19.77 & 0.253 & 0.99 & 20.09 & 0.242 \\
\cline{2-10} \noalign{\vskip 0.2em}
& Avg-5000 & - & 0.00 & 0.00 & 24.48 & 0.014 & 1.00 & 22.70 & 0.040 \\
\midrule
\multirow{5}{*}{\parbox{1px}{PixArt-$\Sigma$}}                               
& \multirow{4}{*}{\ImprintForgeShort} 
  & 20 & 0.72 & 0.72 & 23.57 & 0.117 & 0.29 & 23.59 & 0.116 \\
& & 50 & 0.91 & 0.91 & 22.12 & 0.161 & 0.55 & 22.15 & 0.160 \\
& & 100 & 0.94 & 0.94 & 20.90 & 0.209 & 0.78 & 20.93 & 0.207 \\
& & 150 & 0.96 & 0.96 & 20.12 & 0.246 & 0.84 & 20.15 & 0.245 \\
\cline{2-10} \noalign{\vskip 0.2em}
& Avg-5000 & - & 0.00 & 0.00 & 16.59 & 0.091 & 0.93 & 16.73 & 0.092 \\
\midrule
\multirow{5}{*}{FLUX.1}                                         
& \multirow{4}{*}{\ImprintForgeShort} 
  & 20 & 0.26 & 0.26 & 23.54 & 0.117 & 0.07 & 23.55 & 0.117 \\
& & 50 & 0.66 & 0.66 & 22.11 & 0.161 & 0.12 & 22.11 & 0.160 \\
& & 100 & 0.86 & 0.86 & 20.91 & 0.209 & 0.20 & 20.91 & 0.205 \\
& & 150 & 0.95 & 0.95 & 20.14 & 0.247 & 0.23 & 20.15 & 0.239 \\
\cline{2-10} \noalign{\vskip 0.2em}
& Avg-5000 & - & 0.00 & 0.00 & 26.42 & 0.016 & 0.90 & 26.16 & 0.017 \\
\bottomrule
\end{tabular}
        }
        \vspace{-0.5em}
        \captionof{table}{\textbf{\ImprintForgeLong Attack} against Gaussian Shading and Tree-Ring applied on different target models. The attacker uses a vanilla SD 2.1 model as proxy.
        Watermark detection (``Det.'') and user attribution (``Attr.'') success are measured using TPR@$X$-FPR.
        We use PSNR and LPIPS~\cite{zhang2018unreasonable} to measure deviation from the original images.
        As baseline, we include the concurrent averaging attack~\cite{yang2024steganalysisdigitalwatermarkingdefense} where 5,000 watermarked reference images are used (Avg-5000).
        See Tab.~\ref{tab:full_forgery} in Sec.~\ref{sec:full_results} for a full baseline comparison.}
        \label{tab:imprint}
\end{minipage}%
\hfill
\begin{minipage}[t]{0.4\textwidth}
    
    \centering
            \includegraphics[width=1.00\linewidth,trim=0em 3em 0em 0em]{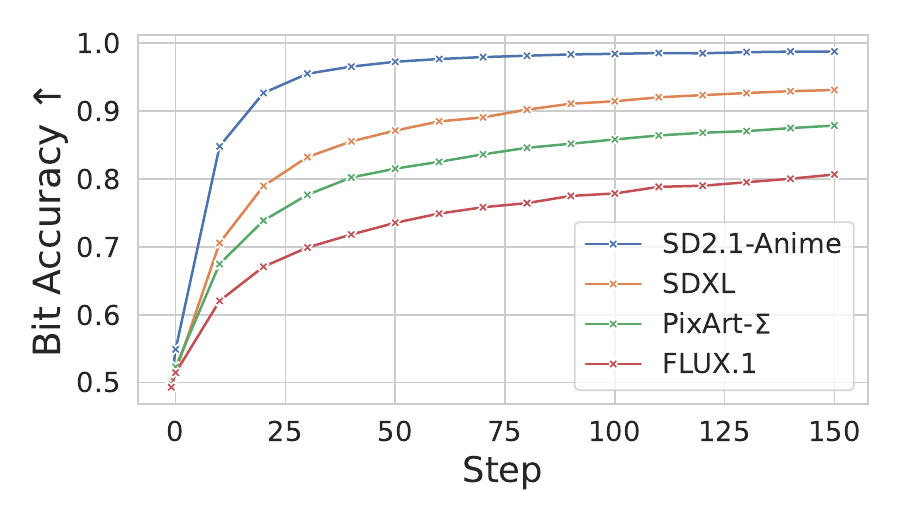}  %
            \caption{Bit accuracy for the \ImprintForgeLong attack on Gaussian Shading at different optimization steps. Attacker aims at higher values.}  %
            \label{fig:imprint:gs:bitacc}
    
    \vspace{0.0cm} %
    
    \centering
            \includegraphics[width=1.00\linewidth,trim=0em 3em 0em 0em]{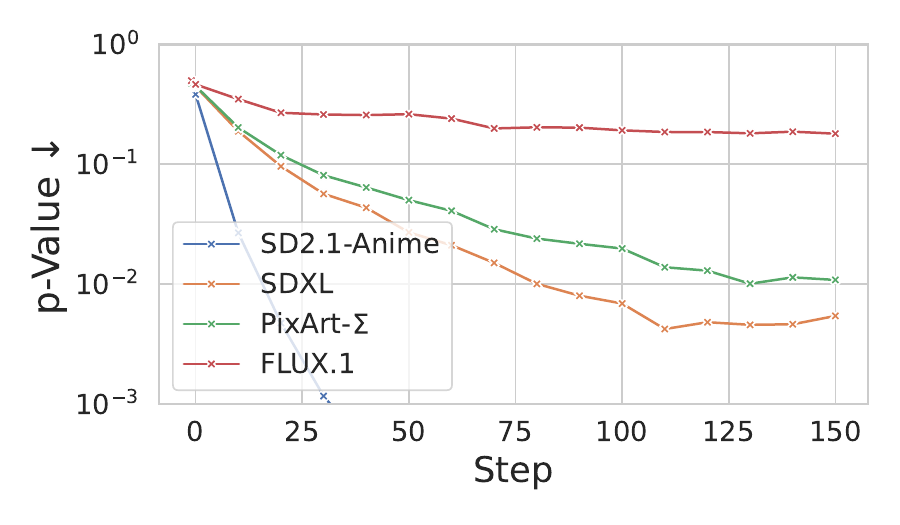}  %
            \captionof{figure}{P-values for the \ImprintForgeLong attack on Tree-Ring at different optimization steps. Attacker aims at smaller values.}  %
            \label{fig:imprint:tr:pvalues}
    
\end{minipage}
\end{figure*}

\subsection{Experimental Setting}
We use Stable Diffusion 2.1~\cite{RomBlaLor22} as proxy model~\attackermodel~and several common models as target model~\model, namely Stable Diffusion XL (SDXL)~\cite{podell2024sdxl}, PixArt-$\Sigma$~\cite{chen2023pixartalpha} and FLUX.1. We also include our own finetune of SD2.1 trained on anime (SD2.1-Anime).
All experiments use $512\times512$ images.

We evaluate the two primary semantic watermarking approaches: Tree-Ring~\cite{Wen2023TreeRing} and Gaussian Shading~\cite{Yang2024GaussianShading}. To ensure consistency with their respective experimental setups, we consider the same scenarios and metrics. %
As Tree-Ring is only applied in the detection scenario, we measure the watermark detection rate using the TPR@1\%FPR. %
The underyling threshold for detecting an image as watermarked is based on the p-value, which reflects the likelihood of observing the watermark pattern by random chance.
Gaussian Shading is evaluated for detection and attribution. 
To this end, the raw bit accuracy~$r(\mess, \messrecov)$ is used to measure how many bits of the recovered message bit string $\messrecov$ from an image under investigation match with the bit string $\mess$ from the target watermark.
In the detection scenario, we count a true positive if $r(\mess,\messrecov)$ exceeds a threshold $\tau$ which is calibrated to achieve a specific FPR. %
For attribution, we compute $r(\mess,\messrecov)$ %
by iterating over all possible bit strings~$s$ from a pool of 100k users.
Like \citet{Yang2024GaussianShading}, we count a true positive as follows.
First, the user id with the highest number of matching bits is retrieved. 
Then, if the number of matching bits exceeds the threshold $\tau$ corresponding to a FPR of $10^{-6}$, the sample is counted if attributed to the correct target user.

Our Supplementary Material provides more information. Sec.~\ref{sec:details:visual:results} shows visual examples. 
Sec.~\ref{sec:full_results} provides a detailed comparison with previously proposed attacks. 
Sec.~\ref{sec:experiment:details} provides experimental details, such as the prompt and image datasets, model settings, and thresholds. 
Sec.~\ref{sec:experiment:ablation} provides an ablation study on the choice of samplers and the number of inversion steps.

\begin{figure*}[t]
\centering
\begin{minipage}[t]{0.550\textwidth}
   
    \centering
    
    \resizebox{1.00\linewidth}{!}{%
        \begin{tabular}{@{}lllrrrrrrr@{}}
\toprule
& & & \multicolumn{4}{c}{Gaussian Shading} & \multicolumn{3}{c}{Tree-Ring} \\
\cmidrule(lr){4-7} \cmidrule(lr){8-10}
& & & \multicolumn{4}{c}{\small (FPR=$10^{-6}$)} & \multicolumn{3}{c}{\small (FPR=$1\%$)} \\
Model & Attack & Step & Det. & Attr. & PSNR & LPIPS & Det. & PSNR & LPIPS \\ 
\midrule
\multirow{5}{*}{\parbox{1px}{SD2.1-Anime}} 
& \multirow{4}{*}{\ImprintRemovalShort} 
  & 20 & 0.04 & 0.04 & 25.88 & 0.051 & 0.50 & 25.94 & 0.051 \\
& & 50 & 0.00 & 0.00 & 22.75 & 0.096 & 0.14 & 22.82 & 0.095 \\
& & 100 & 0.00 & 0.00 & 20.62 & 0.145 & 0.04 & 20.67 & 0.144 \\
& & 150 & 0.00 & 0.00 & 19.31 & 0.184 & 0.03 & 19.35 & 0.182 \\
\cline{2-10} \noalign{\vskip 0.2em}
& Avg-5000 & - & 1.00 & 1.00 & 31.87 & 0.007 & 0.00 & 28.89 & 0.014 \\
\midrule
\multirow{5}{*}{SDXL}                                          
& \multirow{4}{*}{\ImprintRemovalShort} 
  & 20 & 0.66 & 0.66 & 25.27 & 0.081 & 1.00 & 26.80 & 0.082 \\
& & 50 & 0.00 & 0.00 & 22.83 & 0.136 & 0.64 & 24.32 & 0.137 \\
& & 100 & 0.00 & 0.00 & 21.09 & 0.185 & 0.33 & 22.56 & 0.185 \\
& & 150 & 0.00 & 0.00 & 19.91 & 0.224 & 0.12 & 21.47 & 0.219 \\
\cline{2-10} \noalign{\vskip 0.2em}
& Avg-5000 & - & 1.00 & 1.00 & 24.14 & 0.013 & 0.01 & 22.37 & 0.046 \\
\midrule
\multirow{5}{*}{\parbox{1px}{PixArt-$\Sigma$}}                               
& \multirow{4}{*}{\ImprintRemovalShort} 
  & 20 & 0.94 & 0.94 & 26.12 & 0.085 & 0.96 & 26.45 & 0.084 \\
& & 50 & 0.18 & 0.18 & 23.62 & 0.145 & 0.75 & 23.90 & 0.145 \\
& & 100 & 0.00 & 0.00 & 21.74 & 0.206 & 0.41 & 21.98 & 0.206 \\
& & 150 & 0.00 & 0.00 & 20.59 & 0.248 & 0.14 & 20.81 & 0.248 \\
\cline{2-10} \noalign{\vskip 0.2em}
& Avg-5000 & - & 1.00 & 1.00 & 16.22 & 0.075 & 0.19 & 16.34 & 0.075 \\
\midrule
\multirow{5}{*}{FLUX.1}                                         
& \multirow{4}{*}{\ImprintRemovalShort} 
  & 20 & 0.75 & 0.75 & 25.86 & 0.112 & 0.07 & 25.22 & 0.108 \\
& & 50 & 0.00 & 0.00 & 24.10 & 0.176 & 0.00 & 23.67 & 0.167 \\
& & 100 & 0.00 & 0.00 & 22.67 & 0.243 & 0.00 & 22.33 & 0.232 \\
& & 150 & 0.00 & 0.00 & 21.71 & 0.291 & 0.00 & 21.48 & 0.275 \\
\cline{2-10} \noalign{\vskip 0.2em}
& Avg-5000 & - & 1.00 & 1.00 & 26.74 & 0.013 & 0.11 & 26.47 & 0.013 \\
\bottomrule
\end{tabular}

        }
        \captionof{table}{\textbf{\ImprintRemovalLong attack} against Gaussian Shading and Tree-Ring.
        We use the same target models and metrics from the \ImprintForgeLong attacks (see \cref{tab:imprint}). 
        However, attack success for removal is now assessed by the decrease in detection and attribution metrics.
        For reference, we include the best removal baseline (Avg-5000)~\cite{yang2024steganalysisdigitalwatermarkingdefense}.
        See Tab.~\ref{tab:full_removal} in Sec.~\ref{sec:full_results} for a full baseline comparison.
        }
        \label{tab:removal}

\end{minipage}%
\hfill
\begin{minipage}[t]{0.400\textwidth}

    \centering
            \includegraphics[width=1.00\linewidth,trim=0em 3em 0em 0em]{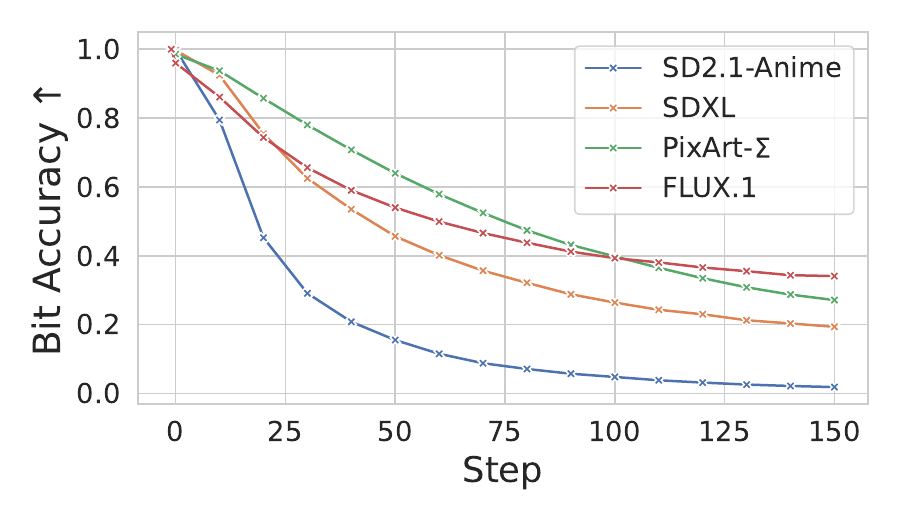}  %
            \captionof{figure}{Bit accuracy of removal attack against Gaussian Shading at different optimization steps. Attacker aims at smaller values.}  %
            \label{fig:removal:gs:bitacc}
    
    \vspace{0.0cm} %
    
    \centering
            \includegraphics[width=1.00\linewidth,trim=0em 3em 0em 0em]{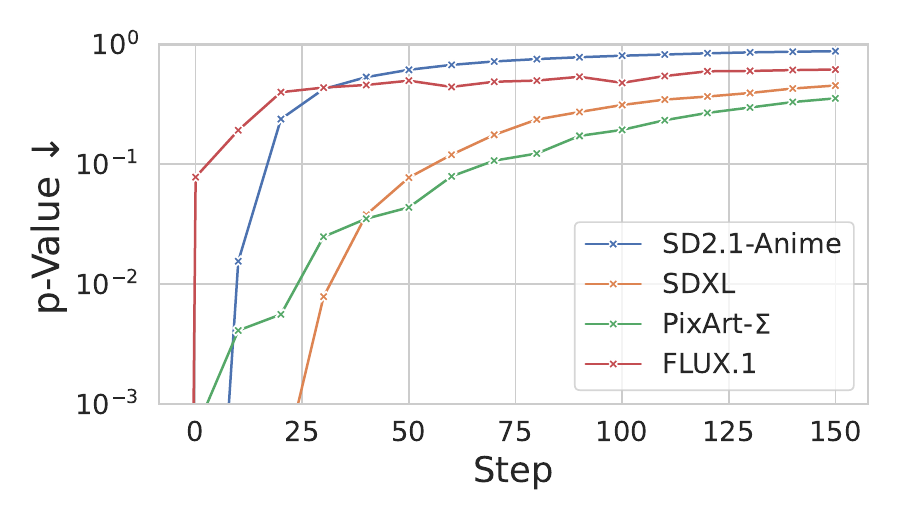}  %
            \captionof{figure}{P-values of removal attack against Tree-Ring at different optimization steps. Attacker aims at higher values.}  %
            \label{fig:removal:tr:pvalues}
    
\end{minipage}

\end{figure*}

\subsection{\ImprintForgeLong Attack}
\label{sec:evaluation:imprinting}
First, we investigate whether an attacker can \imprint a semantic watermark on existing images by using our \ImprintForgeLong attack. 
For pairs of 100 watermarked images and natural cover images, we minimize \cref{eq:attack2:loss} for 150 steps with a learning rate of $0.01$, and evaluate at every tenth~step.

\paragraph{Results.} 
\cref{tab:imprint} presents the performance of the attack against both watermarking methods.
For Gaussian Shading, the attack successfully implants a watermark that is detected in almost all images across different models in the detection and the attribution scenarios. 
For Tree-Ring, the attack achieves more than 84\% detection rate for three out of four models. 
We account the lower success rate on FLUX.1 to the fact that Tree-Ring embeds a weaker watermark signal than Gaussian Shading while FLUX.1 is a completely unrelated model with a different architecture and auto-encoder.
This makes successful transfer more difficult. Still, more than a fifth is attributed, which is still enough to discredit the watermarking method.

We observe that the attacks often retain overall picture content and quality, as highlighted by the examples in Sec.~\ref{sec:details:visual:results} in our Supplementary Material. 
\cref{tab:imprint} reports the image quality in terms of PSNR\footnote{We have updated the PSNR numbers due to a data-processing error in previous versions (see our \href{https://github.com/and-mill/semantic-forgery}{Github repository} for details).} and LPIPS. We observe acceptable LPIPS scores up to 100 optimization steps, while the PSNR scores are all low. We attribute this to our optimization in the latent space, causing shifting edges and large pixel differences in some positions, severely affecting the PSNR.

\cref{fig:imprint:gs:bitacc,fig:imprint:tr:pvalues} show the evolution of the bit accuracies and p-values as optimization progresses, with clear trends towards higher bit accuracies and lower p-values.
We can also see the different degrees of transferability, where more similar models (SD2.1-Anime) can be attacked more easily.

As baseline, Tab.~\ref{tab:imprint} includes the concurrent averaging attack~\cite{yang2024steganalysisdigitalwatermarkingdefense}.
Unlike our attack, the averaging attack cannot forge Gaussian Shading watermarks. It also 
requires that an attacker has access to a large number of images with the same watermark (here: 5,000 images) while our proposed attack only needs one reference image. 
We refer to Sec.~\ref{sec:full_results} in the Supplementary Material for a detailed baseline comparison.

\subsection{\ImprintRemovalLong Attack}
\label{sec:evaluation:removal}
In our next experiment, we test the effectiveness of the proposed watermark removal technique.
We minimize \cref{eq:removal:loss} for 150 steps with a learning rate of $0.01$ for 100 watermarked images, and evaluate at every tenth step.

\paragraph{Results.}
\cref{tab:removal} shows the results for the removal attack.
For Gaussian Shading, after 100 steps, watermarks are no longer detectable in the detection and attribution setting.
Removal of Tree-Ring watermarks is more difficult but TPR@1\%FPR eventually drops below 15\% for all target models.

In addition, \cref{tab:removal} shows results for the averaging attack as best baseline across all tested removal attacks (see Sec.~\ref{sec:full_results}). This baseline performs similarly to our attack for Tree-Ring, but fails on Gaussian shading.
Finally, \cref{fig:removal:gs:bitacc,fig:removal:tr:pvalues} show similar trends in transferability as seen for the \ImprintForgeLong attack. An exception is FLUX.1, which shows a direct sharp increase in the p-values. 
This is likely caused by significant auto-encoder differences with SD2.1.

\subsection{\Reprompting Attack}
\label{sec:evaluation:reprompting}
As outlined in \cref{sec:attack}, we have a basic variant that generates an image with another prompt, and an augmented variant that tries $k$ different prompts, and for Gaussian Shading $l$ bin resamplings in addition.
For the basic variant (``\Reprompt''), for each target model, we use 1000 watermarked images and re-generate an attack image using a harmful prompt.
For the augmented variant (``\RepromptPlus''), for each target model, we use 100 target images and set $k=3$ and $l=3$. That is, we have 3~attack candidates for each target image with Tree-Ring, and 9~attack candidates with Gaussian Shading. If the watermark is valid in one of the candidates, we count the target image as successfully forged. 

\paragraph{Results.} 
\cref{tab:generation} reports the attack performance. 
Both watermarking methods can be successfully tricked by \reprompting. 
With Gaussian Shading, the attack achieves a near-perfect detection and identification rate on all models. With Tree-Ring, three out of four models detect the forged watermark in almost all cases. Similar to the \ImprintForgeLong attack, FLUX.1 is harder to attack; however, 35\% of the images are still detected as watermarked which is enough to undermine the method. 
Comparing the two attack variants, even the basic variant achieves a high success rate in different settings, which can often be increased to (almost) 100\% using the augmented~variant.

\begin{table}[t!]
\centering
\resizebox{0.90\linewidth}{!}{%
\begin{tabular}{llccc}
\toprule
 &  & \multicolumn{2}{c}{G. Shad.} & \multicolumn{1}{c}{Tree-Ring} \\
 &  & \multicolumn{2}{c}{\small (FPR=$10^{-6}$)} & \multicolumn{1}{c}{\small (FPR=$1\%$)} \\
 Model &  & Det. & Attr. & Det. \\
\midrule

\multirow[c]{2}{*}{SD2.1-Anime} & \Reprompt & 0.983 & 0.983 & 0.896 \\
 & \RepromptPlus & 1.000 & 1.000  & 1.000 \\
\midrule 
\multirow[c]{2}{*}{SDXL} & \Reprompt & 0.995 & 0.995 & 0.968 \\
 & \RepromptPlus & 1.000 & 1.000 & 1.000 \\
\midrule
\multirow[c]{2}{*}{PixArt-$\Sigma$} & \Reprompt & 0.934 & 0.934 & 0.875  \\
 & \RepromptPlus & 1.000 & 1.000 & 0.990 \\
\midrule
\multirow[c]{2}{*}{FLUX.1} & \Reprompt & 0.656 & 0.656 & 0.128 \\
 & \RepromptPlus & 0.880 & 0.880 & 0.350  \\

\bottomrule
\end{tabular}

}
\caption{%
\textbf{\Reprompting attack} against Gaussian Shading and Tree-Ring.
The metrics are computed as in the experiments on the \ImprintForgeLong attack (see \cref{tab:imprint}). 
\textit{\RepromptShort} is the default \Reprompting attack, %
\textit{\RepromptPlusShort} is the enhanced \Reprompting attack.
}
\label{tab:generation}
\end{table}

\subsection{Transferability Analysis}
\label{sec:evaluation:transferability}

\begin{figure*}[ht]
    \centering
    % Left Side: 2x2 Layout
    %\begin{minipage}{0.67\textwidth} % Adjust width as needed
    \begin{minipage}{0.64\textwidth} % Adjust width as needed
        \centering
        \begin{subfigure}{0.49\textwidth}
            \includegraphics[width=\textwidth]{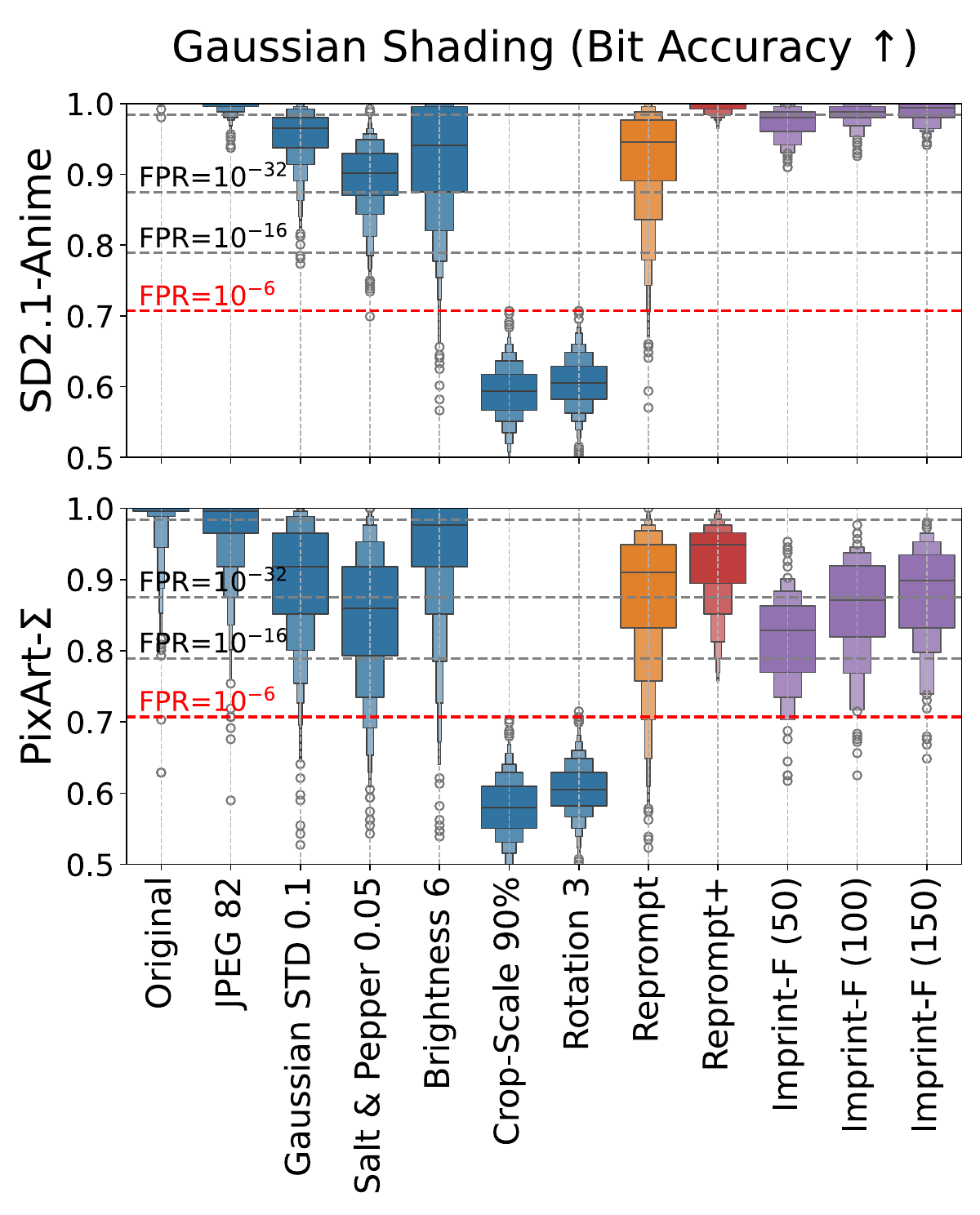}
            \caption{Bit Accuracy ($\uparrow$) for Gaussian Shading on SD2.1-Anime (top) and PixArt-$\Sigma$ (bottom).}
            \label{fig:gs:perturb:anime}
        \end{subfigure}%
        \hfill
        \begin{subfigure}{0.49\textwidth}
            \centering
            \includegraphics[width=\textwidth]{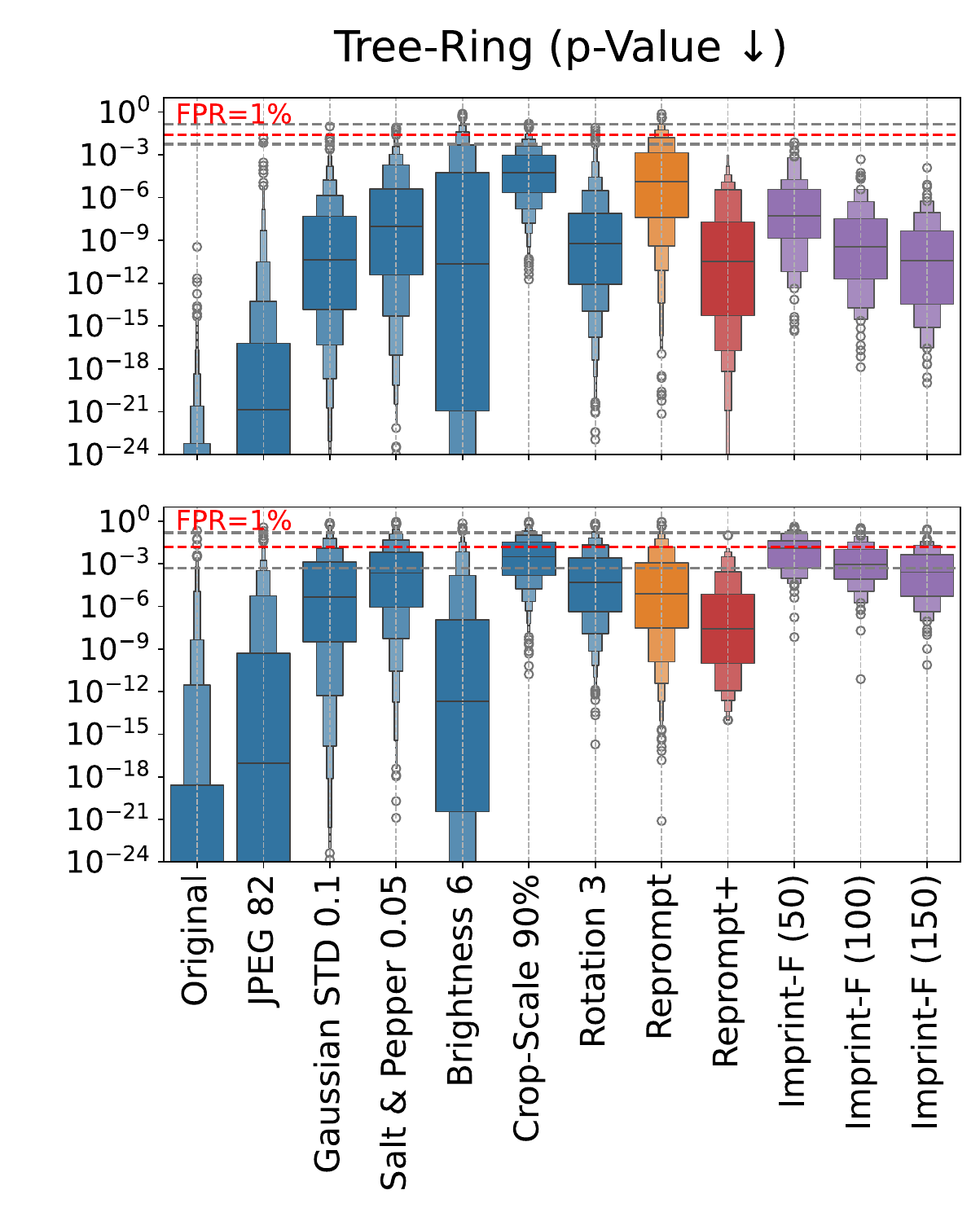}
            \caption{P-values ($\downarrow$) for Tree-Ring on SD2.1-Anime (top) and PixArt-$\Sigma$ (bottom).}
            \label{fig:gs:perturb:pixart}
        \end{subfigure}
        
        % \centering
        % \begin{subfigure}{0.5\textwidth}
        %     \includegraphics[width=\textwidth]{images/experiments/plot_baselines_vs_attacks/stabilityai/stable-diffusion-2-1-base/DDIM/512/1girl/fulltune/step=10000/GS/baseline_vs_attacks.pdf}
        %     \caption{Gaussian Shading: SD2.1-Anime.}
        %     \label{fig:gs:perturb:anime}
        % \end{subfigure}%
        % \hfill
        % \begin{subfigure}{0.5\textwidth}
        %     \centering
        %     \includegraphics[width=\textwidth]{images/experiments/plot_baselines_vs_attacks/PixArt-alpha/PixArt-Sigma-XL-2-512-MS/DPM/512/Pixart/notune/step=0/GS/baseline_vs_attacks.pdf}
        %     \caption{Gaussian Shading: PixArt-$\Sigma$}
        %     \label{fig:gs:perturb:pixart}
        % \end{subfigure}

        % \vspace{0.5em} % Space between rows

        % \begin{subfigure}{0.5\textwidth}
        %     \centering
        %     \includegraphics[width=\textwidth]{images/experiments/plot_baselines_vs_attacks/stabilityai/stable-diffusion-2-1-base/DDIM/512/1girl/fulltune/step=10000/TR/baseline_vs_attacks.pdf}
        %     \caption{Tree-Ring: SD2.1-Anime}
        %     \label{fig:tr:perturb:anime}
        % \end{subfigure}%
        % \hfill
        % \begin{subfigure}{0.5\textwidth}
        %     \centering
        %     \includegraphics[width=\textwidth]{images/experiments/plot_baselines_vs_attacks/PixArt-alpha/PixArt-Sigma-XL-2-512-MS/DPM/512/Pixart/notune/step=0/TR/baseline_vs_attacks.pdf}
        %     \caption{Tree-Ring: PixArt-$\Sigma$}
        %     \label{fig:tr:perturb:pixart}
        % \end{subfigure}
        
        \caption{
        Impact of common transformations (blue) and our forgery attacks (orange, red, purple) on bit accuracy and p-values, and thus on suitable thresholds for watermarking. 
        No threshold can effectively separate images from common transformations and attacks.
        See Sec.~\ref{app:robustness_full} in the Supplementary Material for a more complete set of transformations.
        }
        \label{fig:perturb}
    \end{minipage}%
    \hfill
    % Right Side: Stacked Images
    \begin{minipage}{0.28\textwidth} % Adjust width as needed
        \centering
        
        % \begin{subfigure}{\textwidth}
            \centering
            \includegraphics[width=0.95\linewidth]{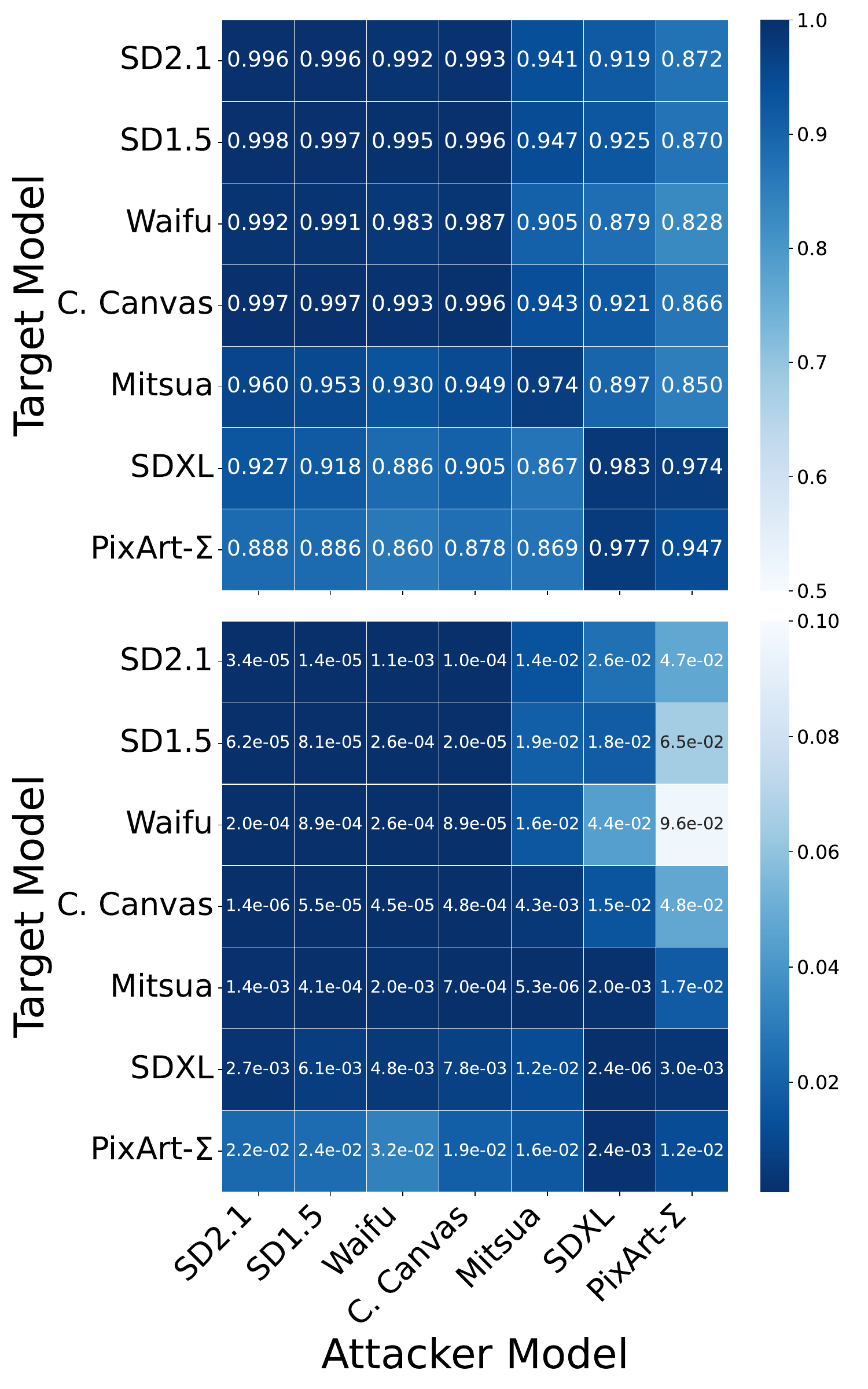}
        % \end{subfigure}
        \caption{Transferability in terms of bit accuracy (Gaussian Shading, top) and p-value (Tree-Ring, bottom) of the \Reprompting attack across models.}
        
        \label{fig:transferability}
    \end{minipage}
    
    % \caption{A block with four images in a 2x2 grid on the left and two stacked images on the right}
    % \label{fig:images_block}
\end{figure*}

The previous experiments show the effectiveness of our proposed attacks. In the next step, we further analyze the transferability across different models to understand the extent to which models are vulnerable.  
To this end, we examine our \Reprompting attack in the basic variant across different pairs of target and attacker model. In addition to SD2.1, SDXL and \mbox{PixArt-$\Sigma$}, we include SD1.5, Mitsua Diffusion One, Common Canvas S-C~\cite{gokaslan2023commoncanvasopendiffusionmodel} and Waifu Diffusion. We consider Mitsua and Common Canvas as they share the same architecture as Stable Diffusion models but are trained from scratch on different datasets.
Waifu Diffusion is an extensive community fine-tune of SD1.4.
Additional model information is provided in Sec.~\ref{sec:experiment:details} in the Supplementary Material.
For each target model, we use 100 watermarked images from this model to create respective forgeries on the attacker model. 

\paragraph{Results.} 
\cref{fig:transferability} illustrates the transferability in terms of bit accuracy (Gaussian Shading) and p-value (Tree-Ring). 
Similar models (SD1.5, SD2.1) and their fine-tunes (Waifu) show nearly-100\% bit accuracy and extremely low p-values across all combinations.
When considering models that have been trained independently \emph{from scratch} (Mitsua, Common Canvas) on public domain data but have a similar architecture, we also observe a high transferability. Thus, even models with different training data and possibly different training protocols are severely impacted by our attack.

\paragraph{Analysis.} 
We continue by analyzing the unexpected transferability of inversion that enables our attacks.
We find a correlation between the success of attacks on different target and proxy models and the functional similarity of their corresponding auto-encoders.
This indicates that our attacks benefit from high similarity in the latent spaces between target and proxy models.
We encourage readers to consult Sec.~\ref{app:analysis} in the Supplementary Material for more details.

\subsection{No Defense by Adjusting Thresholds}
\label{sec:evaluation:robustness}
Finally, we demonstrate that it is not possible to defend against our forgery attacks by simply making the thresholds tighter (lower p-value for Tree-Ring, higher bit accuracy for Gaussian Shading).
The first problem is that a watermark must be resilient to common perturbations, such as JPEG compression and Gaussian noise, in order to be practical. This requirement significantly limits the feasible thresholds.
The second problem is that our attack images achieve p-values and bit accuracies that already overlap with those of the original, unperturbed watermarked images.

\paragraph{Setup.} 
We adopt the attack setup as outlined in previous sections, and report results for SD2.1-Anime and \mbox{PixArt-$\Sigma$} as target models. 
We consider the standard perturbations from related work that are applied to watermarked images before watermark verification: JPEG compression (82 quality), Gaussian Noise ($\sigma$=0.1), Salt-and-Pepper noise (p=0.05), brightness jitter, Rotation (3 degrees), 90\% crop-and-scale. 
We measure the p-value for Tree-Ring and the bit accuracy for Gaussian Shading on watermarked images that have undergone one of these perturbations, and on attack images from our different forgery attacks.
We provide examples of the transformations in Sec.~\ref{sec:experiment:details} in the Supplementary Material.

\paragraph{Results.} 
\cref{fig:perturb} depicts the p-values for Tree-Ring and bit accuracies for Gaussian Shading. 
It is not possible to tighten the detection threshold enough to distinguish attack images (orange, red and purple bars) while maintaining robustness to common perturbations (blue bars).
For example, Salt-and-Pepper noise results in higher p-values for Tree-Ring and lower bit-accuracies for Gaussian Shading, which are comparable to those of our attack images.
We can conclude that tightening thresholds is not a viable defense.

\section{Related Work}
\label{sec:relatedwork}
In the following, we recap related work that examines security aspects of inversion-based semantic watermarks.

\paragraph{Forgery Attacks.}
While there is prior work on the forgery of post-hoc watermarks~\cite{kinakh2024evaluation, wang2021watermark}, only few works study semantic watermarks forgery.
These works mostly focus on Tree-Ring and operate under unrealistic assumptions.
\citet{saberi2023robustness} propose a method against Tree-Ring where the attacker requests the SP to generate a white-noise image containing a watermark, and then blends this image into a clean cover image. However, obtaining white-noise images from a black-box API through prompting is challenging and easy to defend against. 
In their code implementation, watermarked reference images are obtained by having the SP artificially sample white-noise pixels, applying DDIM inversion, embedding a Tree-Ring, and regenerating the image.
Furthermore, the attack does not work for watermarks that do attribution. As the attacker requests watermarked images from the SP, all forged images are traceable to the attacker.
\citet{yang2024steganalysisdigitalwatermarkingdefense} propose to average watermarked images---either in the pixel space or their inverted latents---to get a watermark pattern which is then added on clean cover images (or removed from watermarked images). 
This averaging attack, however, requires access to the original model for the inversion or a large number of watermarked images if working in the pixel space. 
Neither is the attack applicable against Gaussian Shading, which draws a unique
watermark key for each image.

In contrast, our forgery attacks work with just a single watermarked image, require no knowledge of the target model, are applicable under detection and attribution, and work for Tree-Ring and Gaussian Shading.

\paragraph{Removal Attacks.}
Concurrently to our work, \citet{controllableregen} proposes a training-based controllable regeneration approach.
In comparison, our methods are 
training-free and instead rely on cleaning the initial latent using gradient descent.
Simpler regeneration attacks have also been studied in recent work on watermark removal~\cite{ZhaZhaSu2024invisibleimagewatermarksprovably,AnDinRab2024Benchmarking}, but were found ineffective for semantic watermark removal~\cite{ZhaZhaSu2024invisibleimagewatermarksprovably}.
\citet{lukDiaFen2024leveraging} present a general approach for creating removal attacks specifically designed for a particular watermarking scheme at hand, assuming detailed knowledge of internal parameters. In the case of Tree-Ring, the attacker needs exact knowledge of the ring-embedding parameters. It is unclear how this method would translate to Gaussian Shading.
Finally, the Averaging attack ~\cite{yang2024steganalysisdigitalwatermarkingdefense} and the Surrogate attack~\cite{saberi2023robustness} require training with multiple samples.
Again, these attacks are not applicable against Gaussian Shading.

In comparison, our removal attack works with a single watermarked image without knowledge of the target model, and is applicable against both Tree-Ring and Gaussian Shading.

\section{Discussion and Conclusion}
The attacks described in this paper expose fundamental issues in the practical use of current inversion-based semantic watermarking approaches. In a realistic scenario, where the attacker can simply leverage their own model, we show that the attacker can forge and remove semantic watermarks with almost perfect success rate. 
Concerningly, even a large difference between the attacker and the target watermarked model still permits an attack, as the high attack success rates on completely unrelated and independently trained models show. For instance, our attacks succeed even when the attacker uses an SD2.1 model against \mbox{PixArt-$\Sigma$} and FLUX.1, which are both DiT~\cite{dit} models trained from scratch.
Given these findings, our proposed attacks should be used for the security auditing of semantic watermarks.

From a defense perspective, we find that adjusting the watermark detection thresholds is not effective.
We believe that %
fundamental improvements of semantic watermarking techniques are required %
to prevent watermarks %
from being easily imposed on new images.

\FloatBarrier
\section*{Acknowledgements}
This work was funded by the Deutsche Forschungsgemeinschaft (DFG, German Research Foundation) under Germany’s Excellence Strategy -- EXC 2092 CASA -- 390781972 and by the Ministry of Culture and Science of Northrhine-Westphalia as part of the Lamarr Fellow Network.

{
    \small
    \bibliographystyle{abbrvnat}  %

}

\FloatBarrier

\clearpage
\appendix

\onecolumn
    \begin{center}
        {\Large \bfseries Supplementary Material for}\\
        {\Large \bfseries \thetitle}
        \vspace{1em}
    \end{center}

\section*{Table of Contents for the Supplementary Material}

\startcontents[appendix] 

\printcontents[appendix]{l}{1}{\setcounter{tocdepth}{2}} 

\newpage
\section{Semantic Watermarking}
\label{sec:appendix:semantic-watermarks}
In this section, we present more details on the examined semantic watermarking methods.

\begin{figure}[b]
    \centering
  \begin{subfigure}[t]{0.454\textwidth}
    \includegraphics[width=0.875\textwidth]{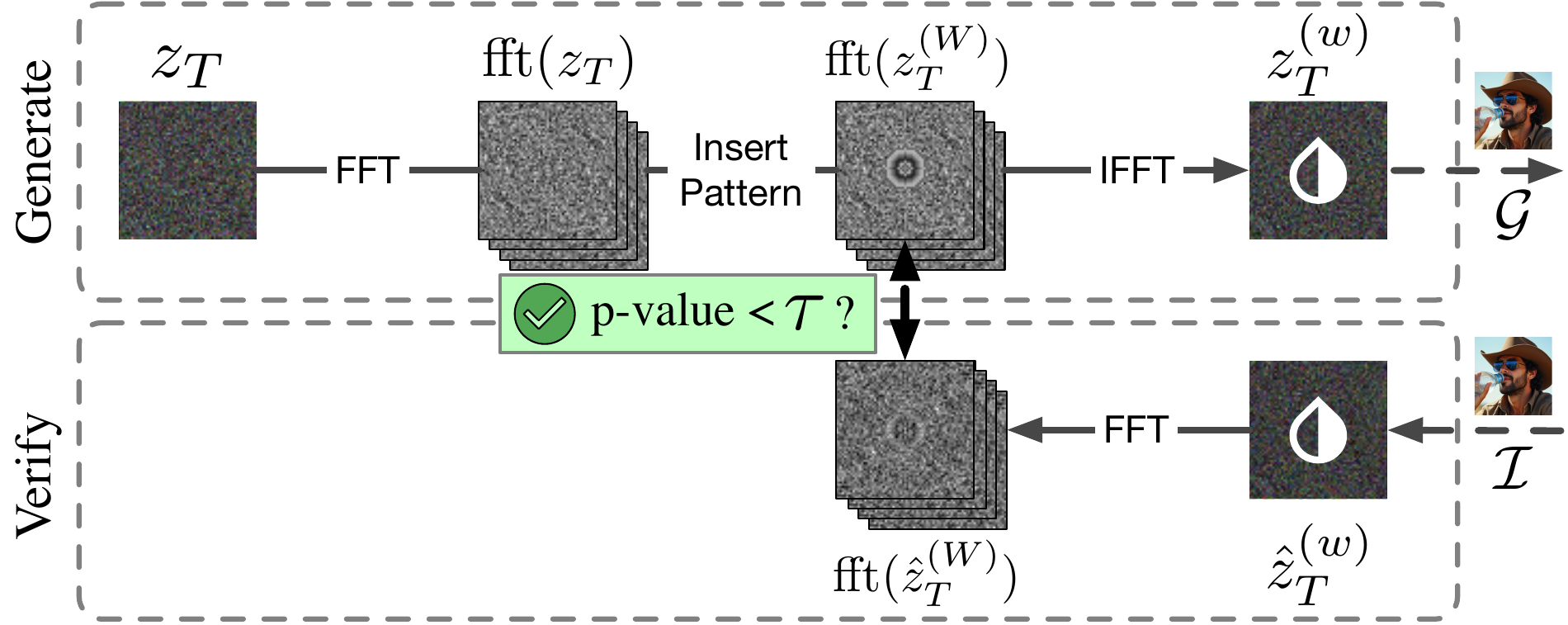}
    \caption{Tree-Ring}
    \label{fig:diagram:basic:treering}
  \end{subfigure} 
  \begin{subfigure}[t]{0.496\textwidth}
    \includegraphics[width=\textwidth]{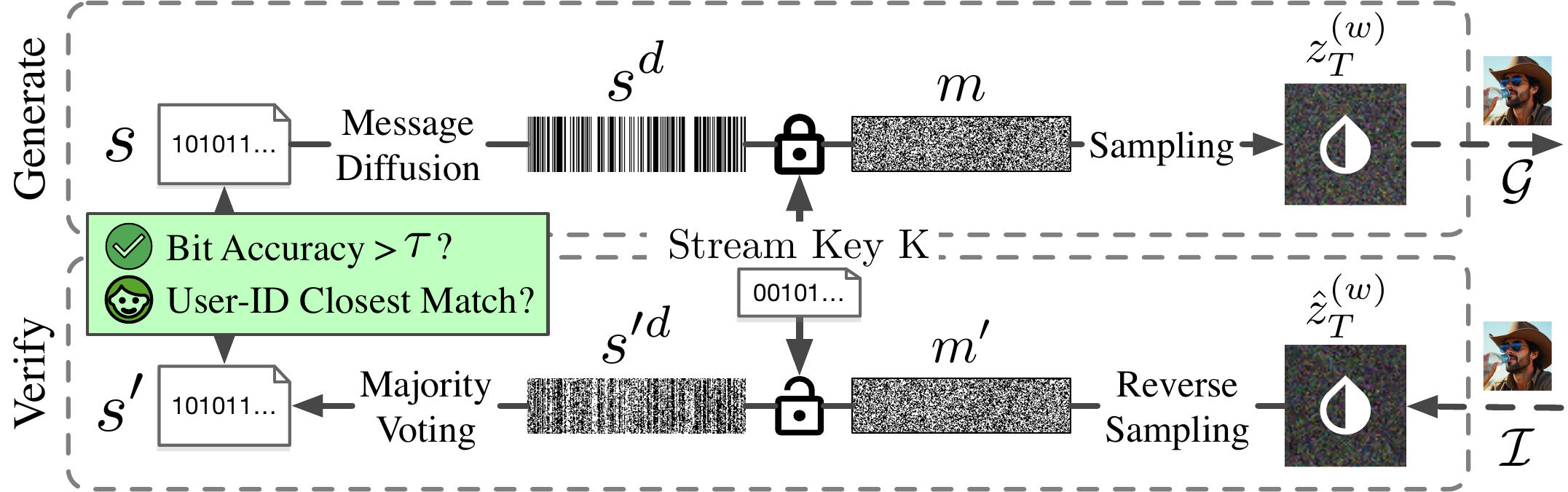}
    \caption{Gaussian Shading}
    \label{fig:diagram:basic:gaussianshading}
  \end{subfigure}
  \caption{Concept of the two inversion-based semantic watermarking approaches} 
  \label{fig:diagram:basic:semanticwatermarks-supplementarymaterial}
\end{figure}

\vspace{0.75em}
\paragraph{Tree-Ring.} 
\citet{Wen2023TreeRing} have initially presented the concept of inversion-based semantic watermarking. \cref{fig:diagram:basic:treering} illustrates the proposed Tree-Ring watermark approach, which can be divided into two phases. 
\begin{itemize}
    \vspace{0.6em}
    \setlength{\itemsep}{0.5em} %
    
    \item \emph{Generation.} During image generation, Tree-Ring modifies a clean initial latent $\zT \sim \mathcal{N}(\textbf{O}, \textbf{I})$ by adding a concentric circular pattern into its frequency representation. This step produces the watermarked latent noise~$\zwat{T}$. This noise vector is then used as usual in the further generation process (denoising + decoding) to finally obtain the watermarked image: $\xwat = \decoder (\mathcal{G}_{T \rightarrow 0}(\zwat{T};\denoising) )$.
    
    \item \emph{Verification.} The inverse $\zwathat{T} = \inv{\encoder(\xwat)}$ is computed, and then the existence of the original circular pattern in the frequency spectrum of~$\zwathat{T}$ is checked. 
    The final detection is based on a statistical test. 
    The test statistic aggregates the squared absolute difference between the observed and the expected frequency values from each respective ring.
    The null hypothesis $H_0$ is that an input image is not watermarked. In this case, its inverted initial latent noise should follow a Gaussian distribution with an unknown variance. 
    Under $H_0$, the test statistic leads to a noncentral $\chi^2$ distribution. 
    This allows computing a p-value that reflects the probability of observing the test-statistic output under the assumption that the input image is not watermarked. 
    If the p-value is below a pre-defined threshold $\tau$, $H_0$ is rejected and the watermark's presence is assumed.      
    \vspace{0.6em}
\end{itemize}
Tree-Ring is designed as a so-called zero-bit watermarking scheme, as only watermark presence or absence can be detected. 
Note that the modification to the sampling procedure changes which images will be generated (it is not distribution-preserving), but appears to still have sufficient variety in generated images while retaining image quality.

A subsequent work, RingID~\cite{CiYanSon24RingID}, improves Tree-Ring by making it more robust to perturbations, ensuring that the distribution of $\zwat{T}$ is closer to $\mathcal{N}(\textbf{O}, \textbf{I})$, and providing so-called multi-bit watermarking which can carry a multiple bit long message and thus allows distinguishing between different users.

\vspace{0.75em}
\paragraph{Gaussian Shading.}
\citet{Yang2024GaussianShading} expand the previous concept by relying on cryptographic primitives to achieve distribution-preserving generation.
\cref{fig:diagram:basic:gaussianshading} illustrates the two phases of Gaussian Shading: 
\begin{itemize}
    \vspace{0.6em}
    \setlength{\itemsep}{0.5em} %
    
    \item \emph{Generation.} 
    A message $\mess$ of length $k$ is created. This message $\mess$ is first repeated (``diffused'') $\rep$ times to obtain $\mess^d$, which is then encrypted using the symmetric stream chipher ChaCha20~\cite{bernstein2008chacha}. %
    This stream cipher takes a secret key and a nonce as input, which are completely random for each image.
    The encrypted message $m$ is then used to steer the sampling of $\zwat{T}$.
    To this end, Gaussian Shading splits a Gaussian distribution into $2^\ell$ bins with equal probability. 
    In the following, we focus on the standard setting with $\ell = 1$. This means we have two bins, the negative and the positive area of the Gaussian curve. 
    The bit in the encrypted message $m[i] \in \lbrace 0, 1 \rbrace$ specifies if $\zwat{T}[i]$ is sampled from the negative or the positive area of a Gaussian distribution. 
    Due to the encryption, the bits in $m$ are uniformly distributed, ensuring a similar number of positive and negative samples from the Gaussian distribution. Hence, $\zwat{T}$ still follows a Gaussian distribution. 
    After this sampling step, the usual generation process continues with $\xwat = \decoder (\mathcal{G}_{T \rightarrow 0}(\zwat{T};\denoising) )$.
    
    \item \emph{Verification.}
    To verify the watermark, Gaussian Shading also relies on a full inversion using $\inv{\zwathat{0}}$ to obtain an estimated $\zwathat{T}$. This inverted latent noise $\zwathat{T}$ is quantized to retrieve encrypted message bits $m'$, that is, we assume $m'[i] = 0$ if $\zwathat{T}[i] < 0$ and $m'[i]=1$ if $\zwathat{T}[i] \geq 0$. Next, $m'$ is decrypted to reconstruct the message ${\messrecov}^d$. Since each bit was repeated $\rep$ times, its value is recovered by performing a majority voting on all its duplicates to finally obtain $\messrecov$. This step corrects errors and makes the scheme more robust to noise.

    \vspace{0.6em}
\end{itemize}

Gaussian Shading is applicable for both zero-bit watermarking and multi-bit watermarking. In both cases, $\messrecov$ is compared to some $\mess$ bit-wise and the number of matches per message length is recorded as bit-accuracy $r(\mess,\messrecov)$.
Like Tree-Ring, a statistical test is performed, but this time the distribution is assumed to be binomial if the image is not watermarked and therefore the test is computed using a regularized
incomplete beta function. 
In zero-bit watermarking, $\messrecov$ is compared to the predefined message $\mess$ used by the SP. It is checked if a certain fraction of at least $\tau \, \%$ of the bits match, such that it is unlikely to observe this by chance with a predefined FPR, which is denoted as $\operatorname{FPR}(\tau)$. 
In multi-bit watermarking, $\messrecov$ is a specific user~id. It is compared with all user ids known to the SP. The match with the best accuracy $r(\mess,\messrecov)$ is selected and then checked if the value is above the threshold $\tau$. As this resembles a multiple test, in order to maintain a certain FPR, the threshold $\tau$ is adapted depending on the number of users $N$ by
\begin{equation}
\label{eq:app:fprgsmulti}
\operatorname{FPR}(\tau,N) = 1- (1- \operatorname{FPR}(\tau))^N    .
\end{equation}
For further information, we refer to \citet[][Section 7.1]{Yang2024GaussianShading}.

\paragraph{Deployment Ambiguity of Gaussian Shading.}
Finally, we note that there is a considerable ambiguity on how Gaussian Shading~\cite{Yang2024GaussianShading} is deployed and evaluated.
This is examined in detail by \citet{Thietke2025Towards}. In a nutshell, the correct and secure way to use Gaussian Shading is to draw a \emph{new} nonce (and/or a new secret key) for the stream cipher for each generated image. 
This is how Gaussian Shading is implemented in its original \href{https://github.com/bsmhmmlf/Gaussian-Shading}{Github repository}. 
However, this key-nonce aspect and the resulting need for key management were not clearly addressed in the original publication, leading to some misconceptions.
For example, prior work~\cite{yang2024steganalysisdigitalwatermarkingdefense,Gunn2024Undetectable} has used the \emph{same} key and nonce for the generated images.
This reduces image diversity and also makes the watermarking scheme vulnerable to attacks that leverage multiple watermarked reference images to learn a common pattern, specifically the \textit{Averaging}~\cite{yang2024steganalysisdigitalwatermarkingdefense} and the \textit{Surrogate} Attack~\cite{saberi2023robustness, Gunn2024Undetectable}. 
As shown in our baseline comparison in Sec.~\ref{sec:full_results}, these attacks are no longer effective if Gaussian Shading is implemented securely. In all our experiments, we draw new nonces and keys for every generated image.

\vspace{2em} %
\section{Example Images and Image Quality Trade-off}
\label{sec:details:visual:results}
We provide image examples to give more intuition on the attacks and the resulting visual quality. Note that all image examples are of size $512\times512$, and the attacker's proxy model is SD2.1.
We examine different aspects in the following figures:
\begin{itemize}
    \vspace{0.6em}
    \setlength{\itemsep}{0.5em} %
    
    \item \cref{fig:visual:imprint:progression} shows the progression of the \ImprintForgeLong attack with respect to the number of optimization steps. 
    
    \item \cref{fig:visual:imprint:models:1} shows successful examples of the \ImprintForgeLong attack on different target models and both watermarking approaches (Tree-Ring and Gaussian Shading).

    \item \cref{fig:visual:removal:models:1} depicts successful examples of the \ImprintRemovalLong attack. %

    \item \cref{fig:visual:reprompting} shows successful examples of the \Reprompting attack.

    \item Finally, \cref{fig:visual:trade-off} shows the {trade-off between detection rate by the target model and image quality} for four combinations of settings: \ImprintForgeLong \& \ImprintRemovalLong, and Tree-Ring \& Gaussian Shading. 
    
    \vspace{0.6em}
\end{itemize}

\newcommand{\setupfig}{} 
\newcommand{\resultfig}{\emph{Observation:}\xspace}

\begin{figure*}[t]
    \centering
    \includegraphics[width=0.77\linewidth]{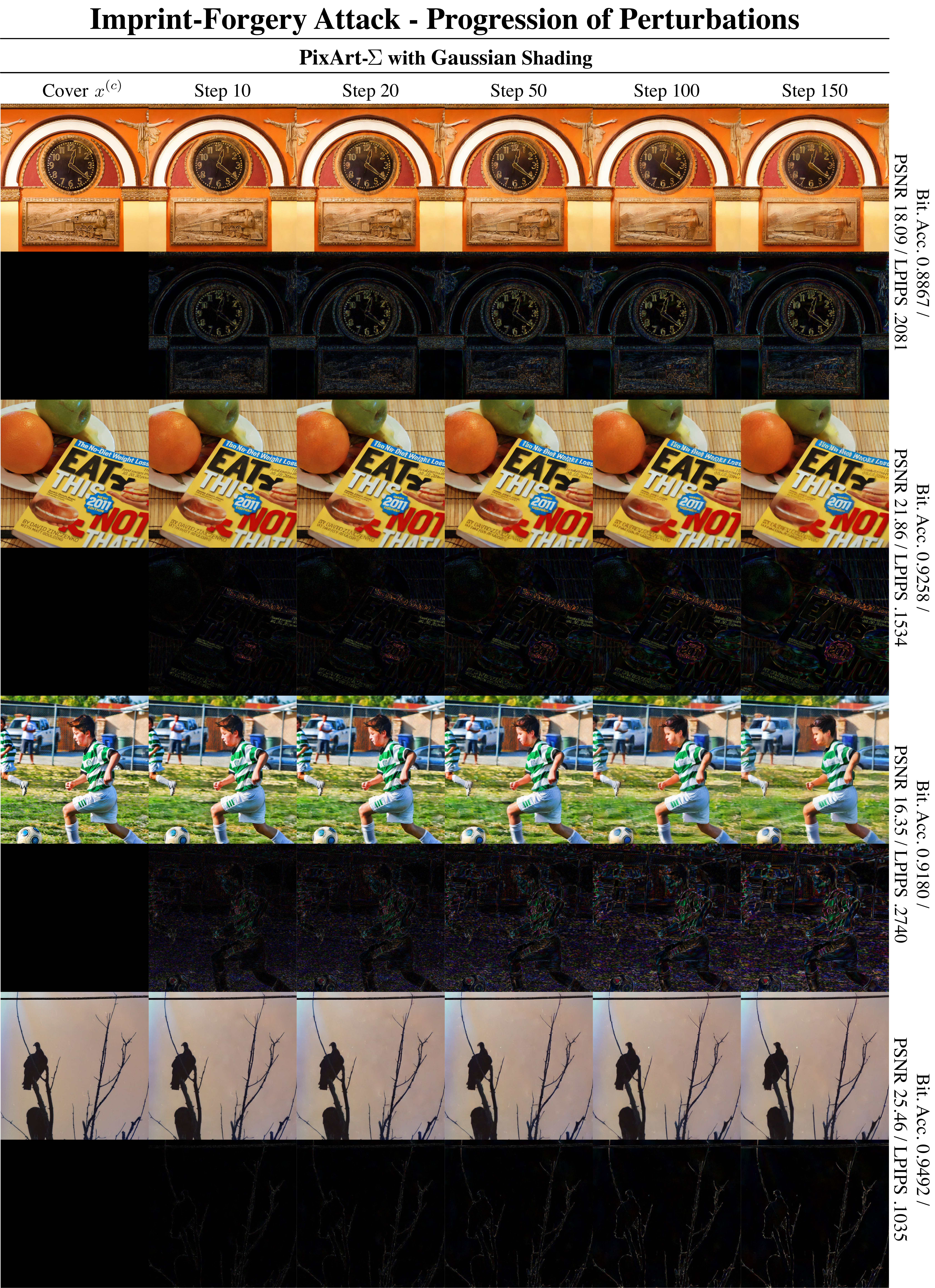}
    \caption{Progression of our \ImprintForgeLong attack with respect to optimization steps. 
    \setupfig
    The target model is PixArt-$\Sigma$, and the forged watermark is Gaussian Shading.
    Results are presented for four cover images~$\xcover$. For each cover image, the top row shows the initial cover and then the attack versions after different numbers of optimization steps.
    The bottom row illustrates the absolute pixel difference between the attack image and its initial cover version.  
    \resultfig
    The \imprinting procedure is targeting edges that carry semantic meaning and reflect unique features in the initial watermarked latent $\zwat{T}$ which we are optimizing towards. Some images with small objects defined by fine characteristics are noticeably changed (small text, small faces), while other features such as the outline of dark objects on bright backgrounds remain almost indistinguishable. 
    The removal attack introduces very similar changes to the images.
    Note that such critical changes can be avoided simply by using masks during optimization, as described in \cref{sec:approach}.
    }
    \label{fig:visual:imprint:progression}
\end{figure*}

\begin{figure*}[b]
    \centering
    \includegraphics[width=0.75\linewidth]{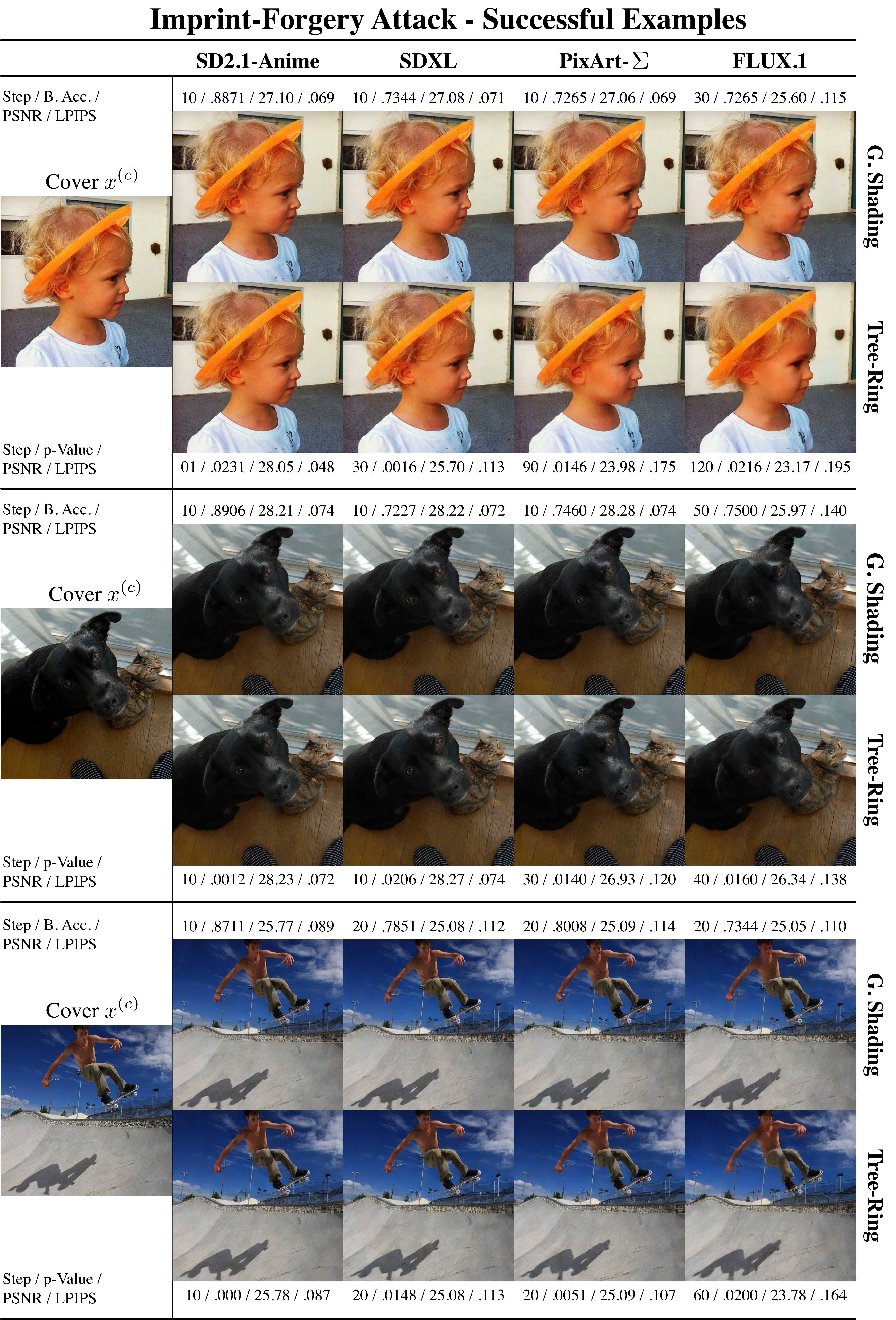}
    \caption{Examples of our \ImprintForgeLong attack on different target models and both watermark approaches.
    \setupfig
    For each cover image~\xcover, during the different optimization steps, we show the first successful attack image that passes the detection threshold $\tau$ of 
    Gaussian Shading with an FPR of $10^{-6}$ (top row) or Tree-Ring with an FPR of $1\%$ (bottom row).
    For each image, we report the step, the bit accuracy (for Gaussian Shading) or the p-value (for Tree-Ring), as well as the PSNR and LPIPS between cover images and attack example. 
    \resultfig
    Attacks against PixArt-$\Sigma$ and FLUX.1 usually have stronger visual perturbations, reflecting a higher \imprinting difficulty. 
    }
    \label{fig:visual:imprint:models:1}
\end{figure*}

\begin{figure*}[t]
    \centering
    \includegraphics[width=0.62\linewidth]{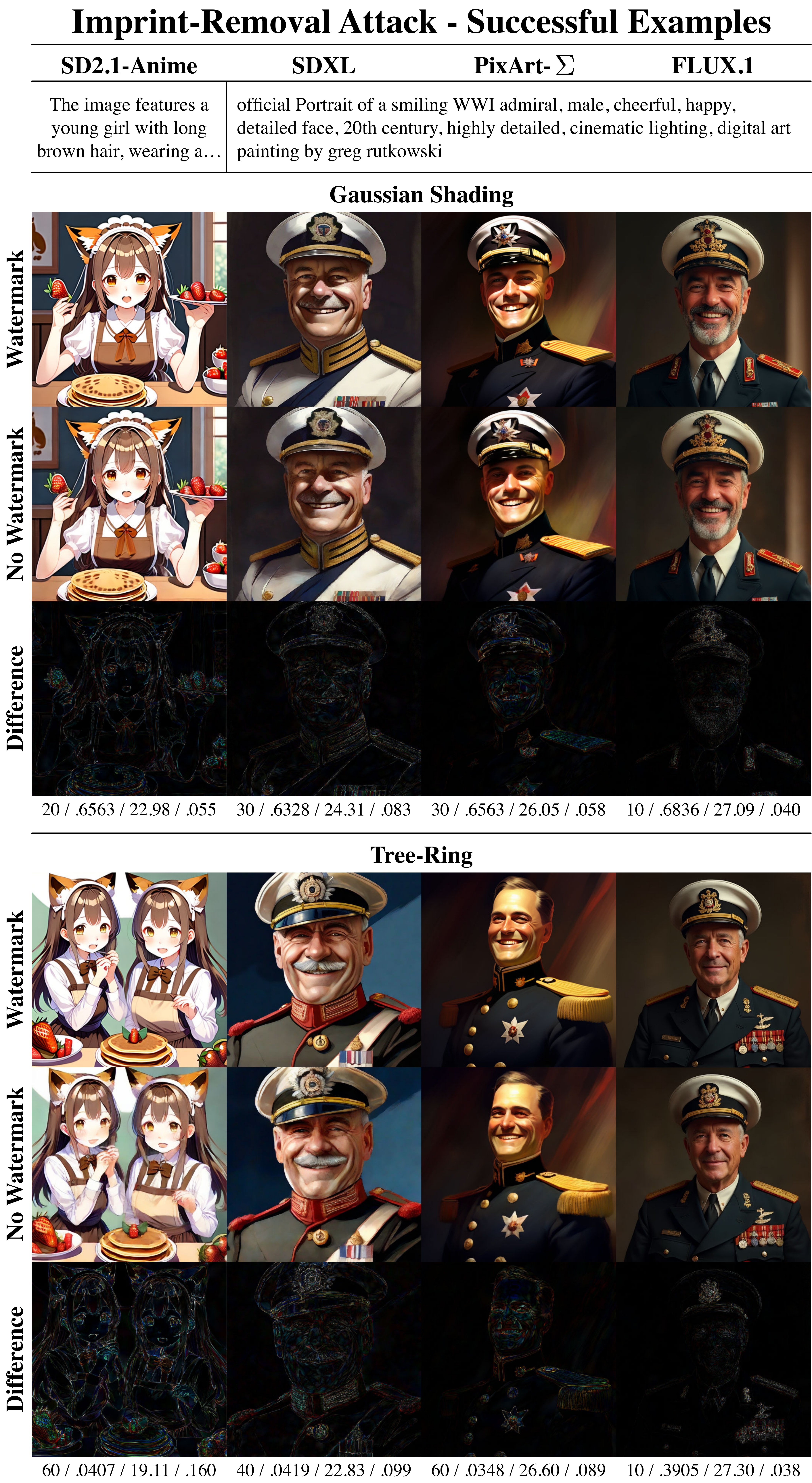}
    \caption{Examples of our watermark removal attack on different target models and both watermark approaches. 
    \setupfig
    For each target model (in the different columns), the 1st row shows the generated, watermarked image~\xwat, the 2nd row the attack example, and the 3rd row the absolute pixel difference between the two.
    Each attack example is the first one during generation to pass the detection threshold $\tau$ of Gaussian Shading (FPR=$10^{-6}$) or Tree-Ring (FPR=$1\%$). We report the step, the bit accuracy {(for Gaussian Shading)} or the p-value {(for Tree-Ring)}, as well as the PSNR and LPIPS score between the original and the attack examples. 
    \resultfig   
    Attacks against FLUX.1 generally display weaker visual perturbations, reflecting a lower removal difficulty.}
    \label{fig:visual:removal:models:1}
\end{figure*}

\begin{figure*}[t]
    \centering
    \includegraphics[width=0.90\linewidth]{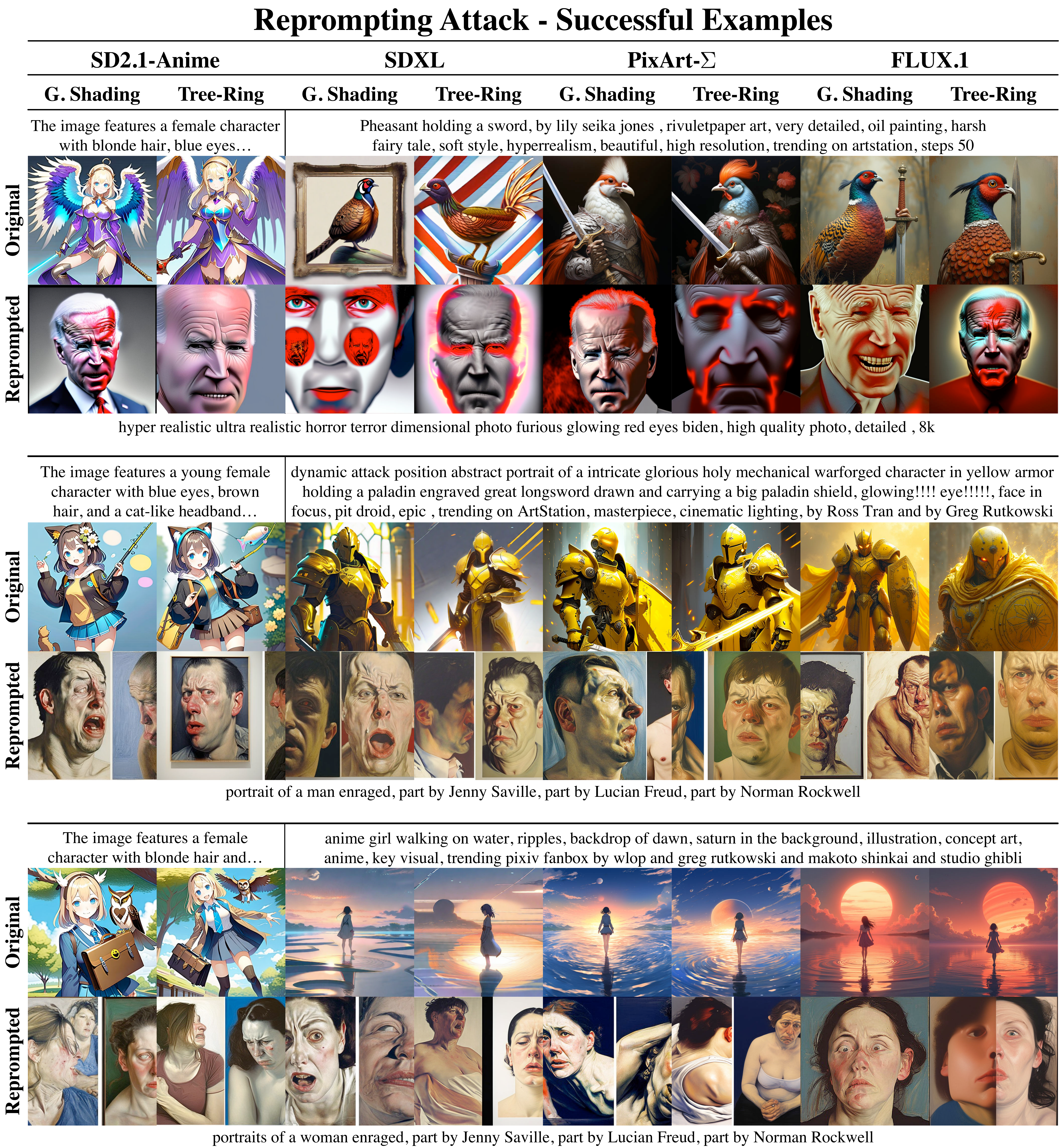}
    \caption{Examples of the \Reprompting attack on different target models and both watermarking approaches. 
    \setupfig
    In each section, the top row shows original images which carry a semantic watermark generated with prompts from the \href{https://huggingface.co/datasets/Gustavosta/Stable-Diffusion-Prompts}{Stable-Diffusion-Prompts} dataset using the respective target model indicated at the top of the figure. 
    The bottom row shows successful attack instances which our \Reprompting attack generated from each original reference image (using SD2.1 as proxy model) using prompts from the \href{https://huggingface.co/datasets/AIML-TUDA/i2p}{Inappropriate Image Prompts (I2P)} dataset.
    Note that we tried to choose less disturbing prompts (without blood, violence, or nudity) for this figure.
    \resultfig   
    These results illustrate how an attacker can potentially generate arbitrary harmful images (within the limits of the proxy model) with semantic watermarks spoofed from retrieved watermarked images to appear as generated by a service or user using the watermarking method. 
    }
    \label{fig:visual:reprompting}
\end{figure*}

\begin{figure}[t]
    \centering

    \begin{subfigure}{0.45\textwidth}
        \centering
        \includegraphics[width=0.70\linewidth,trim=0em 2em 0em 1em]{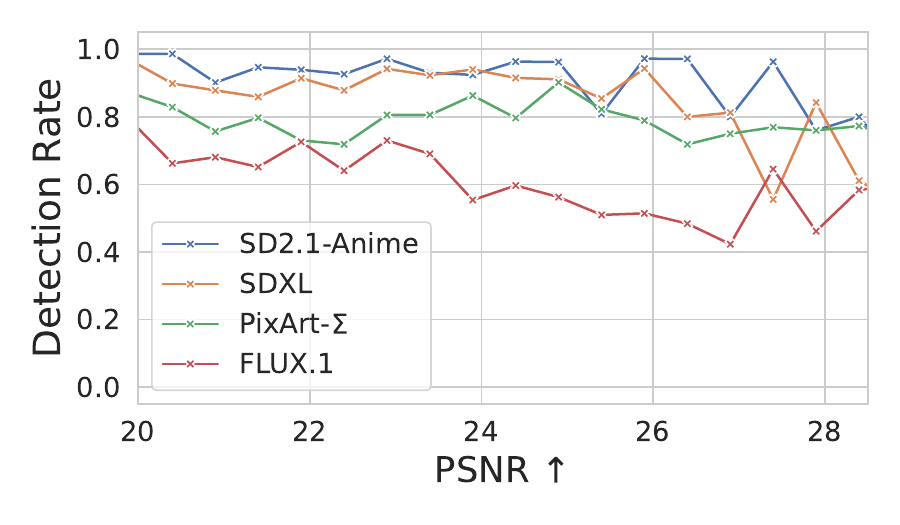}  %
        \caption{\ImprintForgeLong attack on {Gaussian Shading}.
        }  
        \label{fig:visual:trade-off:imprint:GS}
    \end{subfigure}
    \hfill
    \begin{subfigure}{0.45\textwidth}
    \centering
            \includegraphics[width=0.70\linewidth,trim=0em 2em 0em 1em]{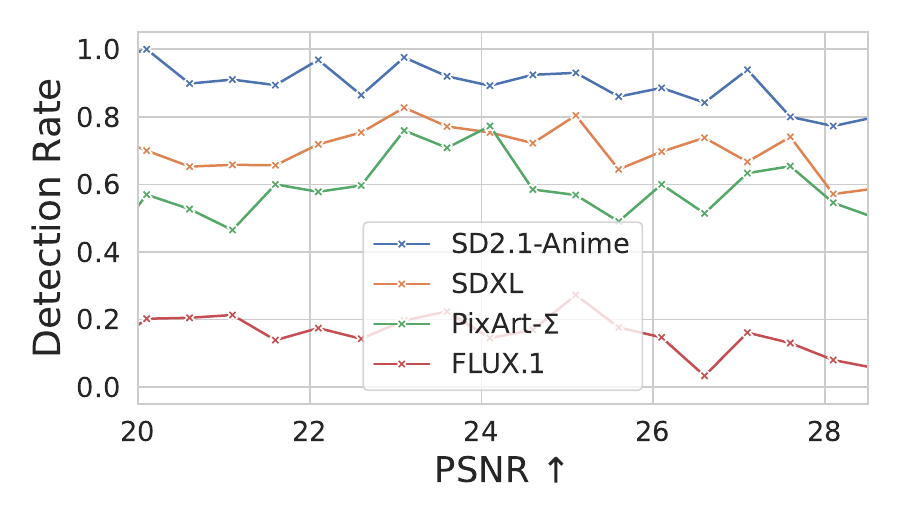} 
            \caption{\ImprintForgeLong attack on {Tree-Ring}.
            }  
            \label{fig:visual:trade-off:imprint:TR}
    \end{subfigure}

    \medskip
    \hspace{-0.2em}
    \begin{subfigure}{0.45\textwidth}
    \centering
            \includegraphics[width=0.70\linewidth,trim=0em 2em 0em 1em]{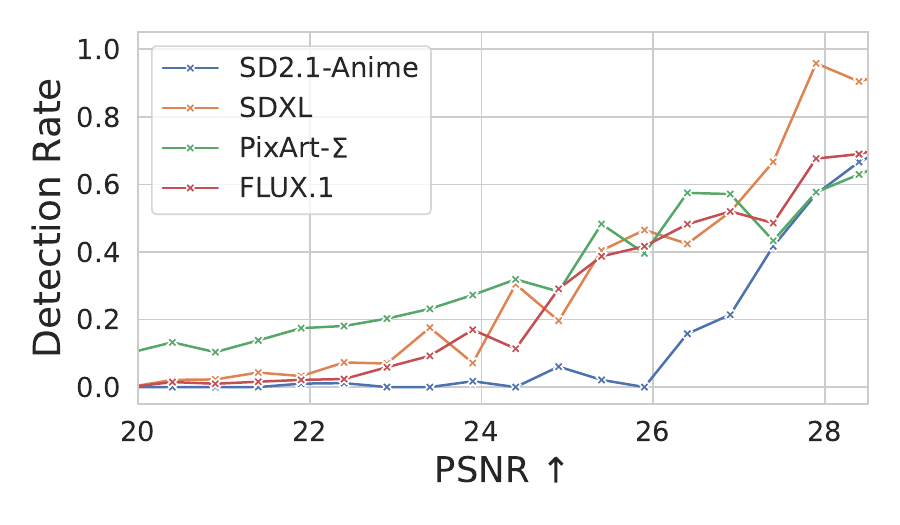} 
            \caption{\ImprintRemovalLong attack on {Gaussian Shading}.
            }  
            \label{fig:visual:trade-off:removal:GS}
    \end{subfigure}
    \hfill
    \begin{subfigure}{0.45\textwidth}
    \centering
            \includegraphics[width=0.70\linewidth,trim=0em 2em 0em 1em]{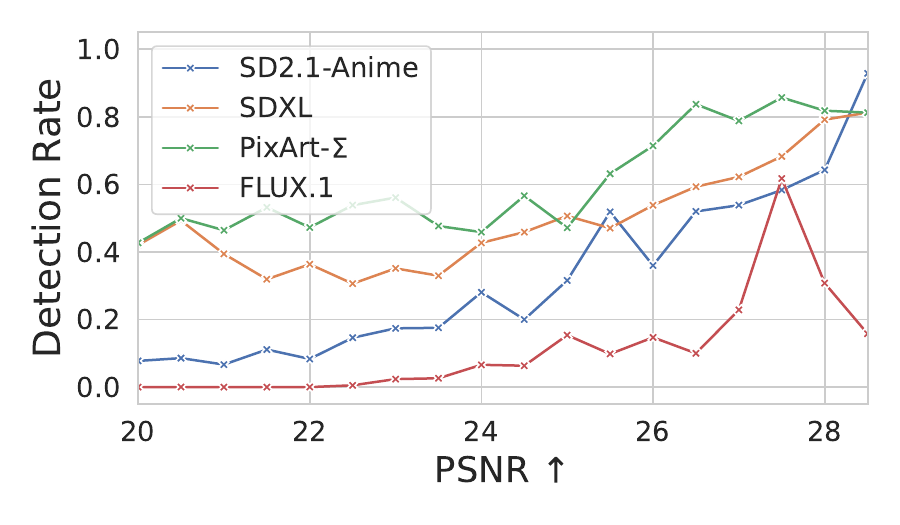}  %
            \caption{\ImprintRemovalLong attack on {Tree-Ring}.
            }  
            \label{fig:visual:trade-off:removal:TR}
    \end{subfigure}

\caption{Trade-off between attack success and image quality. To this end, we plot the detection rate vs. PSNR for watermark forgery (top) and removal (bottom) for both watermark approaches (left and right) on different target models. An attacker aims at a higher detection rate in the forgery case, and at a lower detection rate in the removal case.
For Gaussian Shading (left), the single-bit detection accuracy is reported with a threshold $\tau$ corresponding to an FPR of $10^{-6}$.
For Tree-Ring (right), the detection accuracy is reported with a threshold corresponding to an FPR of $1\%$.
PSNR is computed between the attack image and its initial starting image: the cover image for forgery, and watermarked image for removal. 
\resultfig
In the case of forgery, we observe a consistent trend where the detection rate increases as image quality decreases. For removal, the detection rate tends to decline rapidly as PSNR decreases.
}
\label{fig:visual:trade-off}
\end{figure}

\begin{figure}[t]
    \centering

    \includegraphics[width=0.78\linewidth]{images/visual/Baseline_comparison.pdf}%
    \caption{
    Visual comparison of our \Imprint attacks against baselines for watermark forgery (top) and removal (bottom) against a service provider using FLUX.1. This model is the one against which most of the baseline attacks are relatively effective, while our attacks have higher difficulty because we use SD2.1 as proxy model, which is very dissimilar to FLUX.1.
    Green check marks indicate successful detection of a watermark while red crossed indicate no detection (against threshold $\tau$ of Gaussian Shading (FPR=$10^{-6}$) or Tree-Ring (FPR=$1\%$)).
    \ImprintForgeLong and \ImprintRemovalLong quickly achieve forgery/removal success, while obtaining competitive PSNR. Notice however, that PSNR does not perfectly correlate with image quality, as exemplified by the high PSNR values for AdvEmb and Surrogate baselines, which introduce noticeable patterns in the image.
    }
\label{fig:baseline_comparison}
\end{figure}

\clearpage

\section{Full Experimental Results and Baselines}
\label{sec:full_results}
Here we compare our attacks with related work. 

\paragraph{Baselines.}
We use the \textit{Averaging Attack}~\cite{yang2024steganalysisdigitalwatermarkingdefense} as baseline for watermark forgery, and the \textit{Averaging}~\cite{yang2024steganalysisdigitalwatermarkingdefense}, \textit{Regeneration}~\cite{ZhaZhaSu2024invisibleimagewatermarksprovably}, \textit{Adversarial Embedding (AdvEmb)}~\cite{AnDinRab2024Benchmarking}, and \textit{Surrogate}~\cite{saberi2023robustness} Attacks as baselines for watermark removal.  
As implementation, we use the provided Github repositories\footnote{
The Github repositories for the Averaging (\url{https://github.com/showlab/watermark-steganalysis}), 
Regeneration (\url{https://github.com/XuandongZhao/WatermarkAttacker}), 
AdvEmb~(\url{https://github.com/umd-huang-lab/WAVES}), 
and Surrogate (\url{https://github.com/umd-huang-lab/WAVES}) attacks.}. 
Similar to the main attack experiments via \imprinting, the forgery baselines are evaluated using 100 \textit{MS-COCO-2017}~\cite{coco} cover images and the removal baselines are evaluated on 100 watermarked images generated using the \textit{Stable-Diffusion-Prompts} dataset.
We use the strongest attack settings available. Our setup is as follows:

\begin{itemize}
    \vspace{0.6em}
    \setlength{\itemsep}{0.3em} %
    \item The \textit{Averaging Attack}~\cite{yang2024steganalysisdigitalwatermarkingdefense} is applicable for forgery and removal. To obtain sets of averaged images, we randomly draw 30 different sets of size 10, 50, 100, 1k, and 5k from a set of 10k watermarked images which are then added to cover images (forgery) or subtracted from generated images (removal).

    \item The \textit{Regeneration Attack}~\cite{ZhaZhaSu2024invisibleimagewatermarksprovably} is deployed with maximum settings (i.e. 200 noising steps) using SD2.1.

    \item The \textit{Adversarial Embedding (AdvEmb) Attack}~\cite{AnDinRab2024Benchmarking} is deployed with maximum settings as well, i.e. a $L_{\infty}$-perturbation budget of $\frac{8}{255}$, using the auto-encoder of the SD2.1 model for 200 optimization steps.

    \item For the \textit{Surrogate Attack}~\cite{saberi2023robustness}, we follow the original training procedure. In particular, we train a binary classifier (ResNet-18) with 7500 real images from the ImageNet validation dataset~\cite{russakovsky2015imagenet} as negative class and 7500 watermarked images generated with each respective model using randomly drawn ImageNet class names as positive class. Optimization is performed with the maximum setting for the $L_{\infty}$-perturbation budget of $\frac{8}{255}$.
\end{itemize}

\paragraph{Results and Comparison.}
\cref{tab:full_forgery} shows detailed results for watermark forgery while \cref{tab:full_removal} shows the outcomes for watermark removal. In addition, \cref{fig:baseline_comparison} compares our attacks with all baselines \emph{visually}.

\vspace{0.5em}
\noindent \textit{1) Watermark Forgery.} 
For Tree-Ring, the Averaging Attack performs similarly to our attack on three out of four target models. With FLUX.1 as target model, the baseline outperforms our attack in terms of detection rate (of the forged watermarks). To avoid visible distortions, however, this attack requires a large number of reference images with the same watermark pattern. The cake example in \cref{fig:baseline_comparison} shows that 10 reference images lead to visible distortions, and more images for averaging are actually required. A large number of references images is, however, a rather strong assumption. In contrast, our attack requires only one reference image. For FLUX.1, we can still achieve up to 35\% forgery rate in our realistic setup.

Regarding Gaussian Shading, the Averaging Attack fails in all cases. This is in contrast to reported results in \cite{yang2024steganalysisdigitalwatermarkingdefense}. 
We attribute this discrepancy to an incorrect implementation of Gaussian Shading in the initial evaluation of the Averaging Attack, where the key and nonce of Gaussian Shading were reused.
If Gaussian Shading is used correctly, the attack fails as expected. We explain this issue in more detail in Sec.~\ref{sec:appendix:semantic-watermarks}.

\vspace{0.5em}
\noindent \textit{2) Watermark Removal.} 
For Tree-Ring, the Averaging Attack is the only removal attack that consistently works on all four target models---again with the assumption to have a large number of images with the same watermark key. 
For Gaussian Shading, our proposed attacks are the only attacks that are able to remove the watermarks.

Note that \citet{Gunn2024Undetectable} demonstrated that the Surrogate Attack is effective against Gaussian Shading. Like the Averaging Attack, this success relied on an incorrect instantiation of Gaussian Shading where key and nonce were re-used. Moreover, the Surrogate Attack also fails for Tree-Ring in the realistic setting where the negative class consists of real images (in contrast to unwatermarked images obtained by the same service). This is in line with the observation by~\citet{AnDinRab2024Benchmarking}. 

\paragraph{Summary.}
In terms of watermark forgery and removal, our attacks outperform or perform similarly to previously proposed methods while only requiring one watermarked target image. Importantly, we observe that previous attacks fail on Gaussian Shading, whereas our proposed attacks are effective. %

\clearpage

\begin{minipage}{1.0\textwidth}
    \centering
    
    \resizebox{1.0\linewidth}{!}{%
        \begin{tabular}{llllrrrrrrrrrrr}
\toprule
 &  &  &  & \multicolumn{6}{c}{{Gaussian Shading (FPR=$10^{-6}$)}} & \multicolumn{5}{c}{{Tree-Ring (FPR=$1\%$)}} \\
 &  &  &  & Bit Acc. & Det. & Attr. & PSNR & MS-SSIM & LPIPS & p-Value & Det. & PSNR & MS-SSIM & LPIPS \\
Model & Attk & Ref.Img. & Step &  &  &  &  &  &  &  &  &  &  &  \\
\midrule
\parbox{1px}{SD2.1-Anime} & Cover & - & - & ${0.50}_{\pm0.03}$ & 0.00 & 0.00 & inf & ${1.000}_{\pm0.000}$ & ${0.000}_{\pm0.000}$ & ${\num{4.97e-01}}_{\pm\num{2.54e-01}}$ & 0.00 & inf & ${1.000}_{\pm0.000}$ & ${0.000}_{\pm0.000}$ \\
\cline{2-15} \noalign{\vskip 0.2em}
 & \ImprintForgeShort & 1 & 10 & ${0.85}_{\pm0.05}$ & 1.00 & 1.00 & ${24.519}_{\pm4.204}$ & ${0.904}_{\pm0.052}$ & ${0.092}_{\pm0.046}$ & ${\num{2.68e-02}}_{\pm\num{7.19e-02}}$ & 0.85 & ${24.517}_{\pm4.192}$ & ${0.904}_{\pm0.053}$ & ${0.092}_{\pm0.047}$ \\
 &  &  & 20 & ${0.93}_{\pm0.04}$ & 1.00 & 1.00 & ${23.581}_{\pm3.978}$ & ${0.877}_{\pm0.061}$ & ${0.116}_{\pm0.050}$ & ${\num{4.74e-03}}_{\pm\num{1.81e-02}}$ & 0.94 & ${23.592}_{\pm3.968}$ & ${0.878}_{\pm0.062}$ & ${0.116}_{\pm0.051}$ \\
 &  &  & 50 & ${0.97}_{\pm0.02}$ & 1.00 & 1.00 & ${22.129}_{\pm3.713}$ & ${0.829}_{\pm0.081}$ & ${0.160}_{\pm0.058}$ & ${\num{1.83e-04}}_{\pm\num{8.89e-04}}$ & 1.00 & ${22.162}_{\pm3.707}$ & ${0.830}_{\pm0.081}$ & ${0.160}_{\pm0.058}$ \\
 &  &  & 100 & ${0.98}_{\pm0.02}$ & 1.00 & 1.00 & ${20.900}_{\pm3.487}$ & ${0.775}_{\pm0.101}$ & ${0.208}_{\pm0.066}$ & ${\num{6.59e-06}}_{\pm\num{4.87e-05}}$ & 1.00 & ${20.946}_{\pm3.502}$ & ${0.778}_{\pm0.101}$ & ${0.209}_{\pm0.067}$ \\
 &  &  & 150 & ${0.99}_{\pm0.01}$ & 1.00 & 1.00 & ${20.114}_{\pm3.329}$ & ${0.736}_{\pm0.114}$ & ${0.246}_{\pm0.073}$ & ${\num{1.45e-06}}_{\pm\num{1.25e-05}}$ & 1.00 & ${20.156}_{\pm3.350}$ & ${0.739}_{\pm0.113}$ & ${0.248}_{\pm0.074}$ \\
\cline{2-15} \noalign{\vskip 0.2em}
 & \Reprompt & 1 & - & ${0.92}_{\pm0.07}$ & 0.98 & 0.98 & - & - & - & ${\num{1.40e-02}}_{\pm\num{5.45e-02}}$ & 0.90 & - & - & - \\
 & \RepromptPlus & 1 & - & ${1.00}_{\pm0.01}$ & 1.00 & 1.00 & - & - & - & ${\num{3.83e-05}}_{\pm\num{1.69e-04}}$ & 1.00 & - & - & - \\
\cline{2-15} \noalign{\vskip 0.2em}
 & Averaging & 10 & 0 & ${0.50}_{\pm0.03}$ & 0.00 & 0.00 & ${18.504}_{\pm0.822}$ & ${0.724}_{\pm0.095}$ & ${0.318}_{\pm0.128}$ & ${\num{2.65e-04}}_{\pm\num{1.13e-03}}$ & 1.00 & ${18.326}_{\pm0.966}$ & ${0.718}_{\pm0.102}$ & ${0.325}_{\pm0.133}$ \\
 &  & 50 & 0 & ${0.50}_{\pm0.03}$ & 0.00 & 0.00 & ${24.426}_{\pm1.289}$ & ${0.893}_{\pm0.051}$ & ${0.138}_{\pm0.087}$ & ${\num{7.05e-07}}_{\pm\num{6.45e-06}}$ & 1.00 & ${23.986}_{\pm1.051}$ & ${0.885}_{\pm0.055}$ & ${0.142}_{\pm0.091}$ \\
 &  & 100 & 0 & ${0.50}_{\pm0.03}$ & 0.00 & 0.00 & ${27.398}_{\pm0.837}$ & ${0.937}_{\pm0.032}$ & ${0.086}_{\pm0.065}$ & ${\num{4.46e-09}}_{\pm\num{3.70e-08}}$ & 1.00 & ${25.861}_{\pm0.808}$ & ${0.926}_{\pm0.037}$ & ${0.090}_{\pm0.065}$ \\
 &  & 1000 & 0 & ${0.50}_{\pm0.03}$ & 0.00 & 0.00 & ${31.908}_{\pm0.904}$ & ${0.990}_{\pm0.006}$ & ${0.017}_{\pm0.013}$ & ${\num{3.13e-10}}_{\pm\num{3.09e-09}}$ & 1.00 & ${28.829}_{\pm0.649}$ & ${0.975}_{\pm0.014}$ & ${0.027}_{\pm0.021}$ \\
 &  & 5000 & 0 & ${0.50}_{\pm0.03}$ & 0.00 & 0.00 & ${32.779}_{\pm0.483}$ & ${0.996}_{\pm0.003}$ & ${0.010}_{\pm0.006}$ & ${\num{4.67e-10}}_{\pm\num{4.66e-09}}$ & 1.00 & ${29.272}_{\pm0.389}$ & ${0.980}_{\pm0.011}$ & ${0.021}_{\pm0.015}$ \\
\midrule

SDXL & Cover & - & - & ${0.50}_{\pm0.03}$ & 0.00 & 0.00 & inf & ${1.000}_{\pm0.000}$ & ${0.000}_{\pm0.000}$ & ${\num{4.72e-01}}_{\pm\num{2.47e-01}}$ & 0.01 & inf & ${1.000}_{\pm0.000}$ & ${0.000}_{\pm0.000}$ \\
\cline{2-15} \noalign{\vskip 0.2em}
 & \ImprintForgeShort & 1 & 10 & ${0.71}_{\pm0.04}$ & 0.44 & 0.44 & ${24.189}_{\pm4.291}$ & ${0.892}_{\pm0.092}$ & ${0.101}_{\pm0.067}$ & ${\num{1.89e-01}}_{\pm\num{1.85e-01}}$ & 0.21 & ${24.500}_{\pm4.186}$ & ${0.903}_{\pm0.052}$ & ${0.093}_{\pm0.048}$ \\
 &  &  & 20 & ${0.79}_{\pm0.04}$ & 0.99 & 0.99 & ${23.272}_{\pm4.029}$ & ${0.865}_{\pm0.096}$ & ${0.125}_{\pm0.068}$ & ${\num{9.63e-02}}_{\pm\num{1.30e-01}}$ & 0.43 & ${23.552}_{\pm3.952}$ & ${0.877}_{\pm0.062}$ & ${0.117}_{\pm0.052}$ \\
 &  &  & 50 & ${0.87}_{\pm0.03}$ & 1.00 & 1.00 & ${21.858}_{\pm3.703}$ & ${0.817}_{\pm0.104}$ & ${0.168}_{\pm0.071}$ & ${\num{2.70e-02}}_{\pm\num{5.44e-02}}$ & 0.72 & ${22.090}_{\pm3.679}$ & ${0.827}_{\pm0.080}$ & ${0.161}_{\pm0.059}$ \\
 &  &  & 100 & ${0.91}_{\pm0.02}$ & 1.00 & 1.00 & ${20.510}_{\pm3.758}$ & ${0.756}_{\pm0.135}$ & ${0.219}_{\pm0.095}$ & ${\num{6.89e-03}}_{\pm\num{2.82e-02}}$ & 0.95 & ${20.863}_{\pm3.459}$ & ${0.775}_{\pm0.100}$ & ${0.207}_{\pm0.066}$ \\
 &  &  & 150 & ${0.93}_{\pm0.03}$ & 1.00 & 1.00 & ${19.770}_{\pm3.572}$ & ${0.718}_{\pm0.140}$ & ${0.253}_{\pm0.096}$ & ${\num{5.43e-03}}_{\pm\num{3.67e-02}}$ & 0.99 & ${20.087}_{\pm3.309}$ & ${0.736}_{\pm0.112}$ & ${0.242}_{\pm0.073}$ \\
\cline{2-15} \noalign{\vskip 0.2em}
 & \Reprompt & 1 & - & ${0.91}_{\pm0.06}$ & 0.99 & 0.99 & - & - & - & ${\num{5.20e-03}}_{\pm\num{3.46e-02}}$ & 0.97 & - & - & - \\
 & \RepromptPlus & 1 & - & ${0.95}_{\pm0.02}$ & 1.00 & 1.00 & - & - & - & ${\num{2.57e-04}}_{\pm\num{1.73e-03}}$ & 1.00 & - & - & - \\
\cline{2-15} \noalign{\vskip 0.2em}
 & Averaging & 10 & 0 & ${0.50}_{\pm0.03}$ & 0.00 & 0.00 & ${17.433}_{\pm1.482}$ & ${0.693}_{\pm0.108}$ & ${0.364}_{\pm0.143}$ & ${\num{3.25e-04}}_{\pm\num{1.94e-03}}$ & 1.00 & ${17.123}_{\pm1.218}$ & ${0.683}_{\pm0.107}$ & ${0.363}_{\pm0.139}$ \\
 &  & 50 & 0 & ${0.49}_{\pm0.03}$ & 0.00 & 0.00 & ${22.085}_{\pm1.301}$ & ${0.870}_{\pm0.060}$ & ${0.166}_{\pm0.101}$ & ${\num{2.18e-06}}_{\pm\num{1.93e-05}}$ & 1.00 & ${20.449}_{\pm1.099}$ & ${0.860}_{\pm0.059}$ & ${0.166}_{\pm0.096}$ \\
 &  & 100 & 0 & ${0.50}_{\pm0.03}$ & 0.00 & 0.00 & ${22.757}_{\pm1.613}$ & ${0.920}_{\pm0.038}$ & ${0.105}_{\pm0.073}$ & ${\num{3.49e-06}}_{\pm\num{3.49e-05}}$ & 1.00 & ${21.650}_{\pm1.246}$ & ${0.901}_{\pm0.045}$ & ${0.113}_{\pm0.073}$ \\
 &  & 1000 & 0 & ${0.50}_{\pm0.03}$ & 0.00 & 0.00 & ${24.244}_{\pm0.771}$ & ${0.984}_{\pm0.009}$ & ${0.023}_{\pm0.014}$ & ${\num{3.99e-08}}_{\pm\num{2.72e-07}}$ & 1.00 & ${22.611}_{\pm0.596}$ & ${0.954}_{\pm0.021}$ & ${0.046}_{\pm0.029}$ \\
 &  & 5000 & 0 & ${0.50}_{\pm0.03}$ & 0.00 & 0.00 & ${24.482}_{\pm0.457}$ & ${0.992}_{\pm0.007}$ & ${0.014}_{\pm0.008}$ & ${\num{3.28e-08}}_{\pm\num{2.16e-07}}$ & 1.00 & ${22.699}_{\pm0.490}$ & ${0.959}_{\pm0.018}$ & ${0.040}_{\pm0.024}$ \\
\midrule

PixArt-$\Sigma$ & Cover & - & - & ${0.50}_{\pm0.03}$ & 0.00 & 0.00 & inf & ${1.000}_{\pm0.000}$ & ${0.000}_{\pm0.000}$ & ${\num{4.74e-01}}_{\pm\num{2.63e-01}}$ & 0.01 & inf & ${1.000}_{\pm0.000}$ & ${0.000}_{\pm0.000}$ \\
\cline{2-15} \noalign{\vskip 0.2em}
 & \ImprintForgeShort & 1 & 10 & ${0.67}_{\pm0.06}$ & 0.28 & 0.28 & ${24.504}_{\pm4.219}$ & ${0.904}_{\pm0.052}$ & ${0.092}_{\pm0.046}$ & ${\num{2.03e-01}}_{\pm\num{2.00e-01}}$ & 0.12 & ${24.522}_{\pm4.200}$ & ${0.904}_{\pm0.053}$ & ${0.093}_{\pm0.047}$ \\
 &  &  & 20 & ${0.74}_{\pm0.07}$ & 0.72 & 0.72 & ${23.570}_{\pm3.991}$ & ${0.877}_{\pm0.061}$ & ${0.117}_{\pm0.050}$ & ${\num{1.19e-01}}_{\pm\num{1.67e-01}}$ & 0.29 & ${23.590}_{\pm3.972}$ & ${0.878}_{\pm0.062}$ & ${0.116}_{\pm0.051}$ \\
 &  &  & 50 & ${0.82}_{\pm0.07}$ & 0.91 & 0.91 & ${22.125}_{\pm3.730}$ & ${0.829}_{\pm0.080}$ & ${0.161}_{\pm0.058}$ & ${\num{5.02e-02}}_{\pm\num{9.19e-02}}$ & 0.55 & ${22.145}_{\pm3.708}$ & ${0.829}_{\pm0.081}$ & ${0.160}_{\pm0.058}$ \\
 &  &  & 100 & ${0.86}_{\pm0.08}$ & 0.94 & 0.94 & ${20.902}_{\pm3.523}$ & ${0.777}_{\pm0.100}$ & ${0.209}_{\pm0.067}$ & ${\num{1.98e-02}}_{\pm\num{5.38e-02}}$ & 0.78 & ${20.925}_{\pm3.487}$ & ${0.777}_{\pm0.102}$ & ${0.207}_{\pm0.066}$ \\
 &  &  & 150 & ${0.88}_{\pm0.08}$ & 0.96 & 0.96 & ${20.121}_{\pm3.384}$ & ${0.738}_{\pm0.112}$ & ${0.246}_{\pm0.074}$ & ${\num{1.08e-02}}_{\pm\num{3.65e-02}}$ & 0.84 & ${20.149}_{\pm3.335}$ & ${0.739}_{\pm0.114}$ & ${0.245}_{\pm0.074}$ \\
\cline{2-15} \noalign{\vskip 0.2em}
 & \Reprompt & 1 & - & ${0.88}_{\pm0.09}$ & 0.93 & 0.93 & - & - & - & ${\num{1.64e-02}}_{\pm\num{6.91e-02}}$ & 0.88 & - & - & - \\
 & \RepromptPlus & 1 & - & ${0.92}_{\pm0.06}$ & 1.00 & 1.00 & - & - & - & ${\num{1.54e-03}}_{\pm\num{1.03e-02}}$ & 0.99 & - & - & - \\
\cline{2-15} \noalign{\vskip 0.2em}
 & Averaging & 10 & 0 & ${0.50}_{\pm0.03}$ & 0.00 & 0.00 & ${14.969}_{\pm2.120}$ & ${0.749}_{\pm0.082}$ & ${0.308}_{\pm0.126}$ & ${\num{3.73e-02}}_{\pm\num{8.14e-02}}$ & 0.69 & ${15.392}_{\pm2.057}$ & ${0.752}_{\pm0.083}$ & ${0.308}_{\pm0.124}$ \\
 &  & 50 & 0 & ${0.49}_{\pm0.03}$ & 0.00 & 0.00 & ${16.203}_{\pm1.346}$ & ${0.877}_{\pm0.043}$ & ${0.164}_{\pm0.079}$ & ${\num{1.61e-02}}_{\pm\num{4.70e-02}}$ & 0.87 & ${16.323}_{\pm1.491}$ & ${0.874}_{\pm0.045}$ & ${0.166}_{\pm0.081}$ \\
 &  & 100 & 0 & ${0.50}_{\pm0.03}$ & 0.00 & 0.00 & ${16.362}_{\pm1.344}$ & ${0.903}_{\pm0.035}$ & ${0.133}_{\pm0.064}$ & ${\num{1.35e-02}}_{\pm\num{4.87e-02}}$ & 0.85 & ${16.769}_{\pm1.395}$ & ${0.902}_{\pm0.040}$ & ${0.132}_{\pm0.070}$ \\
 &  & 1000 & 0 & ${0.50}_{\pm0.03}$ & 0.00 & 0.00 & ${16.635}_{\pm1.132}$ & ${0.934}_{\pm0.030}$ & ${0.093}_{\pm0.042}$ & ${\num{1.14e-02}}_{\pm\num{5.22e-02}}$ & 0.91 & ${16.729}_{\pm1.178}$ & ${0.931}_{\pm0.030}$ & ${0.096}_{\pm0.045}$ \\
 &  & 5000 & 0 & ${0.49}_{\pm0.03}$ & 0.00 & 0.00 & ${16.587}_{\pm1.112}$ & ${0.937}_{\pm0.030}$ & ${0.091}_{\pm0.041}$ & ${\num{1.18e-02}}_{\pm\num{6.07e-02}}$ & 0.93 & ${16.731}_{\pm1.165}$ & ${0.934}_{\pm0.030}$ & ${0.092}_{\pm0.044}$ \\
\midrule

FLUX.1 & Cover & - & - & ${0.49}_{\pm0.04}$ & 0.00 & 0.00 & inf & ${1.000}_{\pm0.000}$ & ${0.000}_{\pm0.000}$ & ${\num{4.99e-01}}_{\pm\num{2.65e-01}}$ & 0.00 & inf & ${1.000}_{\pm0.000}$ & ${0.000}_{\pm0.000}$ \\
\cline{2-15} \noalign{\vskip 0.2em}
 & \ImprintForgeShort & 1 & 10 & ${0.62}_{\pm0.05}$ & 0.02 & 0.02 & ${24.469}_{\pm4.206}$ & ${0.903}_{\pm0.053}$ & ${0.093}_{\pm0.047}$ & ${\num{3.50e-01}}_{\pm\num{2.50e-01}}$ & 0.02 & ${24.487}_{\pm4.164}$ & ${0.904}_{\pm0.052}$ & ${0.092}_{\pm0.047}$ \\
 &  &  & 20 & ${0.67}_{\pm0.05}$ & 0.26 & 0.26 & ${23.540}_{\pm3.992}$ & ${0.877}_{\pm0.063}$ & ${0.117}_{\pm0.051}$ & ${\num{2.69e-01}}_{\pm\num{2.19e-01}}$ & 0.07 & ${23.545}_{\pm3.947}$ & ${0.878}_{\pm0.061}$ & ${0.117}_{\pm0.051}$ \\
 &  &  & 50 & ${0.74}_{\pm0.05}$ & 0.66 & 0.66 & ${22.108}_{\pm3.746}$ & ${0.828}_{\pm0.082}$ & ${0.161}_{\pm0.058}$ & ${\num{2.62e-01}}_{\pm\num{2.42e-01}}$ & 0.12 & ${22.107}_{\pm3.695}$ & ${0.829}_{\pm0.081}$ & ${0.160}_{\pm0.058}$ \\
 &  &  & 100 & ${0.78}_{\pm0.06}$ & 0.86 & 0.86 & ${20.905}_{\pm3.545}$ & ${0.776}_{\pm0.102}$ & ${0.209}_{\pm0.068}$ & ${\num{1.92e-01}}_{\pm\num{2.05e-01}}$ & 0.20 & ${20.912}_{\pm3.489}$ & ${0.778}_{\pm0.101}$ & ${0.205}_{\pm0.065}$ \\
 &  &  & 150 & ${0.81}_{\pm0.05}$ & 0.95 & 0.95 & ${20.137}_{\pm3.406}$ & ${0.738}_{\pm0.115}$ & ${0.247}_{\pm0.075}$ & ${\num{1.80e-01}}_{\pm\num{2.22e-01}}$ & 0.23 & ${20.153}_{\pm3.346}$ & ${0.740}_{\pm0.114}$ & ${0.239}_{\pm0.070}$ \\
\cline{2-15} \noalign{\vskip 0.2em}
 & \Reprompt & 1 & - & ${0.74}_{\pm0.08}$ & 0.66 & 0.66 & - & - & - & ${\num{2.37e-01}}_{\pm\num{2.30e-01}}$ & 0.13 & - & - & - \\
 & \RepromptPlus & 1 & - & ${0.79}_{\pm0.06}$ & 0.88 & 0.88 & - & - & - & ${\num{8.99e-02}}_{\pm\num{1.09e-01}}$ & 0.35 & - & - & - \\
\cline{2-15} \noalign{\vskip 0.2em}
 & Averaging & 10 & 0 & ${0.50}_{\pm0.03}$ & 0.00 & 0.00 & ${18.472}_{\pm1.770}$ & ${0.812}_{\pm0.072}$ & ${0.224}_{\pm0.106}$ & ${\num{6.88e-02}}_{\pm\num{1.47e-01}}$ & 0.63 & ${18.935}_{\pm1.744}$ & ${0.810}_{\pm0.073}$ & ${0.228}_{\pm0.114}$ \\
 &  & 50 & 0 & ${0.50}_{\pm0.03}$ & 0.00 & 0.00 & ${23.340}_{\pm2.138}$ & ${0.933}_{\pm0.033}$ & ${0.084}_{\pm0.058}$ & ${\num{3.11e-02}}_{\pm\num{9.06e-02}}$ & 0.80 & ${23.703}_{\pm1.900}$ & ${0.932}_{\pm0.032}$ & ${0.085}_{\pm0.059}$ \\
 &  & 100 & 0 & ${0.50}_{\pm0.03}$ & 0.00 & 0.00 & ${24.794}_{\pm2.072}$ & ${0.960}_{\pm0.020}$ & ${0.052}_{\pm0.039}$ & ${\num{1.84e-02}}_{\pm\num{5.62e-02}}$ & 0.83 & ${24.802}_{\pm1.666}$ & ${0.959}_{\pm0.020}$ & ${0.054}_{\pm0.041}$ \\
 &  & 1000 & 0 & ${0.50}_{\pm0.03}$ & 0.00 & 0.00 & ${26.209}_{\pm1.100}$ & ${0.988}_{\pm0.008}$ & ${0.019}_{\pm0.012}$ & ${\num{9.05e-03}}_{\pm\num{2.90e-02}}$ & 0.90 & ${26.089}_{\pm0.978}$ & ${0.986}_{\pm0.009}$ & ${0.020}_{\pm0.015}$ \\
 &  & 5000 & 0 & ${0.50}_{\pm0.03}$ & 0.00 & 0.00 & ${26.421}_{\pm0.601}$ & ${0.991}_{\pm0.007}$ & ${0.016}_{\pm0.010}$ & ${\num{9.31e-03}}_{\pm\num{3.15e-02}}$ & 0.90 & ${26.162}_{\pm0.667}$ & ${0.989}_{\pm0.008}$ & ${0.017}_{\pm0.014}$ \\
\bottomrule
\end{tabular}
    }
    \captionof{table}{Full experimental results on forgery attacks. \textit{Ref.Img.} and \textit{Step} refer to the number of reference images and the attack step used, respectively. \textit{Bit Acc.} refers to Gaussian Shading bit accuracy. \textit{Det.} and \textit{Att.} refer to mean detection and attribution success. \textit{\ImprintForgeShort} and \textit{\Reprompt/\RepromptPlus} refer to our \ImprintForgeLong and \Reprompting/\RepromptingPlus attacks, respectively. The Averaging attack~\cite{yang2024steganalysisdigitalwatermarkingdefense} is tested with varying amounts of reference images.}
    \label{tab:full_forgery}

\end{minipage}

\begin{minipage}{1.0\textwidth}
    \centering

    \resizebox{1.0\linewidth}{!}{%
        \begin{tabular}{lllllrrrrrrrrrrr}
\toprule
 &  &  &  &  & \multicolumn{6}{c}{{Gaussian Shading (FPR=$10^{-6}$)}} & \multicolumn{5}{c}{{Tree-Ring (FPR=$1\%$)}} \\
 &  &  &  &  & Bit Acc. & Det. & Attr. & PSNR & MS-SSIM & LPIPS & p-Value & Det. & PSNR & MS-SSIM & LPIPS \\
Model & Attk & Ref.Img. & Str. & Step &  &  &  &  &  &  &  &  &  &  &  \\
\midrule
\parbox{1px}{SD2.1-Anime} & Original & - & - & - & ${1.00}_{\pm0.00}$ & 1.00 & 1.00 & inf & ${1.000}_{\pm0.000}$ & ${0.000}_{\pm0.000}$ & ${\num{4.29e-17}}_{\pm\num{3.41e-16}}$ & 1.00 & inf & ${1.000}_{\pm0.000}$ & ${0.000}_{\pm0.000}$ \\
\cline{2-16} \noalign{\vskip 0.2em}
 & \ImprintRemovalShort & 1 & - & 10 & ${0.79}_{\pm0.11}$ & 0.77 & 0.77 & ${28.935}_{\pm1.360}$ & ${0.978}_{\pm0.006}$ & ${0.026}_{\pm0.006}$ & ${\num{1.55e-02}}_{\pm\num{5.63e-02}}$ & 0.87 & ${29.007}_{\pm1.270}$ & ${0.979}_{\pm0.005}$ & ${0.026}_{\pm0.006}$ \\
 &  &  &  & 20 & ${0.45}_{\pm0.13}$ & 0.04 & 0.04 & ${25.876}_{\pm1.330}$ & ${0.957}_{\pm0.012}$ & ${0.051}_{\pm0.011}$ & ${\num{2.39e-01}}_{\pm\num{3.16e-01}}$ & 0.50 & ${25.940}_{\pm1.240}$ & ${0.958}_{\pm0.011}$ & ${0.051}_{\pm0.011}$ \\
 &  &  &  & 50 & ${0.16}_{\pm0.07}$ & 0.00 & 0.00 & ${22.747}_{\pm1.302}$ & ${0.917}_{\pm0.023}$ & ${0.096}_{\pm0.020}$ & ${\num{6.15e-01}}_{\pm\num{4.00e-01}}$ & 0.14 & ${22.815}_{\pm1.199}$ & ${0.920}_{\pm0.021}$ & ${0.095}_{\pm0.020}$ \\
 &  &  &  & 100 & ${0.05}_{\pm0.03}$ & 0.00 & 0.00 & ${20.616}_{\pm1.262}$ & ${0.868}_{\pm0.035}$ & ${0.145}_{\pm0.029}$ & ${\num{8.06e-01}}_{\pm\num{3.29e-01}}$ & 0.04 & ${20.670}_{\pm1.158}$ & ${0.871}_{\pm0.032}$ & ${0.144}_{\pm0.028}$ \\
 &  &  &  & 150 & ${0.02}_{\pm0.01}$ & 0.00 & 0.00 & ${19.310}_{\pm1.245}$ & ${0.823}_{\pm0.046}$ & ${0.184}_{\pm0.033}$ & ${\num{8.79e-01}}_{\pm\num{2.73e-01}}$ & 0.03 & ${19.353}_{\pm1.125}$ & ${0.826}_{\pm0.041}$ & ${0.182}_{\pm0.032}$ \\
\cline{2-16} \noalign{\vskip 0.2em}
 & Averaging & 10 & - & 0 & ${0.91}_{\pm0.05}$ & 1.00 & 1.00 & ${18.672}_{\pm0.789}$ & ${0.789}_{\pm0.055}$ & ${0.247}_{\pm0.066}$ & ${\num{8.15e-01}}_{\pm\num{3.01e-01}}$ & 0.05 & ${18.456}_{\pm0.884}$ & ${0.786}_{\pm0.054}$ & ${0.255}_{\pm0.062}$ \\
 &  & 50 & - & 0 & ${0.96}_{\pm0.04}$ & 1.00 & 1.00 & ${24.370}_{\pm1.310}$ & ${0.919}_{\pm0.024}$ & ${0.108}_{\pm0.046}$ & ${\num{9.46e-01}}_{\pm\num{1.78e-01}}$ & 0.01 & ${24.035}_{\pm1.002}$ & ${0.916}_{\pm0.025}$ & ${0.110}_{\pm0.041}$ \\
 &  & 100 & - & 0 & ${0.97}_{\pm0.03}$ & 1.00 & 1.00 & ${27.276}_{\pm0.963}$ & ${0.953}_{\pm0.015}$ & ${0.068}_{\pm0.035}$ & ${\num{9.75e-01}}_{\pm\num{1.19e-01}}$ & 0.01 & ${25.706}_{\pm0.802}$ & ${0.946}_{\pm0.018}$ & ${0.071}_{\pm0.032}$ \\
 &  & 1000 & - & 0 & ${0.99}_{\pm0.01}$ & 1.00 & 1.00 & ${31.178}_{\pm0.941}$ & ${0.993}_{\pm0.002}$ & ${0.013}_{\pm0.009}$ & ${\num{9.96e-01}}_{\pm\num{3.31e-02}}$ & 0.00 & ${28.503}_{\pm0.641}$ & ${0.982}_{\pm0.008}$ & ${0.019}_{\pm0.012}$ \\
 &  & 5000 & - & 0 & ${1.00}_{\pm0.00}$ & 1.00 & 1.00 & ${31.867}_{\pm0.438}$ & ${0.997}_{\pm0.001}$ & ${0.007}_{\pm0.004}$ & ${\num{9.97e-01}}_{\pm\num{2.32e-02}}$ & 0.00 & ${28.886}_{\pm0.213}$ & ${0.986}_{\pm0.007}$ & ${0.014}_{\pm0.009}$ \\
\cline{2-16} \noalign{\vskip 0.2em}
 & Regen. & 1 & - & 200 & ${0.92}_{\pm0.04}$ & 1.00 & 1.00 & ${21.269}_{\pm1.098}$ & ${0.902}_{\pm0.023}$ & ${0.111}_{\pm0.021}$ & ${\num{4.43e-03}}_{\pm\num{1.14e-02}}$ & 0.95 & ${21.783}_{\pm1.298}$ & ${0.903}_{\pm0.023}$ & ${0.110}_{\pm0.023}$ \\
 & AdvEmb & 1 & 8 & 200 & ${0.86}_{\pm0.06}$ & 0.99 & 0.99 & ${31.299}_{\pm0.192}$ & ${0.961}_{\pm0.008}$ & ${0.221}_{\pm0.056}$ & ${\num{6.41e-03}}_{\pm\num{2.96e-02}}$ & 0.94 & ${31.267}_{\pm0.155}$ & ${0.962}_{\pm0.008}$ & ${0.232}_{\pm0.050}$ \\
\cline{2-16} \noalign{\vskip 0.2em}
 & Surrogate & 7500 & 8 & 200 & ${0.96}_{\pm0.03}$ & 1.00 & 1.00 & ${32.040}_{\pm0.170}$ & ${0.963}_{\pm0.009}$ & ${0.147}_{\pm0.043}$ & ${\num{1.54e-03}}_{\pm\num{1.18e-02}}$ & 0.99 & ${32.025}_{\pm0.162}$ & ${0.963}_{\pm0.008}$ & ${0.151}_{\pm0.039}$ \\
\midrule

SDXL & Original & - & - & - & ${1.00}_{\pm0.00}$ & 1.00 & 1.00 & inf & ${1.000}_{\pm0.000}$ & ${0.000}_{\pm0.000}$ & ${\num{1.10e-21}}_{\pm\num{1.10e-20}}$ & 1.00 & inf & ${1.000}_{\pm0.000}$ & ${0.000}_{\pm0.000}$ \\
\cline{2-16} \noalign{\vskip 0.2em}
 & \ImprintRemovalShort & 1 & - & 10 & ${0.92}_{\pm0.05}$ & 1.00 & 1.00 & ${27.304}_{\pm2.116}$ & ${0.952}_{\pm0.017}$ & ${0.050}_{\pm0.018}$ & ${\num{1.67e-07}}_{\pm\num{1.21e-06}}$ & 1.00 & ${28.883}_{\pm2.192}$ & ${0.956}_{\pm0.015}$ & ${0.050}_{\pm0.017}$ \\
 &  &  &  & 20 & ${0.75}_{\pm0.10}$ & 0.66 & 0.66 & ${25.273}_{\pm2.123}$ & ${0.922}_{\pm0.026}$ & ${0.081}_{\pm0.025}$ & ${\num{2.51e-04}}_{\pm\num{1.02e-03}}$ & 1.00 & ${26.795}_{\pm2.187}$ & ${0.926}_{\pm0.023}$ & ${0.082}_{\pm0.023}$ \\
 &  &  &  & 50 & ${0.46}_{\pm0.09}$ & 0.00 & 0.00 & ${22.834}_{\pm2.164}$ & ${0.868}_{\pm0.041}$ & ${0.136}_{\pm0.036}$ & ${\num{7.75e-02}}_{\pm\num{1.56e-01}}$ & 0.64 & ${24.323}_{\pm2.184}$ & ${0.872}_{\pm0.039}$ & ${0.137}_{\pm0.033}$ \\
 &  &  &  & 100 & ${0.26}_{\pm0.06}$ & 0.00 & 0.00 & ${21.091}_{\pm2.188}$ & ${0.812}_{\pm0.056}$ & ${0.185}_{\pm0.043}$ & ${\num{3.13e-01}}_{\pm\num{3.28e-01}}$ & 0.33 & ${22.562}_{\pm2.211}$ & ${0.817}_{\pm0.056}$ & ${0.185}_{\pm0.039}$ \\
 &  &  &  & 150 & ${0.19}_{\pm0.05}$ & 0.00 & 0.00 & ${19.913}_{\pm2.441}$ & ${0.762}_{\pm0.087}$ & ${0.224}_{\pm0.061}$ & ${\num{4.54e-01}}_{\pm\num{3.44e-01}}$ & 0.12 & ${21.475}_{\pm2.237}$ & ${0.773}_{\pm0.070}$ & ${0.219}_{\pm0.043}$ \\
\cline{2-16} \noalign{\vskip 0.2em}
 & Averaging & 10 & - & 0 & ${0.90}_{\pm0.06}$ & 1.00 & 1.00 & ${17.677}_{\pm1.384}$ & ${0.717}_{\pm0.110}$ & ${0.336}_{\pm0.138}$ & ${\num{5.28e-01}}_{\pm\num{3.61e-01}}$ & 0.17 & ${17.081}_{\pm0.988}$ & ${0.637}_{\pm0.101}$ & ${0.420}_{\pm0.131}$ \\
 &  & 50 & - & 0 & ${0.96}_{\pm0.04}$ & 1.00 & 1.00 & ${22.092}_{\pm1.333}$ & ${0.878}_{\pm0.053}$ & ${0.162}_{\pm0.092}$ & ${\num{7.85e-01}}_{\pm\num{3.27e-01}}$ & 0.05 & ${20.292}_{\pm1.065}$ & ${0.830}_{\pm0.055}$ & ${0.225}_{\pm0.096}$ \\
 &  & 100 & - & 0 & ${0.97}_{\pm0.03}$ & 1.00 & 1.00 & ${22.679}_{\pm1.406}$ & ${0.925}_{\pm0.032}$ & ${0.104}_{\pm0.067}$ & ${\num{8.90e-01}}_{\pm\num{2.55e-01}}$ & 0.04 & ${21.414}_{\pm1.191}$ & ${0.881}_{\pm0.041}$ & ${0.157}_{\pm0.076}$ \\
 &  & 1000 & - & 0 & ${1.00}_{\pm0.01}$ & 1.00 & 1.00 & ${23.938}_{\pm0.720}$ & ${0.983}_{\pm0.008}$ & ${0.022}_{\pm0.014}$ & ${\num{9.58e-01}}_{\pm\num{1.68e-01}}$ & 0.01 & ${22.294}_{\pm0.453}$ & ${0.944}_{\pm0.020}$ & ${0.057}_{\pm0.029}$ \\
 &  & 5000 & - & 0 & ${1.00}_{\pm0.00}$ & 1.00 & 1.00 & ${24.138}_{\pm0.425}$ & ${0.990}_{\pm0.006}$ & ${0.013}_{\pm0.007}$ & ${\num{9.71e-01}}_{\pm\num{1.32e-01}}$ & 0.01 & ${22.367}_{\pm0.313}$ & ${0.950}_{\pm0.018}$ & ${0.046}_{\pm0.023}$ \\
\cline{2-16} \noalign{\vskip 0.2em}
 & Regen. & 1 & - & 200 & ${0.85}_{\pm0.05}$ & 0.99 & 0.99 & ${21.783}_{\pm2.322}$ & ${0.865}_{\pm0.039}$ & ${0.142}_{\pm0.036}$ & ${\num{3.41e-04}}_{\pm\num{1.59e-03}}$ & 1.00 & ${20.895}_{\pm2.599}$ & ${0.867}_{\pm0.031}$ & ${0.141}_{\pm0.035}$ \\
 & AdvEmb & 1 & 8 & 200 & ${0.96}_{\pm0.03}$ & 1.00 & 1.00 & ${31.365}_{\pm0.247}$ & ${0.951}_{\pm0.015}$ & ${0.263}_{\pm0.097}$ & ${\num{1.10e-03}}_{\pm\num{6.21e-03}}$ & 0.99 & ${31.412}_{\pm0.281}$ & ${0.946}_{\pm0.014}$ & ${0.329}_{\pm0.090}$ \\
\cline{2-16} \noalign{\vskip 0.2em}
 & Surrogate & 7500 & 8 & 200 & ${0.97}_{\pm0.03}$ & 1.00 & 1.00 & ${32.171}_{\pm0.303}$ & ${0.961}_{\pm0.012}$ & ${0.164}_{\pm0.065}$ & ${\num{5.89e-05}}_{\pm\num{4.45e-04}}$ & 1.00 & ${32.381}_{\pm0.249}$ & ${0.955}_{\pm0.013}$ & ${0.211}_{\pm0.071}$ \\
\midrule

PixArt-$\Sigma$ & Original & - & - & - & ${1.00}_{\pm0.02}$ & 1.00 & 1.00 & inf & ${1.000}_{\pm0.000}$ & ${0.000}_{\pm0.000}$ & ${\num{1.22e-06}}_{\pm\num{1.21e-05}}$ & 1.00 & inf & ${1.000}_{\pm0.000}$ & ${0.000}_{\pm0.000}$ \\
\cline{2-16} \noalign{\vskip 0.2em}
 & \ImprintRemovalShort & 1 & - & 10 & ${0.94}_{\pm0.06}$ & 1.00 & 1.00 & ${28.034}_{\pm3.333}$ & ${0.952}_{\pm0.026}$ & ${0.053}_{\pm0.022}$ & ${\num{4.10e-03}}_{\pm\num{2.82e-02}}$ & 0.98 & ${28.397}_{\pm3.140}$ & ${0.953}_{\pm0.025}$ & ${0.052}_{\pm0.021}$ \\
 &  &  &  & 20 & ${0.86}_{\pm0.08}$ & 0.94 & 0.94 & ${26.116}_{\pm3.108}$ & ${0.923}_{\pm0.038}$ & ${0.085}_{\pm0.031}$ & ${\num{5.59e-03}}_{\pm\num{3.20e-02}}$ & 0.96 & ${26.447}_{\pm2.925}$ & ${0.924}_{\pm0.036}$ & ${0.084}_{\pm0.030}$ \\
 &  &  &  & 50 & ${0.64}_{\pm0.08}$ & 0.18 & 0.18 & ${23.618}_{\pm2.873}$ & ${0.864}_{\pm0.060}$ & ${0.145}_{\pm0.047}$ & ${\num{4.37e-02}}_{\pm\num{1.36e-01}}$ & 0.75 & ${23.895}_{\pm2.700}$ & ${0.865}_{\pm0.057}$ & ${0.145}_{\pm0.045}$ \\
 &  &  &  & 100 & ${0.40}_{\pm0.07}$ & 0.00 & 0.00 & ${21.735}_{\pm2.699}$ & ${0.794}_{\pm0.083}$ & ${0.206}_{\pm0.058}$ & ${\num{1.94e-01}}_{\pm\num{2.69e-01}}$ & 0.41 & ${21.984}_{\pm2.546}$ & ${0.795}_{\pm0.080}$ & ${0.206}_{\pm0.055}$ \\
 &  &  &  & 150 & ${0.27}_{\pm0.07}$ & 0.00 & 0.00 & ${20.591}_{\pm2.606}$ & ${0.737}_{\pm0.101}$ & ${0.248}_{\pm0.063}$ & ${\num{3.55e-01}}_{\pm\num{3.15e-01}}$ & 0.14 & ${20.812}_{\pm2.466}$ & ${0.736}_{\pm0.099}$ & ${0.248}_{\pm0.061}$ \\
\cline{2-16} \noalign{\vskip 0.2em}
 & Averaging & 10 & - & 0 & ${0.92}_{\pm0.07}$ & 0.99 & 0.99 & ${14.491}_{\pm2.221}$ & ${0.721}_{\pm0.074}$ & ${0.322}_{\pm0.102}$ & ${\num{1.11e-01}}_{\pm\num{2.23e-01}}$ & 0.52 & ${14.901}_{\pm2.138}$ & ${0.719}_{\pm0.072}$ & ${0.332}_{\pm0.110}$ \\
 &  & 50 & - & 0 & ${0.96}_{\pm0.06}$ & 1.00 & 1.00 & ${15.763}_{\pm1.322}$ & ${0.868}_{\pm0.044}$ & ${0.178}_{\pm0.079}$ & ${\num{1.86e-01}}_{\pm\num{2.99e-01}}$ & 0.49 & ${15.823}_{\pm1.269}$ & ${0.865}_{\pm0.044}$ & ${0.180}_{\pm0.082}$ \\
 &  & 100 & - & 0 & ${0.97}_{\pm0.06}$ & 0.99 & 0.99 & ${15.978}_{\pm1.059}$ & ${0.902}_{\pm0.033}$ & ${0.135}_{\pm0.063}$ & ${\num{2.82e-01}}_{\pm\num{3.57e-01}}$ & 0.42 & ${16.316}_{\pm1.164}$ & ${0.902}_{\pm0.032}$ & ${0.135}_{\pm0.065}$ \\
 &  & 1000 & - & 0 & ${0.98}_{\pm0.05}$ & 1.00 & 1.00 & ${16.269}_{\pm0.840}$ & ${0.942}_{\pm0.021}$ & ${0.078}_{\pm0.031}$ & ${\num{5.94e-01}}_{\pm\num{4.20e-01}}$ & 0.20 & ${16.332}_{\pm0.919}$ & ${0.939}_{\pm0.021}$ & ${0.080}_{\pm0.031}$ \\
 &  & 5000 & - & 0 & ${0.99}_{\pm0.03}$ & 1.00 & 1.00 & ${16.218}_{\pm0.802}$ & ${0.946}_{\pm0.019}$ & ${0.075}_{\pm0.029}$ & ${\num{6.76e-01}}_{\pm\num{4.17e-01}}$ & 0.19 & ${16.341}_{\pm0.897}$ & ${0.943}_{\pm0.019}$ & ${0.075}_{\pm0.028}$ \\
\cline{2-16} \noalign{\vskip 0.2em}
 & Regen. & 1 & - & 200 & ${0.87}_{\pm0.06}$ & 0.97 & 0.97 & ${22.546}_{\pm2.441}$ & ${0.850}_{\pm0.057}$ & ${0.158}_{\pm0.042}$ & ${\num{6.40e-03}}_{\pm\num{3.50e-02}}$ & 0.94 & ${22.680}_{\pm2.554}$ & ${0.850}_{\pm0.055}$ & ${0.157}_{\pm0.039}$ \\
 & AdvEmb & 1 & 8 & 200 & ${0.94}_{\pm0.06}$ & 0.99 & 0.99 & ${31.533}_{\pm0.295}$ & ${0.947}_{\pm0.012}$ & ${0.308}_{\pm0.100}$ & ${\num{1.09e-02}}_{\pm\num{6.85e-02}}$ & 0.92 & ${31.494}_{\pm0.307}$ & ${0.946}_{\pm0.012}$ & ${0.313}_{\pm0.100}$ \\
\cline{2-16} \noalign{\vskip 0.2em}
 & Surrogate & 7500 & 8 & 200 & ${0.95}_{\pm0.06}$ & 1.00 & 1.00 & ${31.819}_{\pm0.256}$ & ${0.946}_{\pm0.015}$ & ${0.236}_{\pm0.090}$ & ${\num{2.12e-03}}_{\pm\num{1.32e-02}}$ & 0.97 & ${31.791}_{\pm0.215}$ & ${0.945}_{\pm0.015}$ & ${0.243}_{\pm0.095}$ \\
\midrule

FLUX.1 & Original & - & - & - & ${1.00}_{\pm0.00}$ & 1.00 & 1.00 & inf & ${1.000}_{\pm0.000}$ & ${0.000}_{\pm0.000}$ & ${\num{7.92e-06}}_{\pm\num{7.92e-05}}$ & 1.00 & inf & ${1.000}_{\pm0.000}$ & ${0.000}_{\pm0.000}$ \\
\cline{2-16} \noalign{\vskip 0.2em}
 & \ImprintRemovalShort & 1 & - & 10 & ${0.86}_{\pm0.05}$ & 0.98 & 0.98 & ${27.165}_{\pm2.052}$ & ${0.925}_{\pm0.027}$ & ${0.076}_{\pm0.022}$ & ${\num{1.92e-01}}_{\pm\num{2.00e-01}}$ & 0.22 & ${26.439}_{\pm1.736}$ & ${0.923}_{\pm0.018}$ & ${0.076}_{\pm0.017}$ \\
 &  &  &  & 20 & ${0.74}_{\pm0.06}$ & 0.75 & 0.75 & ${25.861}_{\pm1.948}$ & ${0.888}_{\pm0.036}$ & ${0.112}_{\pm0.030}$ & ${\num{4.00e-01}}_{\pm\num{3.16e-01}}$ & 0.07 & ${25.225}_{\pm1.664}$ & ${0.890}_{\pm0.025}$ & ${0.108}_{\pm0.022}$ \\
 &  &  &  & 50 & ${0.54}_{\pm0.05}$ & 0.00 & 0.00 & ${24.099}_{\pm1.860}$ & ${0.822}_{\pm0.053}$ & ${0.176}_{\pm0.041}$ & ${\num{4.99e-01}}_{\pm\num{2.97e-01}}$ & 0.00 & ${23.673}_{\pm1.549}$ & ${0.833}_{\pm0.039}$ & ${0.167}_{\pm0.030}$ \\
 &  &  &  & 100 & ${0.39}_{\pm0.04}$ & 0.00 & 0.00 & ${22.667}_{\pm1.832}$ & ${0.746}_{\pm0.073}$ & ${0.243}_{\pm0.049}$ & ${\num{4.79e-01}}_{\pm\num{2.62e-01}}$ & 0.00 & ${22.330}_{\pm1.731}$ & ${0.764}_{\pm0.076}$ & ${0.232}_{\pm0.060}$ \\
 &  &  &  & 150 & ${0.34}_{\pm0.05}$ & 0.00 & 0.00 & ${21.710}_{\pm1.824}$ & ${0.683}_{\pm0.089}$ & ${0.291}_{\pm0.054}$ & ${\num{6.18e-01}}_{\pm\num{2.43e-01}}$ & 0.00 & ${21.483}_{\pm1.641}$ & ${0.710}_{\pm0.087}$ & ${0.275}_{\pm0.060}$ \\
\cline{2-16} \noalign{\vskip 0.2em}
 & Averaging & 10 & - & 0 & ${0.99}_{\pm0.02}$ & 1.00 & 1.00 & ${18.270}_{\pm2.046}$ & ${0.743}_{\pm0.061}$ & ${0.296}_{\pm0.087}$ & ${\num{3.00e-01}}_{\pm\num{3.55e-01}}$ & 0.36 & ${18.670}_{\pm1.928}$ & ${0.735}_{\pm0.058}$ & ${0.297}_{\pm0.085}$ \\
 &  & 50 & - & 0 & ${1.00}_{\pm0.01}$ & 1.00 & 1.00 & ${23.267}_{\pm2.160}$ & ${0.909}_{\pm0.027}$ & ${0.121}_{\pm0.051}$ & ${\num{5.09e-01}}_{\pm\num{4.45e-01}}$ & 0.31 & ${23.620}_{\pm1.916}$ & ${0.905}_{\pm0.029}$ & ${0.117}_{\pm0.053}$ \\
 &  & 100 & - & 0 & ${1.00}_{\pm0.01}$ & 1.00 & 1.00 & ${24.836}_{\pm2.125}$ & ${0.947}_{\pm0.018}$ & ${0.073}_{\pm0.035}$ & ${\num{5.94e-01}}_{\pm\num{4.31e-01}}$ & 0.22 & ${24.853}_{\pm1.612}$ & ${0.943}_{\pm0.019}$ & ${0.072}_{\pm0.036}$ \\
 &  & 1000 & - & 0 & ${1.00}_{\pm0.00}$ & 1.00 & 1.00 & ${26.511}_{\pm1.099}$ & ${0.987}_{\pm0.005}$ & ${0.017}_{\pm0.008}$ & ${\num{7.48e-01}}_{\pm\num{3.80e-01}}$ & 0.12 & ${26.355}_{\pm0.980}$ & ${0.985}_{\pm0.007}$ & ${0.017}_{\pm0.008}$ \\
 &  & 5000 & - & 0 & ${1.00}_{\pm0.00}$ & 1.00 & 1.00 & ${26.742}_{\pm0.483}$ & ${0.991}_{\pm0.004}$ & ${0.013}_{\pm0.005}$ & ${\num{7.71e-01}}_{\pm\num{3.73e-01}}$ & 0.11 & ${26.468}_{\pm0.541}$ & ${0.989}_{\pm0.006}$ & ${0.013}_{\pm0.006}$ \\
\cline{2-16} \noalign{\vskip 0.2em}
 & Regen. & 1 & - & 200 & ${0.72}_{\pm0.05}$ & 0.64 & 0.64 & ${23.845}_{\pm2.298}$ & ${0.850}_{\pm0.048}$ & ${0.158}_{\pm0.038}$ & ${\num{2.58e-01}}_{\pm\num{2.39e-01}}$ & 0.13 & ${23.750}_{\pm2.307}$ & ${0.845}_{\pm0.049}$ & ${0.156}_{\pm0.039}$ \\
 & AdvEmb & 1 & 8 & 200 & ${1.00}_{\pm0.02}$ & 1.00 & 1.00 & ${31.266}_{\pm1.187}$ & ${0.940}_{\pm0.011}$ & ${0.303}_{\pm0.081}$ & ${\num{5.13e-02}}_{\pm\num{9.55e-02}}$ & 0.66 & ${31.420}_{\pm0.376}$ & ${0.935}_{\pm0.008}$ & ${0.272}_{\pm0.071}$ \\
\cline{2-16} \noalign{\vskip 0.2em}
 & Surrogate & 7500 & 8 & 200 & ${0.99}_{\pm0.02}$ & 1.00 & 1.00 & ${31.635}_{\pm1.293}$ & ${0.944}_{\pm0.013}$ & ${0.253}_{\pm0.080}$ & ${\num{4.79e-02}}_{\pm\num{1.26e-01}}$ & 0.75 & ${31.606}_{\pm1.633}$ & ${0.943}_{\pm0.016}$ & ${0.244}_{\pm0.075}$ \\
\bottomrule
\end{tabular}
    }
    \captionof{table}{Full experimental results on removal attacks. \textit{Ref.Img.}, \textit{Str.} and \textit{Step} refer to the number of reference images, the strength setting, and the attack step used, respectively. \textit{Bit Acc.} refers to Gaussian Shading bit accuracy. \textit{Det.} and \textit{Att.} refer to mean detection and attribution success. \textit{\ImprintRemovalShort} refers to our \ImprintRemovalLong attack. The Averaging~\cite{yang2024steganalysisdigitalwatermarkingdefense} attack is tested with varying amounts of reference images. The Regeneration~\cite{ZhaZhaSu2024invisibleimagewatermarksprovably} (\textit{Regen.}), AdvEmb~\cite{AnDinRab2024Benchmarking}, and Surrogate~\cite{saberi2023robustness} attacks are set to the strongest settings available in their corresponding Github repositories.
    }
    \label{tab:full_removal}

\end{minipage}

\clearpage

\section{Experimental Details}
\label{sec:experiment:details}
Here, we provide more details on our experimental setup, including the datasets used, details on our finetune of Stable Diffusion 2.1 (SD2.1-Anime), the number of samples used in each experiment, the watermark parameters, and the runtime of our attacks.

\subsection{Attacker and Target Models}
Our default setup is to use Stable Diffusion 2.1 as the attacker model, and SDXL, PixArt-$\Sigma$, FLUX.1, and our SD2.1-Anime model as the target models.
Our transferability analysis (\cref{sec:evaluation:transferability}) is the exception where we test multiple combinations of attacker and target models.
\cref{tab:model_overview} provides an overview of all the models used in our experiments, including the used scheduler and other relevant settings.
The choice of schedulers for each model is based on the default setting in the corresponding pipelines from the Huggingface diffusers library.

\begin{table}[ht]
    \centering
    \resizebox{\linewidth}{!}{%
\begin{tabular}{llllllllll}
\toprule
                     &                                          &       &        & \multicolumn{1}{c}{} &          &          & \multicolumn{3}{l}{G. Shad. Params}  \\
Model                & Huggingface ID                           & Type  & L. Ch. & Scheduler            & Steps & G. Scale & $\ell$       & $\rep$       & k        \\ 
\midrule
SD1.5                & runwayml/stable-diffusion-v1-5           & UNet & 4      & DDIM                 & 50       & 7.5      & 1       & 64                     & 256       \\
SD2.1                & stabilityai/stable-diffusion-2-1-base    & UNet & 4      & DDIM                 & 50       & 7.5      & 1       & 64                     & 256       \\
SD2.1-Anime          & stabilityai/stable-diffusion-2-1-base    & UNet & 4      & DDIM                 & 50       & 7.5      & 1       & 64                     & 256       \\
SDXL                 & stabilityai/stable-diffusion-xl-base-1.0 & UNet & 4      & DDIM                 & 50       & 7.5      & 1       & 64                     & 256       \\
PixArt-$\Sigma$      & PixArt-alpha/PixArt-Sigma-XL-2-512-MS    & DiT   & 4      & DPM                  & 50       & 7.5      & 1       & 64                    & 256       \\
FLUX.1               & black-forest-labs/FLUX.1-dev             & DiT   & 16     & FlowMatchEuler                & 20       & 3.5      & 1       & 256                     & 256     \\
Waifu Diffusion      & hakurei/waifu-diffusion                  & UNet & 4      & DDIM                 & 50       & 7.5      & 1       & 64                     & 256       \\
Mitsua Diffusion One & Mitsua/mitsua-diffusion-one              & UNet & 4      & DDIM                 & 50       & 7.5      & 1       & 64                     & 256       \\
Common Canvas        & common-canvas/CommonCanvas-S-C           & UNet & 4      & DDIM                 & 50       & 7.5      & 1       & 64                     & 256       \\ 
\bottomrule
\end{tabular}
    }
    \caption{Overview of model settings. The settings outlined here are used across all experiments without deviation for both the case that the model at hand is a target model, or the case that it is the proxy model used by the attacker. \textit{L. Ch.} refers to the number of latent channels. \textit{Scheduler} refers to the scheduler during image generation and inversion. \textit{Steps} refers to the number of denoising \& inversion steps during generation and inversion. \textit{G. Scale} refers to the guidance scale during generation. \textit{G. Shad. Params} refers to the Gaussian Shading parameters when the model is deployed as a target model.}
    \label{tab:model_overview}
\end{table}

\subsection{Prompting and Cover Image Datasets}
Except for our SD2.1-Anime model, we use the \textit{Stable-Diffusion-Prompts}\footnote{\href{https://huggingface.co/datasets/Gustavosta/Stable-Diffusion-Prompts}{Stable-Diffusion-Prompts}} dataset to prompt benign watermarked images from the target models for all experiments (\cref{sec:evaluation:imprinting,sec:evaluation:removal,sec:evaluation:reprompting,sec:evaluation:transferability,sec:evaluation:robustness}).
For the cover images used in our \ImprintForgeLong attack (Sec.~\ref{sec:evaluation:imprinting}), we only use the \textit{MS-COCO-2017 Dataset}~\cite{coco}.
In experiments involving our \Reprompting attack (\cref{sec:evaluation:reprompting,sec:evaluation:transferability,sec:evaluation:robustness}), we use the \textit{Inappropriate Image Prompts (I2P)}\footnote{\href{https://huggingface.co/datasets/AIML-TUDA/i2p}{Inappropriate Image Prompts (I2P)}} dataset (sorting the \textit{"inappropriate\_percentage"} column in descending order) to generate harmful images with the attacker model. Across all experiments, images are of size $512\times512$.

\subsection{Finetuned Model SD2.1-Anime}
Our own finetune of SD2.1 is trained
on 10,000 pairs of anime images and captions from the Anime-with-caption-CC0\footnote{\href{https://huggingface.co/datasets/alfredplpl/anime-with-caption-cc0}{Anime-with-caption-CC0}} dataset.
We use the training scripts for Stable Diffusion models from Huggingface with default parameters for full finetuning of the UNet parameters\footnote{\href{https://github.com/huggingface/diffusers/blob/main/examples/text_to_image/train_text_to_image.py}{Huggingface SD training script}} and for LoRA finetuning\footnote{\href{https://github.com/huggingface/diffusers/blob/main/examples/text_to_image/train_text_to_image_lora.py}{Huggingface SD LoRA training script}}.
We add a keyword to each caption in the training set and reuse this keyword when generating images during our experiments.
For every experiment involving SD2.1-Anime as the target model, 
we prompt the model with prompts similar to those in the training phase, by taking prompts from a separate split of {Anime-with-caption-CC0} dataset and adding the keywords as prefix.
By finetuning our own model, we achieve two goals: First, we obtain a baseline for a target model which is equal to the attacker model (SD2.1) in all aspects except for slight changes in the UNet parameters.
This represents a scenario in which a service provider finetunes a publicly available model, %
and considers the resulting model unique enough to securely deploy semantic watermarks.
Second, we are able to quantify the success of forgery attacks at different numbers of training steps for two different finetuning methods (full finetuning and LoRA) (see Sec.~\ref{sec:evaluation:progress}). The model ultimately used in experiments is the fully finetuned variant.

\subsection{Number of Samples in Experiments}
We provide the exact numbers of prompts, generated images, and cover images for our main experiments.
\begin{itemize}
    \vspace{0.6em}
    \setlength{\itemsep}{0.3em} %
    
    \item \emph{\ImprintForgeLong Attack~(\cref{sec:evaluation:imprinting}).}
    We generate 100 images with each target model (using 100 prompts) and apply our \ImprintForgeLong attack on 100 cover images. This adds up to 400 imprinting forgery attacks (100 for 4 target models each) across a fixed set of 100 pairs of prompts and cover images in total. This procedure is performed for both Tree-Ring and Gaussian Shading, doubling the final number of attacks.
    
    \item \emph{\ImprintRemovalLong Attack, \cref{sec:evaluation:removal}.}
    We generate 100 images with each target model (using 100 prompts) and apply our \ImprintRemovalLong on each image. This adds up to 400 removal attacks (100 for 4 target models each) across a fixed set of 100 prompts in total. This procedure is done for both Tree-Ring and Gaussian Shading, doubling the final number of attacks.

    \item \emph{\Reprompting Attack, \cref{sec:evaluation:reprompting}.}
    We evaluate two different variations of our \Reprompting attack. 
    \begin{itemize}
    \item \emph{``\Reprompt''}: Here, we generate 1,000 images with each target model using 1,000 benign prompts and \reprompt each one with the attacker model using 1,000 harmful prompts. This adds up to 4,000 \Reprompting attacks (1,000 for 4 target models each) across a fixed set of 1,000 pairs of benign and harmful prompts. This procedure is performed for both Tree-Ring and Gaussian Shading, doubling the final number of attacks. 
    \item \emph{``\RepromptPlus''}: Here, we generate 100 images with each target model using 100 benign prompts. We \reprompt each one 3 times with the attacker model using 300 harmful prompt, for both Tree-Ring and Gaussian Shading. For Gaussian Shading, we further resample the recovered latent $\zhatatt{T}$ for each target image 3 times before \reprompting. This adds up to 1,200 \Reprompting attacks for Tree-Ring (300 for each target model, across a fixed set of 100 benign and 300 harmful prompts), and 3,600 \Reprompting attacks for Gaussian Shading (900 for each target model, across a set of 100 benign and 300 harmful prompts) in total.
    \end{itemize}

    \item \emph{Transferability Analysis, \cref{sec:evaluation:transferability}.}
    For each pair of attacker and target model, we generate 100 images using 100 benign prompts and \reprompt each one with the attacker model using a set of 100 harmful prompts. This adds up to 4,900 \Reprompting attacks (100 for each of the 49 combination of target/attacker model) across a fixed set of 100 pairs of benign and harmful prompts in total. We apply this procedure for Tree-Ring and Gaussian Shading, doubling the number of attacks.

    \item \emph{No Defense by Adjusting Thresholds, \cref{sec:evaluation:robustness}.}
    We reuse the results from \cref{sec:evaluation:imprinting,sec:evaluation:reprompting} as attack instances. Furthermore, we generate 1,000 images with each target model using 1,000 prompts. These are then transformed using common transformations and again verified by the target model. This adds up to an additional 1,000 image generations (1,000 for each of the two target models). We apply this procedure for Tree-Ring and Gaussian Shading, doubling the number.
\end{itemize}

\subsection{Runtime of Attack Algorithms}
All experiments were performed on 8 NVIDIA A40 GPUs. 
When executed on a single GPU for a single-batch attack example, the approximate runtimes for each attack algorithm are as follows:
\begin{itemize}
    \vspace{0.6em}
    \setlength{\itemsep}{0.3em} %
    \item The \emph{\ImprintForgeLong} (\cref{sec:evaluation:imprinting}) and \emph{\ImprintRemovalLong} (\cref{sec:evaluation:removal}) attacks require between 25 and 40 minutes to perform 150 steps. The most time-consuming part of these algorithms is the gradient-based optimization done by the attacker model (SD2.1). Verification by the target model is set to take place every 10 optimization steps and is comparably fast.
    \item The \emph{\Reprompting attack} (\cref{sec:evaluation:reprompting}) requires between 30 seconds (smaller models including SD2.1, Mitsua Diffusion One) and 2 minutes (FLUX.1). This time includes all steps: generating a single image with a semantic watermark on the target model, inverting and regenerating on the attacker model, and verifying the presence of the watermark with the target model.
\end{itemize}

\subsection{Image Transformations}

\begin{figure}[h]
    \centering
    \includegraphics[width=0.92\linewidth]{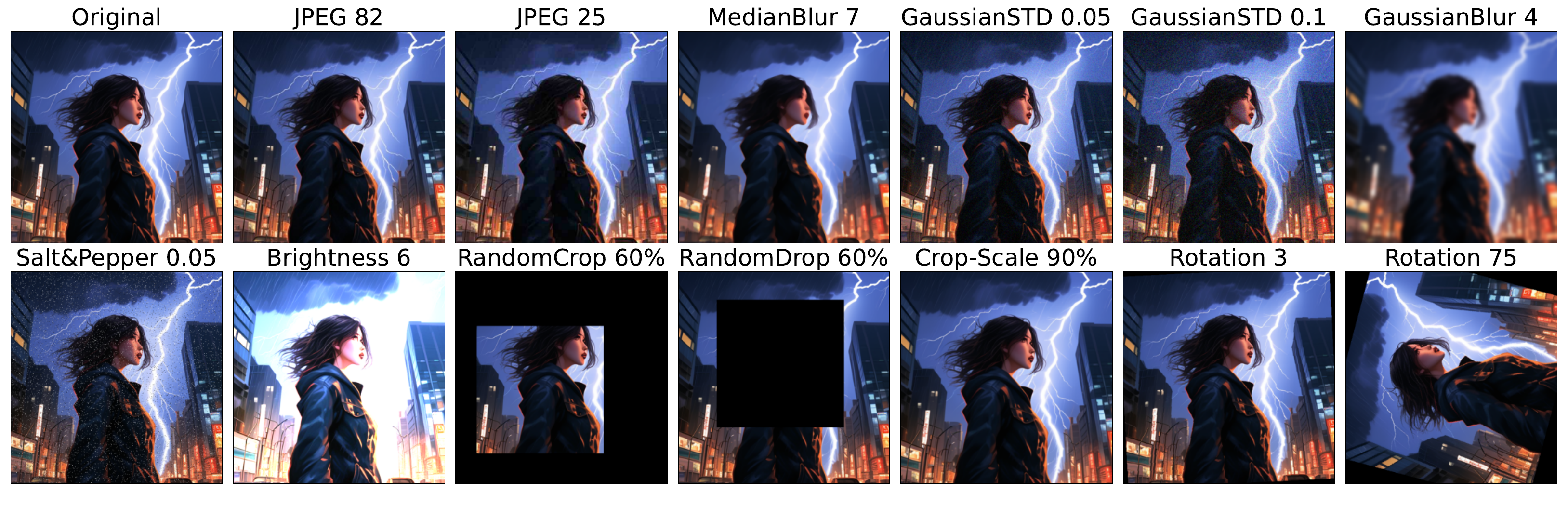}
    \caption{Examples of common image transformations}
    \label{fig:transformations}
\end{figure}

\cref{fig:transformations} shows examples for the standard image transformations that we apply on watermarked images to examine the achievable thresholds of Tree-Ring and Gaussian Shading under these transformations (cf.~\cref{sec:evaluation:robustness} and Sec.~\ref{app:robustness_full}).

\subsection{Parameters for Tree-Ring and Gaussian Shading}
\label{app:paramsforwatermarks}
In order to run Gaussian Shading and Tree-Ring, several parameters need to be selected.

\paragraph{Tree-Ring.}
In the original work of \citet{Wen2023TreeRing}, the threshold for the p-value to assume the watermark's presence is computed by calculating receiver operating characteristics (ROC) curves. For their main results, they report AUROC and TPR@FPR=1\%.
As precise p-values are not reported, we conduct our own experiment similar to the original work. For each model evaluated, we generate 5,000 watermarked images as well as 5,000 clean images to determine the thresholds for desired FPRs.
The thresholds for $10\%$, $1\%$, and $0.1\%$ FPR are reported in \cref{tab:app:fpr_tr}.
We kept all other parameters of the scheme as provided in the {implementation of the authors}\footnote{\href{https://github.com/YuxinWenRick/tree-ring-watermark}{Github Repository of Tree-Ring}}, i.e. we follow their instruction in the readme by inserting the \textit{RING} pattern with default parameters into the 3rd latent channel. For the FLUX.1 model, which uses 16 latent in contrast to 4 latent channels as with the other models, this means that the embedded signal is weaker.

\paragraph{Gaussian Shading} \cite{Yang2024GaussianShading} requires the user to set the message length $k$, the repetition factor $\rep$, and the number of bins $2^\ell$ to sample from a Gaussian distribution. The system was originally evaluated for SD2.1-like 4-channel latents with a scaling factor of 8 (i.e. $4 \times 64 \times 64$ for images of size $512 \times 512$). Ideal parameters for this size were experimentally determined by~\citet{Yang2024GaussianShading} as $k = 256$, $\rep=64$, and $\ell=1$.
In \cref{tab:app:fpr_gs}, we provide detection thresholds $\tau$ for both zero-bit and multi-bit scenario for a desired FPR according to \cref{eq:app:fprgsmulti} (with $N=1$ in case of zero-bit).
Given $k = 256$, the detection threshold for a FPR of $10^{-6}$ is 0.64844 for a zero-bit scenario.
Assuming $k = 256$ and $N = 100,000$ in line with~\citet{Yang2024GaussianShading}, the detection threshold $\tau$ is set at 0.70703 for a multi-bit scenario.
Throughout our work, we always assume this multi-bit scenario, meaning we use a definitive Gaussian Shading detection threshold of $\tau=0.70703$ above which an image is recognized as watermarked.
User attribution was done as described in Sec.~\ref{sec:evaluation}: For each image to be attributed, first, the user id with the highest number of matching bits from a pool of $N = 100,000$ randomly generated users is retrieved.
Then, if the number of matching bits exceeds $\tau$, the sample is counted as attributed to this user. Our results regarding successful user attribution in \cref{tab:imprint,tab:removal,tab:generation} always match the detection success because of small sample sizes, making it improbable for another user to match the bit string at hand and also pass the detection threshold.

In order to adapt the scheme to models like FLUX.1 with 16-channel latents (and the same scaling factor of 8), we have to adjust the parameters. We choose to use the same message length with a higher repetition factor~$\rep$. A higher repetition factor makes the scheme more robust to bit-errors and therefore to perturbations. Using the same message length lets us keep thresholds $\tau$ unchanged regardless of the dimensions of the latent. We consider this to be at least as hard to attack as alternative adjustments, as increasing the message length will also yield lower detection thresholds in return as the probability of randomly drawing correct bit-strings of larger size decreases significantly. 
For example, using $k=512$ would lower the thresholds for an FPR of $10^{-6}$ to 0.60547 (zero-bit) and 0.64648 (multi-bit), which marking removal harder and forgery easier.

In \cref{tab:model_overview}, we report our parameters for Gaussian Shading for multiple models with different latent sizes.

\begin{table}[ht]
\centering
\begin{minipage}[t]{0.5\textwidth}
    \centering
    \begin{tabular}{rcccc}
    \toprule
    {FPR} & {SD2.1-Anime} & {SDXL} & PixArt-$\Sigma$ & FLUX.1 \\
    \midrule
    10\% & 0.14275 & 0.13493 & 0.15809 & 0.14287 \\
    1\% & \textbf{0.02466} & \textbf{0.02610} & \textbf{0.01591} & \textbf{0.02185} \\
    0.1\% & 0.00555 & 0.00553 & 0.00048 & 0.00074 \\
    \bottomrule
    \end{tabular}
    \caption{Thresholds $\tau$ in terms of p-value for Tree-Ring for multiple FPRs and models. The thresholds used in our work are marked in bold.}
    \label{tab:app:fpr_tr}
\end{minipage}
\hfill
\begin{minipage}[t]{0.46\textwidth}
    \centering
    \begin{tabular}{lcc}
    \toprule
    {FPR} & {Zero-bit Threshold} & {Multi-bit Threshold} \\
    \midrule
    $10^{-6}$  & 0.64844 & \textbf{0.70703} \\
    $10^{-16}$ & 0.75000 & 0.78906 \\
    $10^{-32}$ & 0.85156 & 0.87500 \\
    $10^{-64}$ & 0.97266 & 0.98438 \\
    \bottomrule
    \end{tabular}
    \caption{Thresholds $\tau$ in terms of bit accuracy for Gaussian Shading, zero-bit and multi-bit configurations ($N = 100,000$ users) with a watermark capacity $k= 256$ at various FPR levels for all models. The threshold used in our work is marked in bold.}
    \label{tab:app:fpr_gs}
\end{minipage}
\end{table}

\clearpage

\clearpage

\section{Ablation Study of Sampler and Inversion Steps}
\label{sec:experiment:ablation}
\begin{figure*}
    \centering

    \begin{subfigure}{0.85\textwidth}
        \centering
        \includegraphics[width=0.75\linewidth]{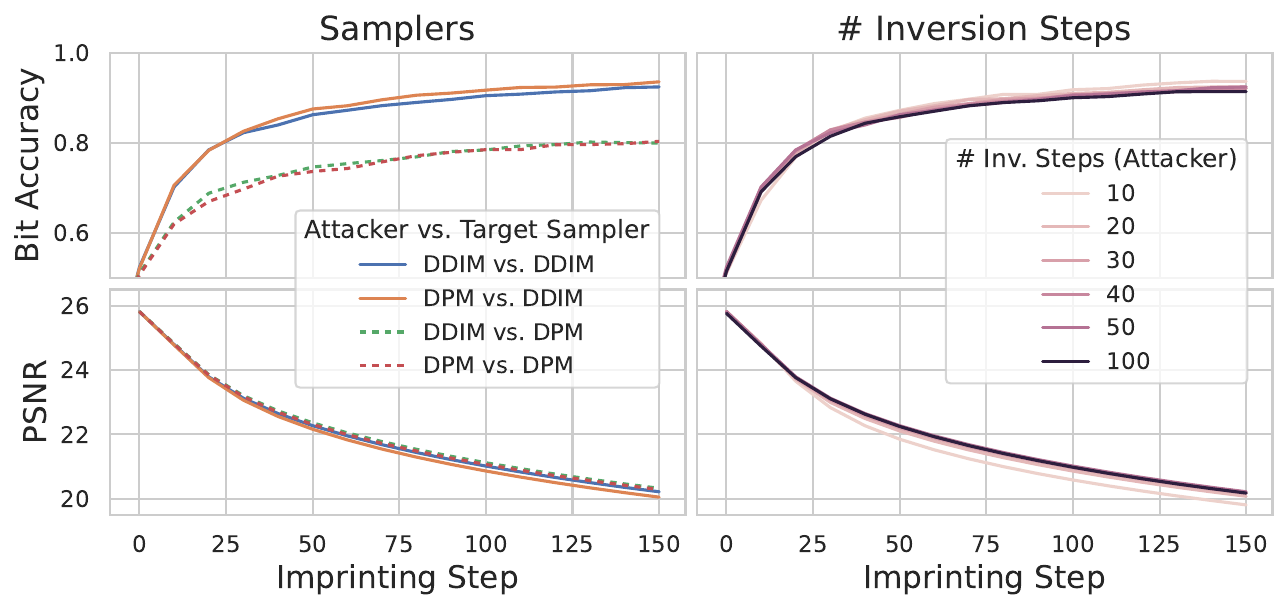}
        \caption{Gaussian Shading}
    \end{subfigure}
    
    \vspace{1em} %

    \begin{subfigure}{0.85\textwidth}
        \centering
        \includegraphics[width=0.80\linewidth]{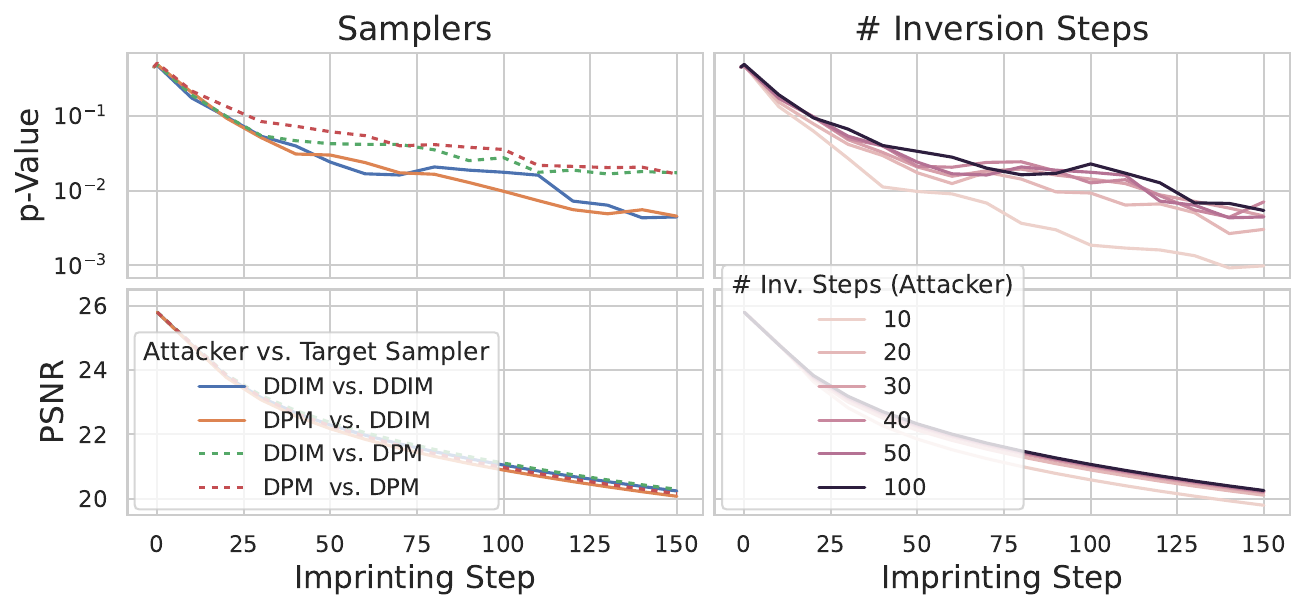}
        \caption{Tree-Ring}
    \end{subfigure}

    \caption{
    Ablation study over samplers used by the attacker and the target model used by the SP (left), as well as inversion steps used by the attacker (right) during the \ImprintForgeLong attack.
    }
    \label{fig:ablation_combined}
    
\end{figure*}

We include ablation experiments to examine the impact of the samplers chosen by the service provider (SP) and the attacker.
For this, we apply the \ImprintForgeLong attack using SD2.1 against SDXL and report averaged results over 30 attack samples. In all cases, the watermarked reference images were generated by the SP with 50 denoising steps.
Fig.~\ref{fig:ablation_combined} shows the results for both Gaussian Shading and Tree-Ring.
For both schemes, we observe similar effects.

First, \textbf{the choice of the sampler} used for both generating watermarked images and retrieving initial latents on the side of the SP affects the attack's effectiveness in terms of bit accuracy and p-values measured on attack examples. If the SP chooses DDIM, the \ImprintForgeLong attack is more effective, regardless of the sampler choice by the attacker. However, this is due to the fact that the SP's choice of sampler affects their general ability to recovery initial latents. 
We confirm this by measuring the average p-value measured for 100 generated watermarked images. While SDXL with DDIM achieves a mean p-value of $1.10\times10^{-21}$ (see Table~\ref{tab:full_removal}), it is much higher for DPM with $1.81\times10^{-7}$. 
This means that DPM is generally worse at retrieving initial latents and detecting the contained watermark. Hence, the SP would have to adjust detection thresholds in order to account for their lowered robustness against common transformation, again increasing their vulnerability against our attacks. We conclude that the choice of sampler on the SP side does not lead to viable defenses. Furthermore, the attacker does not need to know what sampler was used to create the watermarked reference image, as both DDIM and DPM show very similar attack effectiveness.

Second, choosing fewer \textbf{inversion steps} on the attacker's side slightly increases attack success and drastically lowers image quality. Again, we conclude that the attacker again does not require additional knowledge. They can simply choose a reasonably high number of inversions steps that yields satisfactory image quality without much impact on attack effectiveness.

\clearpage

\section{Transferability Analysis}
\label{app:analysis}
In this section, we present additional data for our transferability analysis from \cref{sec:evaluation:transferability} by looking at the similarity of auto-encoders.

We evaluate the similarities of different VAEs, since all our models tested in \cref{sec:evaluation:transferability} deploy VAEs with 4 latent channels and the same order of layers. 
We first examine the weights of the different autoencoders across the different models.
SD2.1 and Common Canvas appear to share the exact same VAE, as all weights of all layers match with a precision of~$10^{-3}$.
Next, we examine if some of the VAEs are functionally similar by comparing their latents for a set of images.
We use a random sample of 100 images from the MS-COCO-2017 dataset~\cite{coco}, for which we compute latent embeddings using a VAE's encoder, and compare the representations obtained by different autoencoders using absolute cosine similarity. 

The functional similarities are reported in \cref{fig:diagram:transfer:cossim}. 
We observe that SD2.1, SD1.5, Waifu Diffusion, and Common Canvas have very similar latents accross all test images.
In addition, we observe that SDXL and \mbox{PixArt-$\Sigma$} share a similar VAE as well. 
Mitsua's autoencoder appears to behave differently from the other VAEs, which supports the claim that Mitsua's VAE was trained from scratch, as stated on the corresponding Huggingface model card\footnote{\href{https://huggingface.co/Mitsua/mitsua-diffusion-one}{Mistua Diffusion One Huggingface model card}}.
Comparing these results to \cref{fig:transferability} (which are also displayed in the figure below (left and middle) for convenience) reveals that latent encoder similarity correlates with high transferability, but does not explain it entirely.
For example, Mitsua and Common Canvas have quite dissimilar autoencoders but still transfer easily in our experiments on the \Reprompting attack.
Also note that the heatmaps are almost symmetrical.

\begin{center}
\vspace{1em}
    \includegraphics[width=1.0\linewidth]{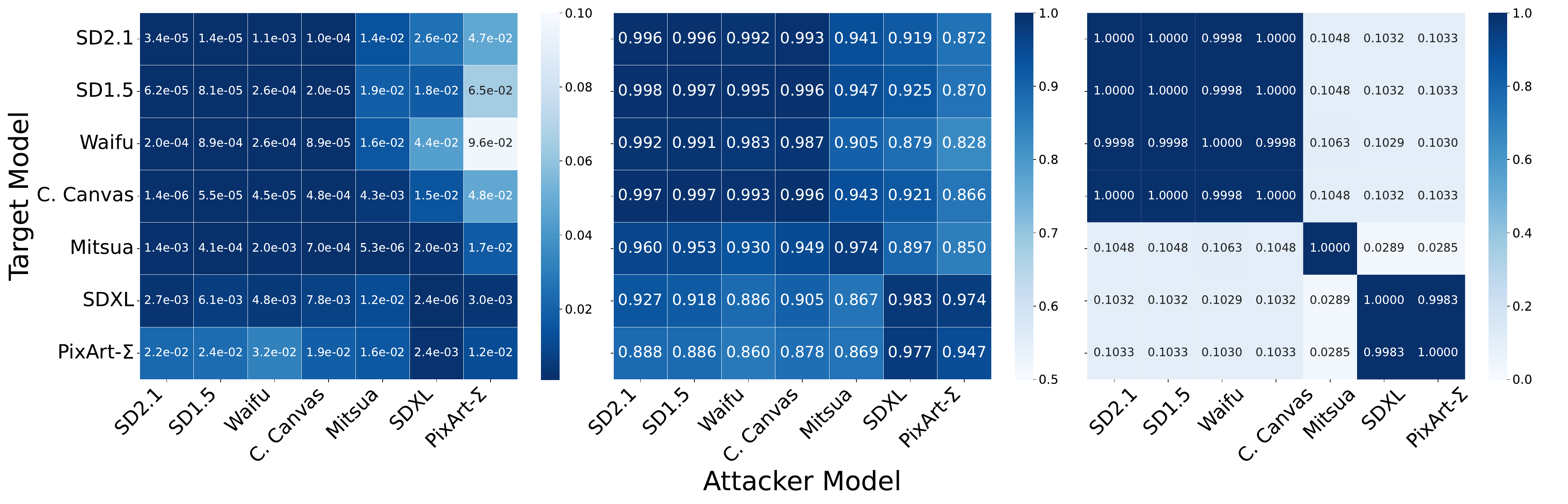}
    \captionof{figure}{Transferability across different pairs of target and attacker model and functional similarity of autoencoders.
    For convenience, the \textbf{left} plot shows the \Reprompting attack transferability for Tree-Ring in terms of p-value~($\downarrow$), and the \textbf{middle} plot for Gaussian Shading in terms of bit accuracy~($\uparrow$). 
    The \textbf{right} plot shows the functional analysis of autoencoders.
    Functional similarity is measured as averaged cosine similarity between latents of different autoencoders for 100 different images.}
    \label{fig:diagram:transfer:cossim}
\vspace{1em}
\end{center}

\clearpage
\onecolumn

\section{A Side-note on Robustness}
\label{app:robustness_full}
\begin{figure*}[h!]
    \centering
    \begin{subfigure}[b]{0.49\textwidth}
        \centering
        \includegraphics[width=0.99\linewidth]{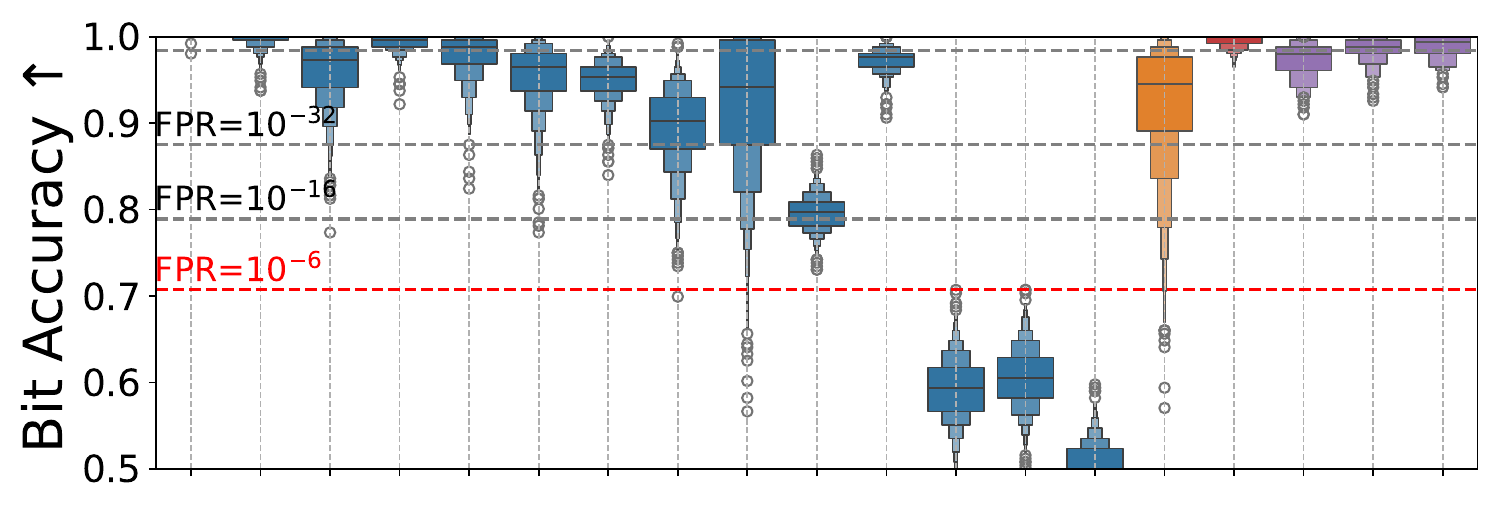}
        \subcaption{Bit Accuracy~($\uparrow$) for Gaussian Shading (SD2.1-Anime)}
        \label{fig:perturb:full:GS:sd21}
        
        \vspace{0.5\floatsep} %
        
        \includegraphics[width=0.99\linewidth]{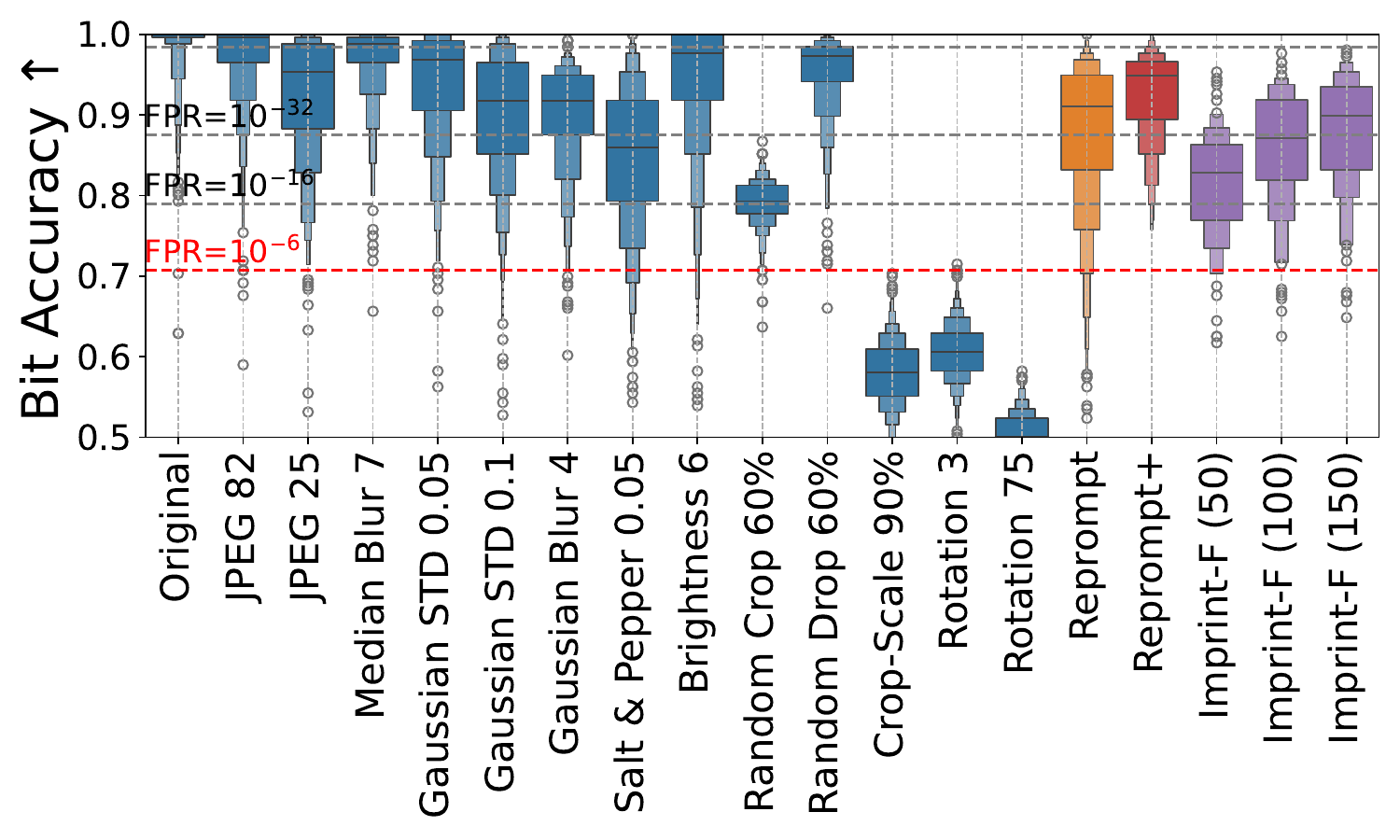}
        \subcaption{Bit Accuracy~($\uparrow$) for Gaussian Shading (PixArt-$\Sigma$)}
        \label{fig:perturb:full:GS:pixart}
    \end{subfigure}
    \hfill
    \begin{subfigure}[b]{0.49\textwidth}
        \centering
        \includegraphics[width=0.99\linewidth]{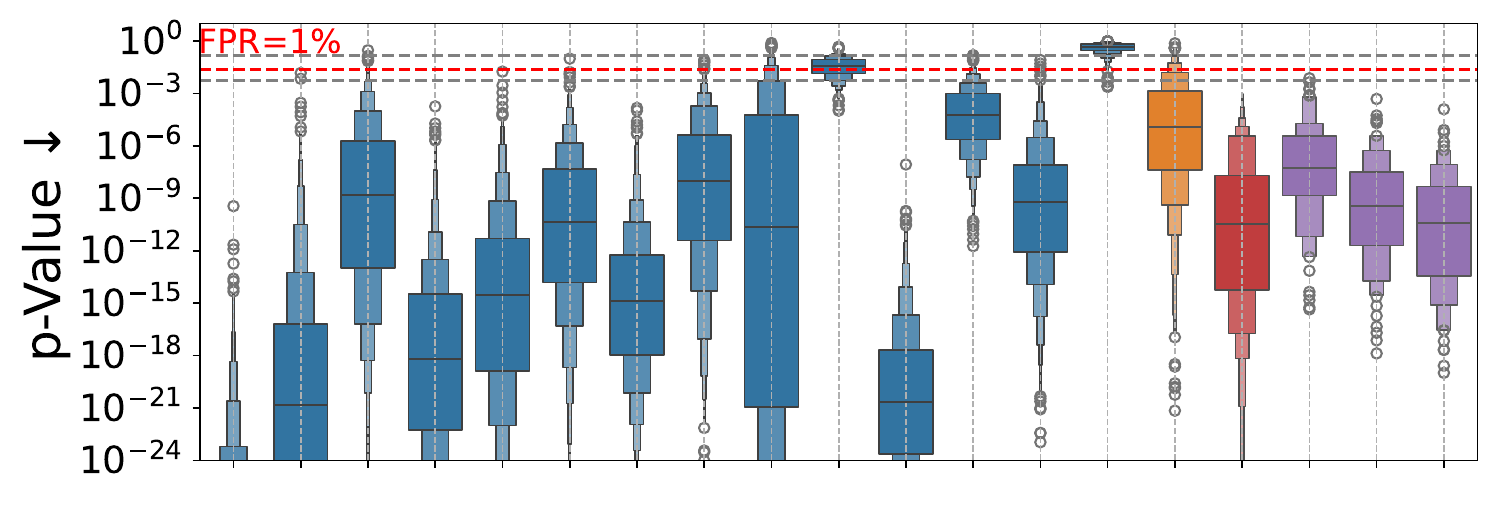}
        \subcaption{P-values~($\downarrow$) for Tree-Ring (SD2.1-Anime)}
        \label{fig:perturb:full:TR:sd21}
        
        \vspace{0.5\floatsep} %
        
        \includegraphics[width=0.99\linewidth]{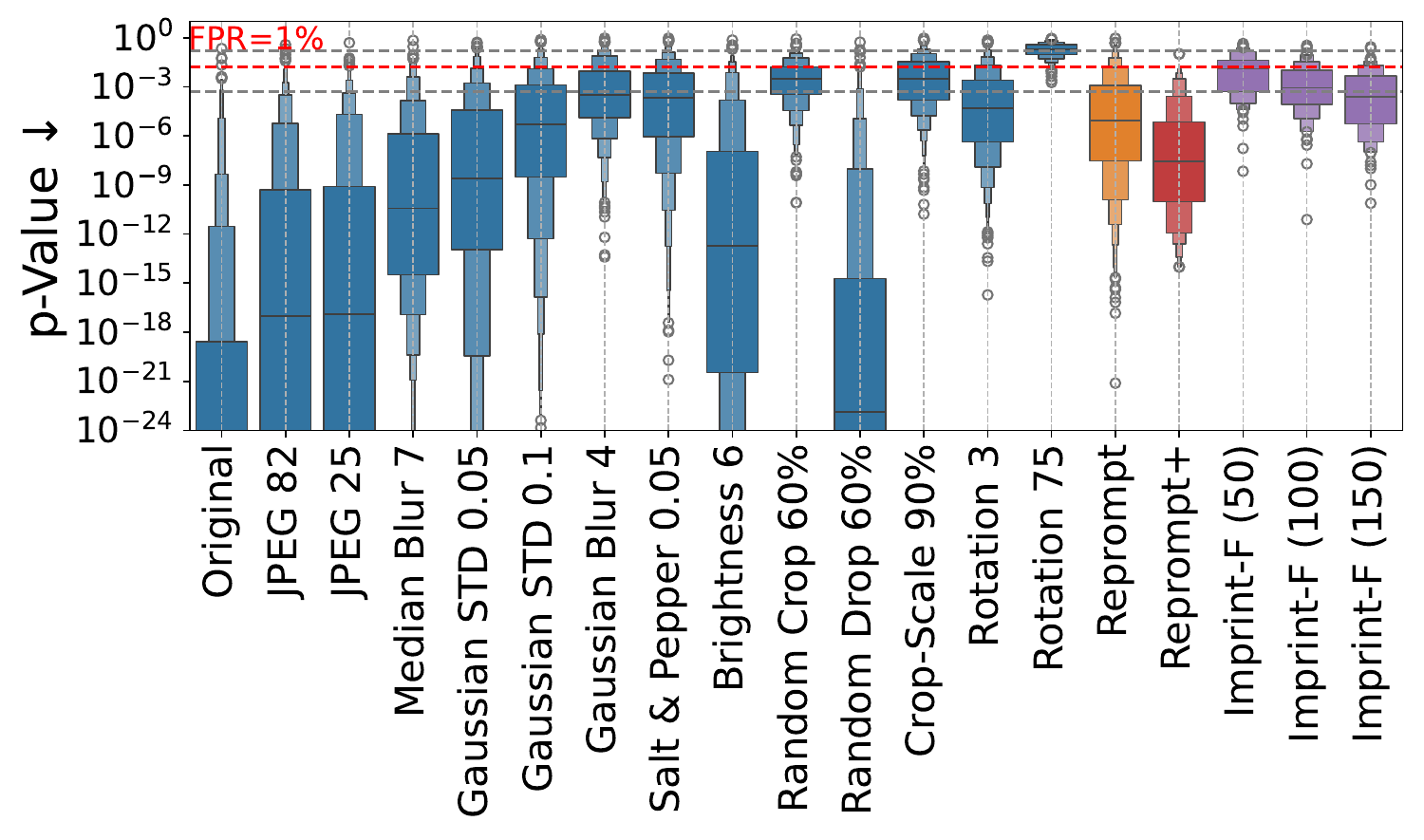}
        \subcaption{P-values~($\downarrow$) for Tree-Ring (PixArt-$\Sigma$)}
        \label{fig:perturb:full:TR:pixart}
    \end{subfigure}
    \caption{
Comparison between image perturbations (blue bars) and our attacks (orange, red and purple bars), for Gaussian Shading (left) and Tree-Ring (right).
Orange bars are the \Reprompting attack, red bars are enhanced \RepromptingPlus attack, purple bars are \ImprintForgeLong for different numbers of optimization steps.
Results are shown for two target models: SD2.1-Anime (top) and PixArt-$\Sigma$ (bottom).
We can observe that no threshold can effectively separate images from common transformations and attacks.
        }
    \label{fig:perturb:full}
\end{figure*}

\cref{fig:perturb:full} provides further intuition on possible thresholds (\cref{fig:perturb}) by showing results for a more complete set of transformations and settings.
We can identify a notable vulnerability of the watermarking methods to some common perturbations.
Gaussian Shading is especially vulnerable to cropping, scaling, and rotation. 
The p-values in Tree-Ring are significantly affected by crop-and-scale operations.
It appears that transformations that change the location of the pixels, such as crop-and-scale and rotation, can significantly disturb both watermarks. %
In contrast, deleting parts of the image (e.g. random drop) preserves the watermark, as enough pixels are still in the correct position.
It appears that in order to decode the watermark, Gaussian Shading requires the image to remain in the same position with which it was generated, which might not be a viable assumption in practical deployment.
This illustrates that these watermarks can be easily removed and that further research into more robust semantic watermarking methods is necessary.

\clearpage
\onecolumn
\section{Attack Success vs Training Steps}
\label{sec:evaluation:progress}
\begin{figure*}[h!]
    \centering
    \begin{subfigure}[b]{0.49\textwidth}
        \centering
        \includegraphics[width=0.99\linewidth]{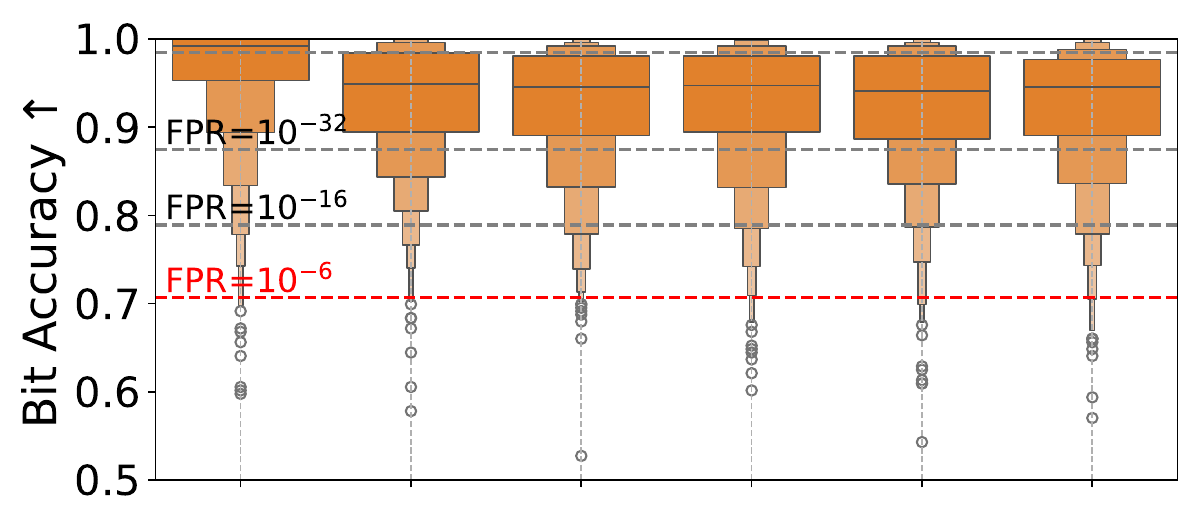}
        \subcaption{Bit Accuracy~($\uparrow$) for Gaussian Shading (Full Finetuning)}
        \label{fig:progress:GS:full}
        
        \vspace{\floatsep} %
        
        \includegraphics[width=0.99\linewidth]{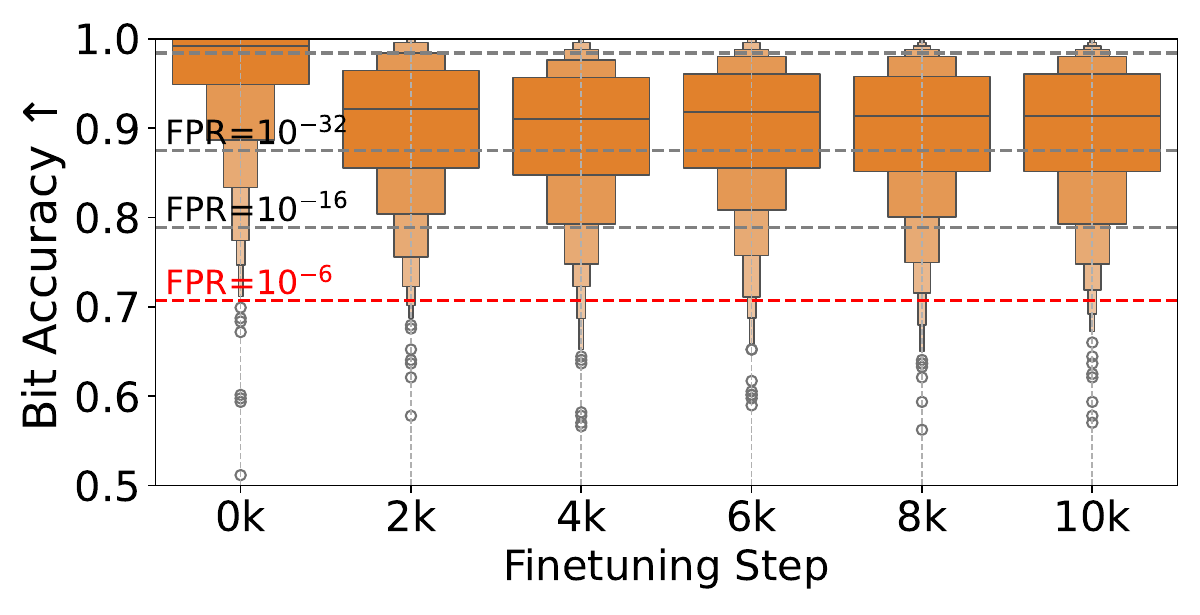}
        \subcaption{Bit Accuracy~($\uparrow$) for Gaussian Shading (LoRA Finetuning)}
        \label{fig:progress:GS:lora}
    \end{subfigure}
    \hfill
    \begin{subfigure}[b]{0.49\textwidth}
        \centering
        \includegraphics[width=0.99\linewidth]{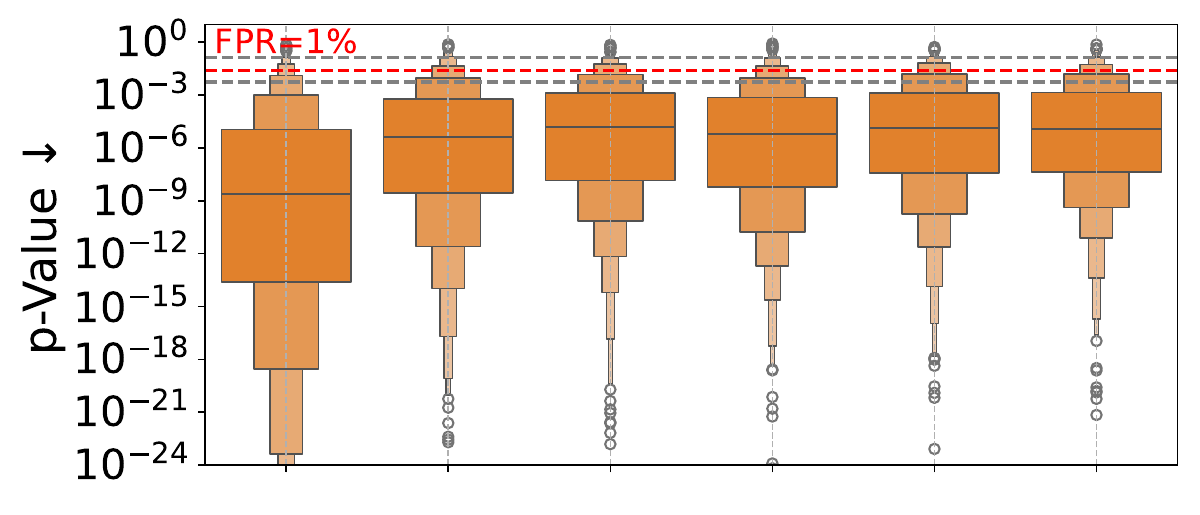}
        \subcaption{P-Values~($\downarrow$) for Tree-Ring (Full Finetuning)}
        \label{fig:progress:TR:full}
        
        \vspace{\floatsep} %
        
        \includegraphics[width=0.99\linewidth]{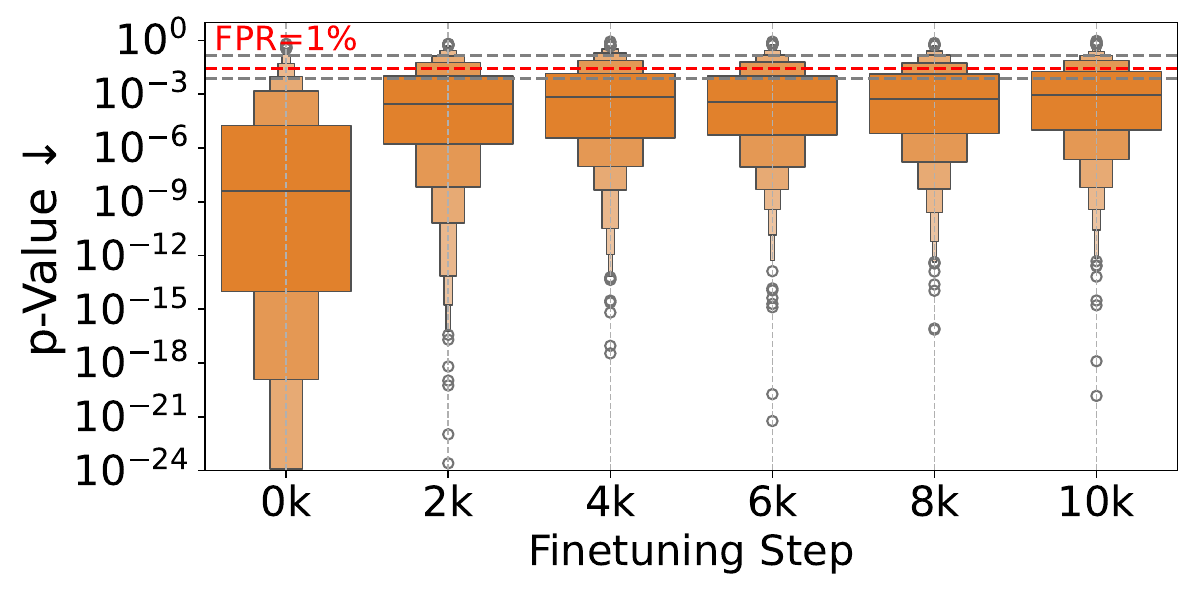}
        \subcaption{P-Values~($\downarrow$) for Tree-Ring (LoRA Finetuning)}
        \label{fig:progress:TR:lora}
    \end{subfigure}
    
    \caption{
    Performance of the \Reprompting attack for Gaussian Shading (left, measured using bit accuracy) and Tree-Ring (right, measured using p-value), after different numbers of finetuning of our SD2.1-Anime model.
    Numbers are reported both for regular full finetuning (top), as well as finetuning using LoRA (bottom).
Marked dashed red lines are the thresholds used in the original papers, corresponding to an FPR of $10^{-6}$ for Gaussian Shading and an FPR of $1\%$ for Tree-Ring watermarks. 
In the case of Gaussian Shading, the theoretical thresholds hold for all the finetuning steps. 
For Tree-Ring, the thresholds displayed are the ones empirically determined on the 10k step for full finetuning and LoRA finetuning, respectively.
        }
    \label{fig:progress}
\end{figure*}

For different amounts of finetuning steps of our SD2.1-Anime finetune, 
\cref{fig:progress} shows bit accuracies and p-values achieved with our \Reprompting attack on Gaussian Shading and Tree-Ring, respectively. The attacker model is SD2.1.

The attack effectiveness decreases with longer finetuning, as expected. 
However, the longer the target model is trained, the slower the rate of decrease.
In our experiment, the attack metrics seems to stabilize after around 4k training steps, still crossing most detection thresholds.
Importantly, the bit accuracy for the attack images exceeds the detection threshold set for FPR=$10^{-6}$ (the threshold from the original paper, marked in red in \cref{fig:progress}).
Similarly, for Tree-Ring, the p-values fall below the detection threshold for the FPR=$1\%$ setting from the original paper (marked in red).

\begin{thebibliography}{51}
\providecommand{\natexlab}[1]{#1}
\providecommand{\url}[1]{\texttt{#1}}
\expandafter\ifx\csname urlstyle\endcsname\relax
  \providecommand{\doi}[1]{doi: #1}\else
  \providecommand{\doi}{doi: \begingroup \urlstyle{rm}\Url}\fi

\bibitem[Al-Haj(2007)]{al2007combined}
A.~Al-Haj.
\newblock Combined dwt-dct digital image watermarking.
\newblock \emph{Journal of computer science}, 3\penalty0 (9):\penalty0
  740--746, 2007.

\bibitem[An et~al.(2024)An, Ding, Rabbani, Agrawal, Xu, Deng, Zhu, Mohamed,
  Wen, Goldstein, and Huang]{AnDinRab2024Benchmarking}
B.~An, M.~Ding, T.~Rabbani, A.~Agrawal, Y.~Xu, C.~Deng, S.~Zhu, A.~Mohamed,
  Y.~Wen, T.~Goldstein, and F.~Huang.
\newblock {WAVES}: benchmarking the robustness of image watermarks.
\newblock In \emph{Proc. of Int. Conference on Machine Learning ({ICML})},
  2024.

\bibitem[Bartz and Hu(2023)]{techcompaniespledgewatermark2023}
D.~Bartz and K.~Hu.
\newblock Openai, google, others pledge to watermark ai content for safety,
  white house says.
\newblock
  \url{https://www.reuters.com/technology/openai-google-others-pledge-watermark-ai-content-safety-white-house-2023-07-21/},
  2023.

\bibitem[Bernstein et~al.(2008)]{bernstein2008chacha}
D.~J. Bernstein et~al.
\newblock Chacha, a variant of salsa20.
\newblock In \emph{Workshop record of SASC}, 2008.

\bibitem[Biden(2023)]{Biden2023ExecutiveOrderAI}
J.~R. Biden.
\newblock Executive order on the safe, secure, and trustworthy development and
  use of artificial intelligence.
\newblock
  \url{https://www.whitehouse.gov/briefing-room/presidential-actions/2023/10/30/executive-order-on-the-safe-secure-and-trustworthy-development-and-use-of-artificial-intelligence/},
  2023.

\bibitem[Chen et~al.(2023)Chen, Yu, Ge, Yao, Xie, Wu, Wang, Kwok, Luo, Lu, and
  Li]{chen2023pixartalpha}
J.~Chen, J.~Yu, C.~Ge, L.~Yao, E.~Xie, Y.~Wu, Z.~Wang, J.~Kwok, P.~Luo, H.~Lu,
  and Z.~Li.
\newblock {PixArt-$\alpha$}: Fast training of diffusion transformer for
  photorealistic text-to-image synthesis.
\newblock arXiv:2310.00426, 2023.

\bibitem[Ci et~al.(2024{\natexlab{a}})Ci, Song, Yang, Xie, and
  Shou]{CiSonYan2024wmadapter}
H.~Ci, Y.~Song, P.~Yang, J.~Xie, and M.~Z. Shou.
\newblock {WMAdapter}: Adding watermark control to latent diffusion models.
\newblock arXiv:2406.08337, 2024{\natexlab{a}}.

\bibitem[Ci et~al.(2024{\natexlab{b}})Ci, Yang, Song, and
  Shou]{CiYanSon24RingID}
H.~Ci, P.~Yang, Y.~Song, and M.~Z. Shou.
\newblock {RingID}: Rethinking tree-ring watermarking for enhanced multi-key
  identification.
\newblock In \emph{Proc. of the European Conference on Computer Vision
  ({ECCV})}, 2024{\natexlab{b}}.

\bibitem[Clegg(2024)]{cleggLabelingAIgeneratedImages2024}
N.~Clegg.
\newblock Labeling {{AI-Generated}} images on {{Facebook}}, {{Instagram}} and
  {{Threads}}.
\newblock
  \url{https://about.fb.com/news/2024/02/labeling-ai-generated-images-on-facebook-instagram-and-threads/},
  2024.

\bibitem[Cox et~al.(2007)Cox, Miller, Bloom, Fridrich, and Kalker]{CoxMilBlo07}
I.~Cox, M.~Miller, J.~Bloom, J.~Fridrich, and T.~Kalker.
\newblock \emph{Digital Watermarking and Steganography}.
\newblock Morgan Kaufmann Publishers Inc., San Francisco, CA, USA, 2 edition,
  2007.

\bibitem[{European Union}(2024)]{EU2024AIAct}
{European Union}.
\newblock Artificial intelligence act: Regulation ({EU}) 2024/1689 of the
  european parliament and of the council, 2024.
\newblock
  \url{https://eur-lex.europa.eu/legal-content/EN/TXT/?uri=CELEX:32024R1689}.

\bibitem[{Europol Innovation Lab}(2024)]{Europol2024}
{Europol Innovation Lab}.
\newblock Facing reality? law enforcement and the challenge of deepfakes.
\newblock
  \url{https://www.europol.europa.eu/publications-events/publications/facing-reality-law-enforcement-and-challenge-of-deepfakes},
  2024.

\bibitem[Fernandez et~al.(2023)Fernandez, Couairon, J{\'e}gou, Douze, and
  Furon]{Fernandez2023Stable}
P.~Fernandez, G.~Couairon, H.~J{\'e}gou, M.~Douze, and T.~Furon.
\newblock The stable signature: Rooting watermarks in latent diffusion models.
\newblock In \emph{Proc. of {IEEE} Conference on Computer Vision and Pattern
  Recognition (CVPR)}, 2023.

\bibitem[Gokaslan et~al.(2024)Gokaslan, Cooper, Collins, Seguin, Jacobson,
  Patel, Frankle, Stephenson, and
  Kuleshov]{gokaslan2023commoncanvasopendiffusionmodel}
A.~Gokaslan, A.~F. Cooper, J.~Collins, L.~Seguin, A.~Jacobson, M.~Patel,
  J.~Frankle, C.~Stephenson, and V.~Kuleshov.
\newblock Commoncanvas: Open diffusion models trained on creative-commons
  images.
\newblock In \emph{Proc. of {IEEE} Conference on Computer Vision and Pattern
  Recognition (CVPR)}, 2024.

\bibitem[Goldstein and Grossman(2021)]{goldsteinHowDisinformationEvolved2021}
J.~A. Goldstein and S.~Grossman.
\newblock How disinformation evolved in 2020, 2021.
\newblock
  \url{https://www.brookings.edu/articles/how-disinformation-evolved-in-2020/}.

\bibitem[{Google DeepMind}(Last visit: Nov.\ 2024)]{GoogleSynthID}
{Google DeepMind}.
\newblock {SynthID}: Identifying ai-generated content with {SynthID}.
\newblock \url{https://deepmind.google/technologies/synthid/}, Last visit:
  Nov.\ 2024.

\bibitem[Gunn et~al.(2025)Gunn, Zhao, and Song]{Gunn2024Undetectable}
S.~Gunn, X.~Zhao, and D.~Song.
\newblock An undetectable watermark for generative image models.
\newblock In \emph{International Conference on Learning Representations
  ({ICLR})}, 2025.

\bibitem[Ho et~al.(2020)Ho, Jain, and Abbeel]{HoJaiAbb20}
J.~Ho, A.~Jain, and P.~Abbeel.
\newblock Denoising diffusion probabilistic models.
\newblock In \emph{Advances in Neural Information Processing Systems}, 2020.

\bibitem[{Hugging Face / Diffusers}(Last visit: Nov.\
  2024)]{HuggingFaceWatermarking}
{Hugging Face / Diffusers}.
\newblock Stable diffusion {XL} repository.
\newblock
  \url{https://github.com/huggingface/diffusers/blob/main/src/diffusers/pipelines/stable_diffusion_xl/pipeline_stable_diffusion_xl.py},
  Last visit: Nov.\ 2024.

\bibitem[Kinakh et~al.(2024)Kinakh, Pulfer, Belousov, Fernandez, Furon, and
  Voloshynovskiy]{kinakh2024evaluation}
V.~Kinakh, B.~Pulfer, Y.~Belousov, P.~Fernandez, T.~Furon, and
  S.~Voloshynovskiy.
\newblock Evaluation of security of {ML}-based watermarking: Copy and removal
  attacks.
\newblock In \emph{{IEEE} International Workshop on Information Forensics and
  Security ({WIFS})}, 2024.

\bibitem[Lin et~al.(2014)Lin, Maire, Belongie, Hays, Perona, Ramanan,
  Doll{\'a}r, and Zitnick]{coco}
T.-Y. Lin, M.~Maire, S.~Belongie, J.~Hays, P.~Perona, D.~Ramanan,
  P.~Doll{\'a}r, and C.~L. Zitnick.
\newblock {Microsoft COCO}: Common objects in context.
\newblock In \emph{Proceedings of the European Conference on Computer Vision
  ({ECCV})}, 2014.

\bibitem[Liu et~al.(2022)Liu, Ren, Lin, and Zhao]{LiuRenLin22}
L.~Liu, Y.~Ren, Z.~Lin, and Z.~Zhao.
\newblock Pseudo numerical methods for diffusion models on manifolds.
\newblock In \emph{International Conference on Learning Representations
  ({ICLR})}, 2022.

\bibitem[Liu et~al.(2025)Liu, Song, Ci, Zhang, Wang, Shou, and
  Bu]{controllableregen}
Y.~Liu, Y.~Song, H.~Ci, Y.~Zhang, H.~Wang, M.~Z. Shou, and Y.~Bu.
\newblock Image watermarks are removable using controllable regeneration from
  clean noise.
\newblock In \emph{International Conference on Learning Representations
  ({ICLR})}, 2025.

\bibitem[Lu et~al.(2022)Lu, Zhou, Bao, Chen, Li, and Zhu]{LuZhoBao22}
C.~Lu, Y.~Zhou, F.~Bao, J.~Chen, C.~Li, and J.~Zhu.
\newblock {DPM}-solver: A fast {ODE} solver for diffusion probabilistic model
  sampling in around 10 steps.
\newblock In \emph{Advances in Neural Information Proccessing Systems
  ({NeurIPS})}, 2022.

\bibitem[Lu et~al.(2025)Lu, Zhou, Lu, Zhu, and Kong]{lu2024}
S.~Lu, Z.~Zhou, J.~Lu, Y.~Zhu, and A.~W.-K. Kong.
\newblock Robust watermarking using generative priors against image editing:
  From benchmarking to advances.
\newblock In \emph{International Conference on Learning Representations
  ({ICLR})}, 2025.

\bibitem[Lukas et~al.(2024)Lukas, Diaa, Fenaux, and
  Kerschbaum]{lukDiaFen2024leveraging}
N.~Lukas, A.~Diaa, L.~Fenaux, and F.~Kerschbaum.
\newblock Leveraging optimization for adaptive attacks on image watermarks.
\newblock In \emph{International Conference on Learning Representations
  ({ICLR})}, 2024.

\bibitem[Mokady et~al.(2023)Mokady, Hertz, Aberman, Pritch, and
  Cohen-Or]{MokHerAbe23}
R.~Mokady, A.~Hertz, K.~Aberman, Y.~Pritch, and D.~Cohen-Or.
\newblock Null-text inversion for editing real images using guided diffusion
  models.
\newblock In \emph{Proc. of {IEEE} Conference on Computer Vision and Pattern
  Recognition (CVPR)}, 2023.

\bibitem[Peebles and Xie(2023)]{dit}
W.~Peebles and S.~Xie.
\newblock Scalable diffusion models with transformers.
\newblock In \emph{Proceedings of the IEEE/CVF International Conference on
  Computer Vision (ICCV)}, 2023.

\bibitem[Podell et~al.(2024)Podell, English, Lacey, Blattmann, Dockhorn,
  M{\"u}ller, Penna, and Rombach]{podell2024sdxl}
D.~Podell, Z.~English, K.~Lacey, A.~Blattmann, T.~Dockhorn, J.~M{\"u}ller,
  J.~Penna, and R.~Rombach.
\newblock {SDXL}: Improving latent diffusion models for high-resolution image
  synthesis.
\newblock In \emph{International Conference on Learning Representations
  ({ICLR})}, 2024.

\bibitem[Rombach et~al.(2022)Rombach, Blattmann, Lorenz, Esser, and
  Ommer]{RomBlaLor22}
R.~Rombach, A.~Blattmann, D.~Lorenz, P.~Esser, and B.~Ommer.
\newblock High-resolution image synthesis with latent diffusion models.
\newblock In \emph{Proceedings of the IEEE/CVF Conference on Computer Vision
  and Pattern Recognition (CVPR)}, 2022.

\bibitem[Russakovsky et~al.(2015)Russakovsky, Deng, Su, Krause, Satheesh, Ma,
  Huang, Karpathy, Khosla, Bernstein, Berg, and
  Fei-Fei]{russakovsky2015imagenet}
O.~Russakovsky, J.~Deng, H.~Su, J.~Krause, S.~Satheesh, S.~Ma, Z.~Huang,
  A.~Karpathy, A.~Khosla, M.~Bernstein, A.~C. Berg, and L.~Fei-Fei.
\newblock {ImageNet} large scale visual recognition challenge.
\newblock \emph{International Journal of Computer Vision ({IJCV})},
  115\penalty0 (3):\penalty0 211--252, 2015.

\bibitem[Saberi et~al.(2024)Saberi, Sadasivan, Rezaei, Kumar, Chegini, Wang,
  and Feizi]{saberi2023robustness}
M.~Saberi, V.~S. Sadasivan, K.~Rezaei, A.~Kumar, A.~M. Chegini, W.~Wang, and
  S.~Feizi.
\newblock Robustness of {AI}-image detectors: Fundamental limits and practical
  attacks.
\newblock In \emph{International Conference on Learning Representations
  ({ICLR})}, 2024.

\bibitem[Salimans and Ho(2022)]{SalHo22}
T.~Salimans and J.~Ho.
\newblock Progressive distillation for fast sampling of diffusion models.
\newblock In \emph{International Conference on Learning Representations
  ({ICLR})}, 2022.

\bibitem[Sohl-Dickstein et~al.(2015)Sohl-Dickstein, Weiss, Maheswaranathan, and
  Ganguli]{SohWeiMah15}
J.~Sohl-Dickstein, E.~Weiss, N.~Maheswaranathan, and S.~Ganguli.
\newblock Deep unsupervised learning using nonequilibrium thermodynamics.
\newblock In \emph{Proc. of Int. Conference on Machine Learning ({ICML})},
  2015.

\bibitem[Song et~al.(2021{\natexlab{a}})Song, Meng, and Ermon]{SonMenErm21}
J.~Song, C.~Meng, and S.~Ermon.
\newblock Denoising diffusion implicit models.
\newblock In \emph{International Conference on Learning Representations
  ({ICLR})}, 2021{\natexlab{a}}.

\bibitem[Song and Ermon(2019)]{SonErm19}
Y.~Song and S.~Ermon.
\newblock Generative modeling by estimating gradients of the data distribution.
\newblock In \emph{Advances in Neural Information Proccessing Systems
  ({NeurIPS})}, 2019.

\bibitem[Song et~al.(2021{\natexlab{b}})Song, Sohl-Dickstein, Kingma, Kumar,
  Ermon, and Poole]{SonSohKin21}
Y.~Song, J.~Sohl-Dickstein, D.~P. Kingma, A.~Kumar, S.~Ermon, and B.~Poole.
\newblock Score-based generative modeling through stochastic differential
  equations.
\newblock In \emph{International Conference on Learning Representations
  ({ICLR})}, 2021{\natexlab{b}}.

\bibitem[Tancik et~al.(2020)Tancik, Mildenhall, and Ng]{tancik2020stegastamp}
M.~Tancik, B.~Mildenhall, and R.~Ng.
\newblock Stegastamp: Invisible hyperlinks in physical photographs.
\newblock In \emph{Proceedings of the IEEE/CVF conference on computer vision
  and pattern recognition}, pages 2117--2126, 2020.

\bibitem[Thietke et~al.(2025)Thietke, M{\"u}ller, Lukovnikov, Fischer, and
  Quiring]{Thietke2025Towards}
J.~Thietke, A.~M{\"u}ller, D.~Lukovnikov, A.~Fischer, and E.~Quiring.
\newblock Towards a correct usage of cryptography in semantic watermarks for
  diffusion models.
\newblock In \emph{ICLR Workshop on GenAI Watermarking}, 2025.
\newblock arXiv: 2503.11404.

\bibitem[Wang et~al.(2021)Wang, Lin, Zhao, and Zhu]{wang2021watermark}
R.~Wang, C.~Lin, Q.~Zhao, and F.~Zhu.
\newblock {Watermark Faker}: towards forgery of digital image watermarking.
\newblock In \emph{IEEE International Conference on Multimedia and Expo
  (ICME)}, 2021.

\bibitem[Wen et~al.(2023)Wen, Kirchenbauer, Geiping, and
  Goldstein]{Wen2023TreeRing}
Y.~Wen, J.~Kirchenbauer, J.~Geiping, and T.~Goldstein.
\newblock {Tree-Ring} watermarks: Invisible fingerprints for diffusion images.
\newblock In \emph{Advances in Neural Information Proccessing Systems
  ({NeurIPS})}, 2023.

\bibitem[Xiao et~al.(2022)Xiao, Kreis, and Vahdat]{XiaKreVah22}
Z.~Xiao, K.~Kreis, and A.~Vahdat.
\newblock Tackling the generative learning trilemma with denoising diffusion
  {GAN}s.
\newblock In \emph{International Conference on Learning Representations
  ({ICLR})}, 2022.

\bibitem[Xiong et~al.(2023)Xiong, Qin, Feng, and Zhang]{XioQinFen23Flexible}
C.~Xiong, C.~Qin, G.~Feng, and X.~Zhang.
\newblock Flexible and secure watermarking for latent diffusion model.
\newblock In \emph{Proceedings of the 31st ACM International Conference on
  Multimedia}, 2023.

\bibitem[Yang et~al.(2024{\natexlab{a}})Yang, Ci, Song, and
  Shou]{yang2024steganalysisdigitalwatermarkingdefense}
P.~Yang, H.~Ci, Y.~Song, and M.~Z. Shou.
\newblock Can simple averaging defeat modern watermarks?
\newblock In A.~Globerson, L.~Mackey, D.~Belgrave, A.~Fan, U.~Paquet,
  J.~Tomczak, and C.~Zhang, editors, \emph{Advances in Neural Information
  Proccessing Systems ({NeurIPS})}, 2024{\natexlab{a}}.

\bibitem[Yang et~al.(2024{\natexlab{b}})Yang, Zeng, Chen, Fang, Zhang, and
  Yu]{Yang2024GaussianShading}
Z.~Yang, K.~Zeng, K.~Chen, H.~Fang, W.~Zhang, and N.~Yu.
\newblock {Gaussian Shading}: Provable performance-lossless image watermarking
  for diffusion models.
\newblock In \emph{Proc. of {IEEE} Conference on Computer Vision and Pattern
  Recognition (CVPR)}, 2024{\natexlab{b}}.

\bibitem[Zhang et~al.(2019)Zhang, Xu, Cuesta-Infante, and
  Veeramachaneni]{rivagan}
K.~A. Zhang, L.~Xu, A.~Cuesta-Infante, and K.~Veeramachaneni.
\newblock Robust invisible video watermarking with attention.
\newblock arXiv:1909.01285, 2019.

\bibitem[Zhang and Chen(2023)]{ZhaChe23}
Q.~Zhang and Y.~Chen.
\newblock Fast sampling of diffusion models with exponential integrator.
\newblock In \emph{International Conference on Learning Representations
  ({ICLR})}, 2023.

\bibitem[Zhang et~al.(2018)Zhang, Isola, Efros, Shechtman, and
  Wang]{zhang2018unreasonable}
R.~Zhang, P.~Isola, A.~A. Efros, E.~Shechtman, and O.~Wang.
\newblock The unreasonable effectiveness of deep features as a perceptual
  metric.
\newblock In \emph{Proc. of {IEEE} Conference on Computer Vision and Pattern
  Recognition (CVPR)}, 2018.

\bibitem[Zhao et~al.(2023)Zhao, Bai, Rao, Zhou, and Lu]{ZhaBaiRao23}
W.~Zhao, L.~Bai, Y.~Rao, J.~Zhou, and J.~Lu.
\newblock Uni{PC}: A unified predictor-corrector framework for fast sampling of
  diffusion models.
\newblock In \emph{Advances in Neural Information Proccessing Systems
  ({NeurIPS})}, 2023.

\bibitem[Zhao et~al.(2024)Zhao, Zhang, Su, Vasan, Grishchenko, Kruegel, Vigna,
  Wang, and Li]{ZhaZhaSu2024invisibleimagewatermarksprovably}
X.~Zhao, K.~Zhang, Z.~Su, S.~Vasan, I.~Grishchenko, C.~Kruegel, G.~Vigna, Y.-X.
  Wang, and L.~Li.
\newblock Invisible image watermarks are provably removable using generative
  {AI}.
\newblock In \emph{Advances in Neural Information Proccessing Systems
  ({NeurIPS})}, 2024.

\bibitem[Zhu et~al.(2018)Zhu, Kaplan, Johnson, and Fei-Fei]{Zhu2018hidden}
J.~Zhu, R.~Kaplan, J.~Johnson, and L.~Fei-Fei.
\newblock {HiDDeN}: Hiding data with deep networks.
\newblock In \emph{Proceedings of the European Conference on Computer Vision
  ({ECCV})}, 2018.

\end{thebibliography}
\end{document}